\documentclass[
reprint,
floatfix,
nofootinbib,
amsmath,amssymb,
aps,
prd,
]{revtex4-2}

\usepackage[utf8]{inputenc}
\usepackage{dcolumn}
\usepackage{bm}
\usepackage{amsmath,mathtools,wasysym}
\usepackage{amsthm}
\usepackage{amssymb}
\usepackage{graphicx}
\usepackage[caption=false]{subfig}
\usepackage{float} 
\usepackage{hyperref}
\usepackage{ulem} 
\usepackage{multirow}
\usepackage{physics}
\usepackage[usenames,dvipsnames]{color}
\usepackage{array} 
\usepackage{xfrac}
\usepackage[nolist,nohyperlinks]{acronym} 
\usepackage{cases} 
\usepackage{afterpage}

\usepackage{tikz}
\usetikzlibrary{decorations.pathmorphing}
\usetikzlibrary{decorations.pathmorphing,calc}
\usetikzlibrary{calc,intersections}

\def\l{\left}
\def\r{\right}
\def\thn{\text{\thorn}}
\newcommand{\teuk}{\ensuremath{\hspace{.05cm}\mathaccent\Box{\text{\tiny \textsc{T}}}\hspace{.05cm}}}     

\graphicspath{{plots/}}

\begin{document}
\title{EFT corrections to scalar and vector quasinormal modes of rapidly rotating black holes}
\author{Filipe S. Miguel}
\email{fsdrm2@cam.ac.uk}
\affiliation{
 Department of Applied Mathematics and Theoretical Physics, University of Cambridge, \\
 Wilberforce Road, Cambridge CB3 0WA, United Kingdom\\
}%

\date{\today}

\begin{abstract}
Quasinormal modes characterize the final stage of a black hole merger. In this regime, spacetime curvature is high, these modes can be used to probe potential corrections to general relativity. In this paper, we utilize the effective field theory framework to compute the leading order correction to massless scalar and electromagnetic quasinormal modes. Proceeding perturbatively in the size of the effective field theory length scale, we describe a general method to compute the frequencies for Kerr black holes of any spin. In the electromagnetic case, we study both parity even and parity odd effective field theory corrections, and, surprisingly, prove that the two have the same spectrum. Furthermore, we find that, the corrected frequencies separate into two families, corresponding to the two polarizations of light. The corrections pertaining to each family are equal and opposite. Our results are validated through several consistency checks.
\end{abstract}


\maketitle

\begin{acronym}
\acro{BC}{boundary condition}
\acro{BH}{black hole}
\acro{BL}{Boyer-Lindquist}
\acro{BVP}{boundary value problem}
\acro{EsGB}{Einstein-scalar-Gauss-Bonnet}
\acro{EFT}{effective field theory}
\acro{EH}{Einstein-Hilbert}
\acro{EM}{Einstein-Maxwell}
\acro{EOM}{equations of motion}
\acro{GB}{Gauss-Bonnet}
\acro{GHP}{Geroch-Held-Penrose}
\acro{GR}{general relativity}
\acro{GW}{Gravitational Wave}
\acro{KG}{Klein-Gordon}
\acro{LHS}{left hand side}
\acro{NP}{Newman-Penrose}
\acro{ODE}{ordinary differential equation}
\acro{PDE}{partial differential equation}
\acro{QNM}{\textit{quasi}normal mode}
\acro{RHS}{right hand side}
\end{acronym}

\tableofcontents

\section{Introduction\label{sec:introduction}}

Studying \ac{BH} spacetimes has proven to be a very fruitful endeavor in understanding the merits and limitations of Einstein's theory of \ac{GR}. These powerful astrophysical objects cause spacetime to heavily curve around them, and thus, we expect possible deviations from \ac{GR} to be noticeable there. The no-hair theorems imply that the mathematical description of these objects is remarkably simple. In a vacuum, asymptotically flat spacetime, all stationary \ac{BH} solutions are described by the two-parameter Kerr family of metrics. The mass and angular momentum $\l(M,a\r)$ of the \ac{BH} are sufficient to specify these highly symmetric solutions of the Einstein equations. In the limit $a\rightarrow0$, we obtain the only static, spherically symmetric solution of the \acp{EOM}, known as the Schwarzschild solution. However, real-world scenarios are dynamic, and thus, solutions will necessarily differ from the stationary ones discussed above.\\

The observation of \acp{GW} by the LIGO/VIRGO experiment offers a key opportunity to experimentally test the dynamical evolution of \ac{BH} spacetimes. In particular, by observing the collision between two \acp{BH}, we can understand how generic \ac{BH} solutions evolve into the stationary Kerr spacetime. The collision of two \acp{BH} is characterized by three phases. Initially, during the inspiral phase, the \acp{BH} orbit each other, with motion that can be described using Newtonian physics. Then, during the merger phase, the \acp{BH} effectively collide and form a single \ac{BH}. This phase is challenging to study analytically, and a numerical relativity approach must be adopted. Finally, during the \ac{BH} ringdown, the final \ac{BH} decays into stationarity, 'vibrating' around a Kerr solution. This phase can be modelled by understanding linear perturbations of the \ac{EOM}.\\

The study of the linear perturbations of Schwarzschild was pioneered in~\cite{Regge:1957td}, where the authors derived the Regge-Wheeler equation, which describes odd parity (axial) perturbations of a Schwarzschild spacetime. This work was complemented in~\cite{Zerilli:1970se} with the Zerilli equation, describing even parity (polar) perturbations of Schwarzschild. Concurrently, Teukolsky demonstrated that in a Kerr background, most fields with linear \acp{EOM} are described by two coupled wave-like \acp{ODE}~\cite{Teukolsky:1973ha}. The Schwarzschild limit of this is known as the Bardeen-Press equation~\cite{Bardeen:1973xb}, which does not coincide with the Zerilli or Regge-Wheeler equation. It was later proved that all three equations are related to each other by ordinary and generalized Darboux transformations~\cite{Chandrasekhar:1975zza,Glampedakis:2017rar}. This implies that the polar and axial \ac{EOM} have the same spectrum, a conclusion that does not necessarily generalize to all physical systems, as we will discuss later. Therefore, conveniently, characterising solutions to the Teukolsky equation is sufficient to understand the behaviour of most linear fields in a Kerr/Schwarzschild background.\\

Solving these equations in general is no simple task, however, similar to quantum mechanical systems, we can decompose general solutions into a set of normal modes and frequencies - the eigenvector/eigenvalue pairs of the system. \acp{BH} are, by definition, absorbent objects: the energy of a field in a \ac{BH} background is gradually dissipated into its interior. Vishveshwara~\cite{Vishveshwara:1970zz} was the first to understand that this implies the frequencies must be complex, with the imaginary part controlling the damping of the fields. Thus, these are \acp{QNM} as was coined in~\cite{Press:1971wr}. See~\cite{Nollert:1999ji,Kokkotas:1999bd,Berti:2009kk,Konoplya:2011qq} for a review.\\

Obtaining the values of the \ac{QNM} frequencies is a non-trivial problem. Analytically, we can use a WKB approach to approximate the values of large \ac{QNM} frequencies. In fact, it can be demonstrated that the frequency and damping of these \acp{QNM} are related to the frequency and Lyapunov exponent of circular null geodesics around the \ac{BH}, as seen in~\cite{Mashhoon:1985cya,Dolan:2010wr,Yang:2012he}. This connection will be vital to make progress in section~\ref{subsubsec:EM_eikonal}. However, to achieve accurate values, we must resort to a numerical approach. Following a strategy to compute the spectrum of the molecular hydrogen ion, Leaver devised the first method that yields accurate results across most of the Kerr parameter space~\cite{Leaver:1985ax}. However, the approach loses accuracy in the extremal limit ($a\rightarrow M$). Modern approaches project the \ac{EOM} onto a discrete grid, finding the \ac{QNM} spectrum with a pseudospectral approach, as detailed in~\cite{Dias:2010eu,Miguel:2020uln,Dias:2015nua,Trefethen:2000,Boyd:2000}.\\

Since \ac{QNM} frequencies are complex-valued, the \ac{EOM} cannot be self-adjoint. Additional complexity arises from the fact that, in general, the \acp{QNM} are not a complete basis for the solutions of the \ac{EOM}; for an arbitrary linear perturbation, there will be a residual in the form of a polynomial tail. Nevertheless, we expect that most of the ringdown waveform can be described with \ac{QNM} contributions, see~\cite{Baibhav:2023clw} for details. By fitting the data to a finite sum of the slowest decaying modes, we can obtain the values of the first few \ac{QNM} frequencies. This is known in the literature as \ac{BH} spectroscopy. In a Kerr background, the frequencies are fully specified by the mass and spin of the \ac{BH}. Theoretically, by extracting the real and imaginary parts of a single \ac{QNM} with sufficient precision, we have enough data to fully specify the parameters of the \ac{BH}. Conversely, by detecting two or more modes, we can test \ac{GR} as a theory~\cite{Dreyer:2003bv}. If the detected spectrum is inconsistent with the predictions from \ac{GR}, we need to consider corrections to \ac{GR}. Modern approaches use the \acp{GW} from the inspiral and merger phases to determine the \ac{BH} parameters, and then check their consistency with the ringdown spectrum~\cite{Brito:2018rfr,Ghosh:2021mrv,Silva:2022srr}.\\

To validate \ac{GR} as a theory, we need a concrete framework to parametrize possible deviations. If we knew the full form of the true theory of quantum gravity, we could make physical predictions at low energy scales by \textit{integrating out} the high energy degrees of freedom. This process maps the Lagrangian into an infinite sum of low-energy interactions. Using dimensional analysis, we know that each spacetime derivative must be preceded by some length scale $L$. Thus, we can organize the sum in ascending order of powers of $L$, its scaling dimension, see~\cite{Weinberg:1996kr,Burgess:2003jk,Burgess:2007pt}. Truncating the sum at some order we obtain an \ac{EFT} description of the system. The approximation is valid as long as $L$ is the smallest length scale of the theory. To consider values of $L$ closer to the spacetime length scale, we must include more terms in the sum. \cite{Reall:2021ebq} provides compelling evidence that at a classical level, the difference between the solutions to \ac{EFT} and UV \ac{EOM} is dominated by the scaling dimension of the next term in the series. Remarkably, making simple assumptions about the theory (e.g., gauge invariance and locality), the form of the interactions becomes heavily constrained. Thus, up to a given accuracy, the space of \ac{EFT} is fully parametrized by a set of finitely many coupling constants. This suggests a bottom-up approach. We include in our Lagrangian all the physically reasonable terms up to a given order, extract physical predictions, and use experiments to constrain the coupling constant values. Then, we eliminate all UV theories that lead to coupling parameters that violate these constraints. Thus, even if we do not observe any deviations from \ac{GR}, we can constrain the UV theory space.\\

Calculating the impact of gravitational \ac{EFT} corrections on \acp{QNM} is tricky. Take some stationary \ac{BH} solution of the uncorrected \ac{GR} \ac{EOM}. \acp{QNM} are encoded in dynamical perturbations of this background, i.e. \acp{GW}. Gravitational \acp{EFT} will correct both the stationary background metric and the \acp{GW} around it. Crucially, the correction to the \acp{GW} will have the same order of magnitude as the correction to the stationary background. Thus, the \ac{EFT} corrections to the \acp{QNM} \ac{EOM} depend on the \ac{EFT} correction to the background metric and pure corrections to the linearised \ac{GR} equations. In the Schwarzschild limit, \cite{Cardoso:2018ptl} computed the correction to the background spacetime, and corresponding correction to gravitational \acp{QNM}. More recently, Cano et al. studied the corrections to Kerr \acp{QNM}. In~\cite{Cano:2020cao}, they obtained the correction to the background metric for \acp{BH} with $a \lesssim 0.7 M$, and used the result to obtain the shifts in the \acp{QNM} of a massless scalar field. Then, in~\cite{Cano:2021myl,Cano:2023tmv,Cano:2023jbk}, the same group obtained the corrections to gravitational \acp{QNM} of \textit{not} rapidly rotating Kerr \acp{BH}, under parity even and parity odd \ac{EFT} terms. Crucially, all of the results are perturbative in $a/M$ (where $a$ is the \ac{BH} spin parameter), with the method breaking down for rapidly rotating \acp{BH}. In fact, in~\cite{Cano:2023tmv}, the authors argue that the method breaks down when considering \acp{BH} with $a \gtrsim 0.7 M$. Astrophysical \acp{BH} can have $a/M$ much closer to $1$, thus, it is essential to develop an approach that is non-perturbative in $a/M$. This is fundamentally more difficult, because the \ac{EOM} cannot be separated into radial and angular \acp{ODE}, the \acp{QNM} are eigenvalues of two dimensional \acp{PDE}. Throughout this paper, we develop a framework to perform this calculation by studying similar, albeit simpler, theories. This is in preparation for future work that will generalize the method to the gravitational case. We will study leading order \ac{EFT} corrections to scalar and electromagnetic \acp{QNM} of Kerr \acp{BH} of any spin.\\

We begin by studying the scalar \acp{QNM} of an \ac{EFT} that couples gravity with a massless scalar field $\Phi$. We require that $\Phi$ is small and that, in the $\Phi\rightarrow0$ limit, we recover classical \ac{GR} solutions. This implies that, to leading order, we must restrict to \ac{EFT} corrections that lead to \ac{EOM} that are linear in $\Phi$. Additionally, for simplicity, we restrict to terms that are parity invariant.\footnote{To achieve full generality, we must include parity odd corrections, however, we will argue in section~\ref{subsec:scalar_paraity_odd} that our method can be swiftly adapted to study this case.} After suitable field redefinitions, the leading order \ac{EFT} can be written as follows~\cite{Weinberg:2008hq}:
\begin{equation}
    S = \frac{1}{16 \pi}\int_{\mathcal M}\dd^4 x\sqrt{-g} \l[R-\frac{1}{2}\nabla_a\Phi\nabla^a\Phi\, +L^2 f(\Phi)\,\mathcal G\r]\,,\label{eq:scalar_EFT_action}
\end{equation}
where
\begin{equation}
    \mathcal G := R_{abcd} R^{abcd} -4R_{ab} R^{ab} + R^2\,, \label{eq:define_GB}  
\end{equation}
is the \ac{GB} term, and $L$ is the \ac{EFT} length scale. This is commonly referred to, in the literature, as the \ac{EsGB} theory. By varying the action with respect to $\Phi$, we derive the \ac{EOM}:
\begin{equation}
\Box \Phi = - L^2 f'(\Phi) \mathcal G\,. \label{eq:EOM_EsGB}
\end{equation}

\begin{figure*}[ht]
    \subfloat[\label{subfig:scalar_QNM_line_plot_Re_background}]{\includegraphics[width=0.457\textwidth]{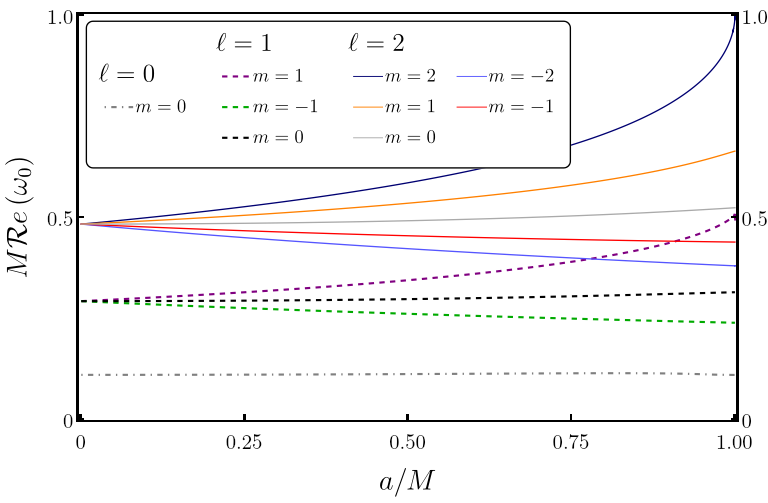}}\hskip 2em
    \subfloat[\label{subfig:scalar_QNM_line_plot_Im_background}]{\includegraphics[width=0.457\textwidth]{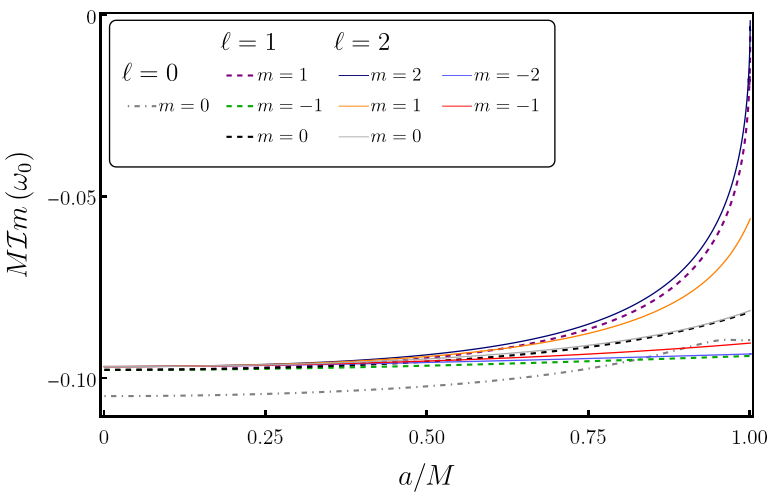}}

    \subfloat[\label{subfig:scalar_QNM_line_plot_Re_correction}]{\includegraphics[width=0.457\textwidth]{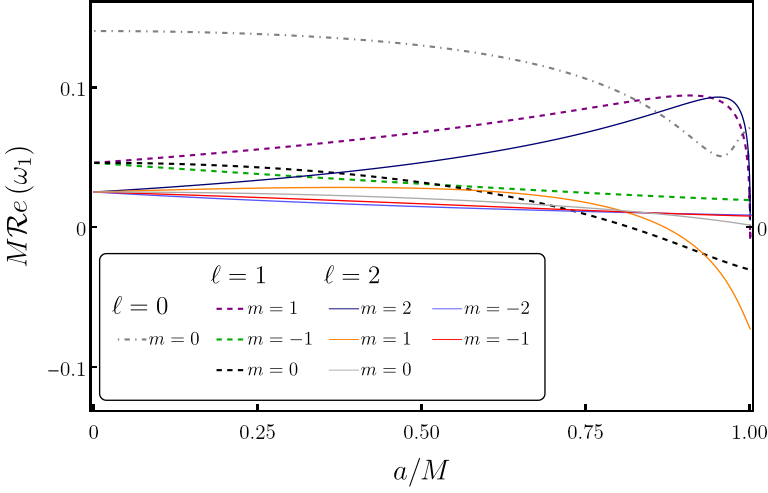}}\hskip 2em
    \subfloat[\label{subfig:scalar_QNM_line_plot_Im_correction}]{\includegraphics[width=0.457\textwidth]{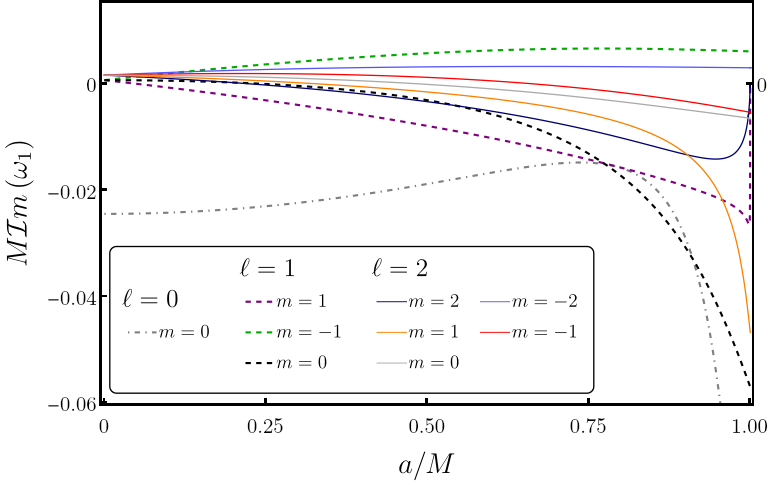}}

\caption{This image summarizes the \ac{EFT} corrections to low $\ell$ \acp{QNM} of a massless scalar field. In panels a) and b) we have the real and imaginary parts of the background \ac{QNM} frequencies, whereas panels c) and d) show the \ac{EFT} correction. To generate these plots, we computed the \acp{QNM} for \acp{BH} with $a \lesssim 0.999 M$ in most cases, or $a \lesssim 0.999975 M$ for $\ell=m\neq0$ modes. We found that, in general, the imaginary part of the background frequency increases with $\ell$. Thus, the lowest lying \acp{QNM} should have $\ell\rightarrow\infty$. This is consistent with the literature, (see table 1 of~\cite{Mamani:2022akq}), but is contrary to the tendency for gravitational \acp{QNM}. As expected, in the Schwarzschild limit, both the background and correction frequencies, are independent of $m$, with the degeneracy being broken for $a>0$. In the near extremal limit, for $\ell=m\neq 0$ modes $M\omega_0 \rightarrow \ell/2$, as predicted in~\cite{Hod:2008zz}. This feature extends to the \ac{EFT} correction: $\omega_1 \rightarrow 0$ as $a\rightarrow M$. \label{fig:scalar_QNM_line_plot}}
\end{figure*}

\begin{figure*}[ht]
    \subfloat[\label{subfig:vector_QNM_line_plot_Re_background}]{
    \includegraphics[width=0.457\textwidth]{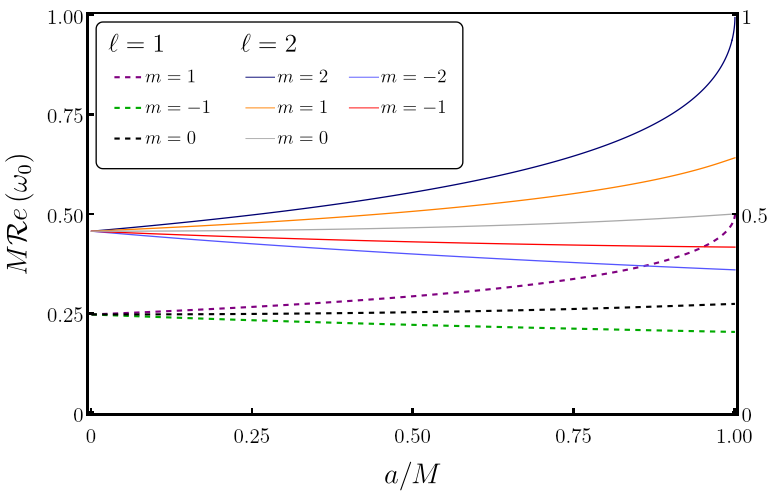}}\hskip 2em
    \subfloat[\label{subfig:vector_QNM_line_plot_Im_background}]{
    \includegraphics[width=0.457\textwidth]{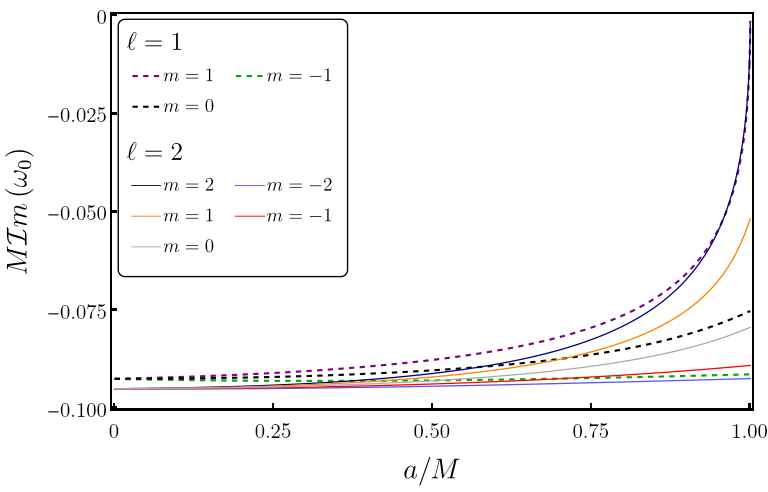}}

    \subfloat[\label{subfig:vector_QNM_line_plot_Re_correction}]{
    \includegraphics[width=0.457\textwidth]{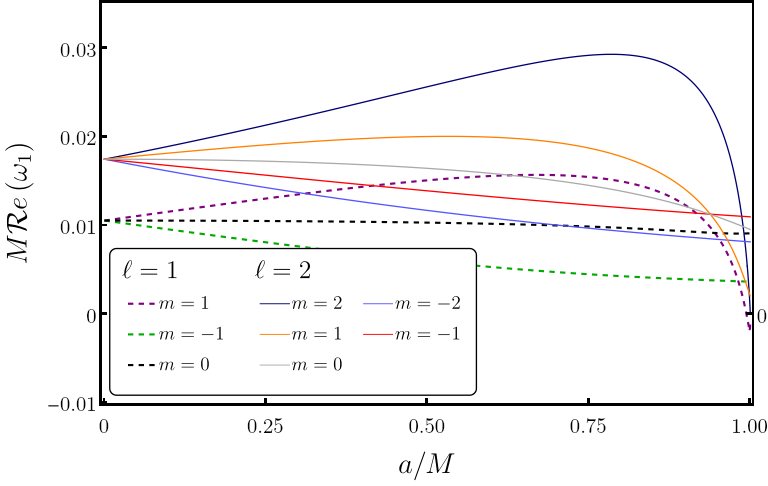}}\hskip 2em
    \subfloat[\label{subfig:vector_QNM_line_plot_Im_correction}]{
    \includegraphics[width=0.457\textwidth]{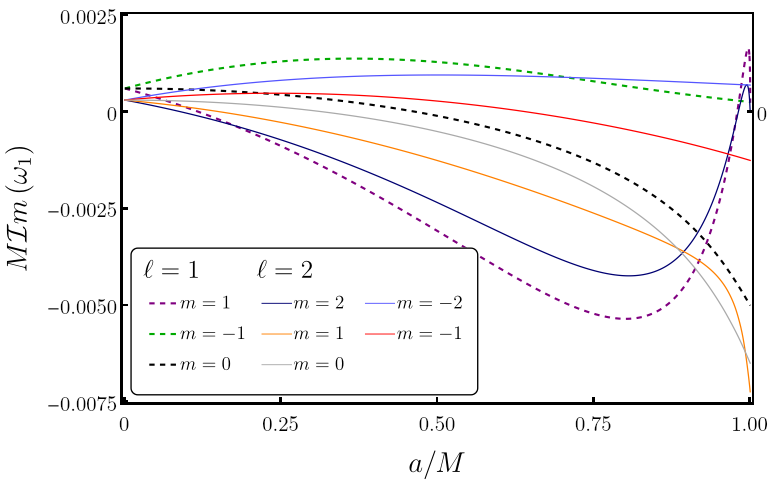}}

\caption{Here, we summarize the \ac{EFT} corrections to the lowest lying electromagnetic \acp{QNM}. The real and imaginary parts of the background \ac{QNM} frequencies can be seen in panels a) and b) whereas the correction can be found in panels c) and d). As expected, in the Schwarzschild limit, modes are independent of $\ell$. Contrary to the scalar case, for most of the parameter space, $\mathcal Im(\omega_0)$ decreases with $\ell$, making the mode with $\ell=1$ the lowest lying mode. This tendency is broken at $a\approx 0.96 M$, where the $\ell=2$ modes become the lowest lying \ac{QNM}. As in the scalar case, for $\ell=m$ modes, in the limit $a \rightarrow M$, we have $M \omega_0\rightarrow \ell/2$. This is in agreement with the prediction in~\cite{Hod:2008zz}. Furthermore, as seen in the scalar case, the corresponding correction obeys $M \omega_1\rightarrow 0$. \label{fig:vector_QNM_line_plot}}
\end{figure*}

Different choices of $f(\Phi)$ have different consequences. If $f(\Phi)$ is constant, the \ac{RHS} vanishes, and we recover the massless \ac{KG} equation. Conversely, if $f(\Phi)\sim \Phi$, the standard \ac{GR} \ac{BH} solutions no longer solve the gravitational \ac{EOM}, resulting in \acp{BH} solutions with scalar hair~\cite{Sotiriou:2013qea}. To avoid this, we must consider second order: $f(\Phi)\sim \Phi^2$. For this theory, $\Phi=0$ solves the scalar \ac{EOM}, and thus, \ac{GR} \acp{BH} are in the solution space. Take $f(\Phi)= \lambda \Phi^2/2$ for some $\mathcal O(1)$ dimensionless coupling parameter $\lambda$, the \ac{EOM} for the scalar field becomes:
\begin{equation}
    \Box \Phi +L^2\,\lambda\,\mathcal G\,\Phi = 0\,. \label{eq:EOM_quadraticEsGB}
\end{equation}
This equation is analogous to the massive \ac{KG} equation $\l(\Box - m^2\r)\Phi = 0$. In a flat background, when $m^2 >0$, $\Phi$ is stable. However, if $m^2 < 0$, the \ac{KG} field is susceptible to the well-known Tachyonic instability. The \ac{GB} coupling acts as a spacetime varying effective mass term, thus, we expect stability to be governed by its sign. \cite{Silva:2017uqg} examined equation~\eqref{eq:EOM_quadraticEsGB} in a Schwarzschild background. The authors discovered that for $\lambda < 0$, the effective mass is positive, and no hairy \ac{BH} solutions can form; all static solutions default to the standard Schwarzschild family of \acp{BH}. On the other hand, for $\lambda > 0$, \acp{BH} with sufficiently small mass ($M$) can be unstable. Indeed, the authors show that for certain mass ranges obeying $\lambda \l(\frac{L}{M}\r)^2\gtrsim0.6$, classical \ac{BH} solutions dynamically evolve into hairy \acp{BH} in a process known as \ac{BH} \textit{scalarization}. When extending this to \acp{BH} with arbitrary spin, we get surprising results. \cite{Dima:2020yac,Herdeiro:2020wei} showed that if the \ac{BH} is sufficiently close to extremality, $\mathcal G$ can be negative, reversing the sign of the effective mass. In this case, the Tachyonic instability emerges for $\lambda<0$. Again, this only occurs for large enough $\abs{\lambda}(L/M)^2$. All of these results occur outside the regime of validity of the \ac{EFT}, and thus, could be suppressed or mitigated by higher-order derivative couplings. Crucially, using the action~\eqref{eq:scalar_EFT_action}, we are only probing corrections to observables that are at most quadratic in $L/M$, or equivalently, linear in $\lambda$. All these papers investigate the instability by time-evolving equation~\eqref{eq:EOM_quadraticEsGB}, and inferring an instability time scale. As far as we know, there are no results in the literature following a standard \ac{QNM} approach.\\

In section~\ref{sec:KG_modes}, we will calculate the corrections to scalar \acp{QNM} for \acp{BH} of any spin, working perturbatively in $L/M$. The frequencies take the form:
\begin{equation}
    \omega = \omega_0\,+\,\lambda\l(\frac{L}{M}\r)^2 \,\omega_1\,.
\end{equation}
For the reader's convenience, we have summarized the main results in figure~\ref{fig:scalar_QNM_line_plot}. We will discuss the results in detail in section~\ref{subsec:KG_results}. For now, we emphasize that these plots reveal non-trivial behaviour for $a/M>0.9$ that would be difficult to capture via an approach perturbative in $a/M$.\\

The electromagnetic sector bears a closer resemblance to the physically relevant gravitational case. In both cases, there are only two degrees of freedom, encoded by two electromagnetic / \ac{GW} polarizations. In the \ac{GR} limit, the two polarizations have the same \ac{QNM} spectrum, i.e. they are isospectral~\cite{Chandrasekhar:1985kt}. Thus, by understanding the corrections to electromagnetic \acp{QNM}, we can develop intuition that will be vital in characterizing the gravitational sector in the future. Consider the following, \ac{EFT} action:
\begin{widetext}
\begin{equation}
    S= \frac{1}{16 \pi}\int_{\mathcal M} \dd^4x\sqrt{-g}\,\l[R-\frac{1}{4}\l(F_{ab} F^{ab} -\frac{1}{3}L^2\l(\lambda_{(e)}\,\mathcal C_{abcd} F^{ab}F^{cd} + \lambda_{(o)}\,\mathcal C_{abcd} \l(\star F\r)^{ab}F^{cd}\r)\r)\r]\,.\label{eq:Maxwell_Weyl_Action}
\end{equation}
\end{widetext}
Here, $L$ is the \ac{EFT} length scale, $F$ represents the electromagnetic tensor, $\mathcal C$ denotes the Weyl tensor, $\star$ denotes the hodge dual operation (see section~\ref{subsec:background_hodge}), and $\lambda_{(e)} / \lambda_{(o)}$ are dimensionless coupling constants. The terms proportional to $\lambda_{(e)}$ and $\lambda_{(o)}$ are, respectively, even / odd under parity transformations, see section~\ref{subsec:background_parity} for details in this. Working perturbatively in the size of $F$, and the \ac{EFT} length scale, we can show that, up to field redefinitions, this is the leading order of the most general \ac{EFT} that couples electromagnetism and gravity, see e.g.~\cite{Davies:2021frz}. The action includes terms up to $\mathcal O(L/M)^2$, thus, as in the scalar case, our results are only valid to that order. Higher order contributions are outside the regime of validity of the \ac{EFT}.\\

Setting $\lambda_{(o)}=0$, the action is invariant under parity transformations. This particular regime has been extensively studied in the literature. \cite{Drummond:1979pp,Daniels:1995yw} focused on the propagation of light in Schwarzschild and Kerr backgrounds, respectively. They concluded that the propagation properties are tied to the photon polarisation, rendering spacetime into a birefringent medium. The observational implications of this have been studied in~\cite{Daniels:1995yw,Dalvit:2000ay,Chen:2016hil,Johnson:2023skw}. \cite{Balakin:2017eur} generalized this result to various algebraically special spacetimes. Notably, for Petrov type-D spacetimes, the authors found two effective metrics that characterize the propagation of each polarization of light. Because \acp{QNM} are related to light propagation in the near extremal limit, we expect that the \ac{QNM} spectrum will no longer be polarization independent. \cite{Chen:2013ysa} confirmed this was the case on a Schwarzschild background by explicitly computing the \acp{QNM} across a wide range of $\lambda$ values.\\

When considering parity invariant \ac{EFT} corrections, we can find a basis for electromagnetic modes, where they are eigenstates of the parity operator. Thus, it is natural, to associate these with the two possible polarizations. We get the parity even / polar polarization and the parity odd / axial polarization. However, when we consider parity odd \ac{EFT} corrections, the parity invariance of the action is broken, and thus, parity even modes will be mixed with parity odd modes, see~\cite{Cano:2021myl}. In section~\ref{sec:EM_modes}, we will consider, both, parity even and parity odd \ac{EFT} corrections. Working perturbatively in $L/M$, we will prove that the shift in the \ac{QNM} spectrum due to parity odd corrections is exactly the same as the one for parity even ones. Furthermore, we find that each background \ac{QNM} frequency has two possible corrections, related by a minus sign. 
\begin{equation}
    \omega^{\pm} = \omega_0 \pm (-1)^{\ell + m +1} \l(\frac{L}{M}\r)^2\abs{\lambda} \,\omega_1\, + \mathcal O \l(\frac{L}{M}\r)^3\,,\label{eq:EM_QNMs_results}
\end{equation}
where:
\begin{equation}
    \abs{\lambda} = \sqrt{\lambda_{(e)}^2 + \lambda_{(o)}^2}\,.
\end{equation}
Cano et. al, obtained similar results for the gravitational case~\cite{Cano:2021myl,Cano:2023tmv,Cano:2023jbk}. The authors found similar mixing between parity even and parity odd corrections. Possibly because the \ac{EFT} corrects the background spacetime, the authors found that $\omega_1^+\neq -\omega_1^-$. The corrected \ac{QNM} frequencies are not centred around $\omega_0$. For the reader's convenience, we summarized the main results in figure~\ref{fig:vector_QNM_line_plot}. As before, the plots reveal non-trivial behaviour for $a/M \gtrsim 0.75$ that would be difficult to capture via a perturbative approach. We will discuss these results in detail in section~\ref{subsubsec:EM_Kerr_results}.\\

In the context of QED in a curved background, when we integrate out the electron degrees of freedom, we get the parity even part of the action~\eqref{eq:Maxwell_Weyl_Action}. In this context, the value of $\lambda_{(e)}$ is known~\cite{Drummond:1979pp}:
\begin{equation}
    \frac{1}{3}\lambda_{(e)} L^2=\frac{1}{1440\pi^2}\l(\frac{M_p^2}{m_e M_\odot}\r)^2M_\odot^2\,\sim\, 4.8 \cdot 10^{-36}M_\odot^2\,.
\end{equation}
Here, $M_p$ is the Planck mass, $m_e$ is the mass of an electron, and $M_\odot$ is the mass of the Sun. For a solar mass \ac{BH}, we anticipate the effects to be extremely small, with corrections only becoming relevant for \acp{BH} of very small masses. However, there might be contributions to $\lambda_{(e)}$ that are unrelated to QED, thus its true value could be much larger. Despite that, this action is still interesting as a toy model for the pure gravity case.\\

In the calculations of this paper, we followed the $(-,+,+,+)$ signature convention. We perform our calculations using geometric units, i.e. $(G_N=c=1)$. This implies that dimensions of any quantity can be expressed as the power of some length scale. Furthermore, we use Latin indices $a,b,c,\cdots$ for abstract index notation, Greek indices $\mu,\nu,\sigma,\cdots$ for quantities in a coordinate basis, and parenthesized Greek indices $(\alpha),(\beta),(\gamma),\cdots$ for quantities in a null tetrad basis.\\

\section{Background Material \label{sec:background}}

\subsection{The Kerr spacetime\label{subsec:kerr}}

Throughout this paper, we will work on a Kerr background. This is an asymptotically flat solution of the vacuum Einstein equations, that describes a rotating \ac{BH}. It is fully specified by two parameters, $(M, a)$, representing the mass and angular momentum of the \ac{BH}, respectively. We will work with modified \ac{BL} coordinates $(t,r,x,\phi)$, that differ from standard \ac{BL} coordinates by the definition $x=\cos\theta$. The line element reads:
\begin{multline}
    \dd s^ 2= -\dd t^2\,-\,\frac{2 M r \l(\dd t - a \,\Delta_x\,\dd \phi\r)^2}{\Sigma}\\
    +\frac{\Sigma\,\dd r^2}{\Delta_r}\,+\,\frac{\Sigma\,\dd x^2}{\Delta_x}\,+\,\Delta_x \l(r^2+a^2\r)\dd \phi^2,\label{eq:line_element_kerr}
\end{multline}
where:
\begin{equation}
\begin{aligned}
    &\Sigma=r^2+ a^2x^2,\\
    &\Delta_r=r^2-2 M r + a^2=(r-r_+)(r-r_-),\\
    &\Delta_x=1-x^2,\\
    &r_\pm =  M\pm \sqrt{M^2-a^2},\\[.2em]
    &t \in \mathbb R, \quad r\in (r_+, +\infty),\quad x \in (-1,1),\quad\phi\in (0, 2\pi).
\end{aligned}
\end{equation}

\begin{figure}[ht]
\resizebox{\columnwidth}{!}{%
    \begin{tikzpicture}
    \node (I)    at (2.7,0)   {};
    \path  
      (I)  +(90:2.7)  coordinate[label=45:{$i^+$}] (Itop)
           +(-90:2.7) coordinate[label=-90:$i^-$]  (Ibot)
           +(0:2.7)   coordinate[label=0:$i^0$]    (Iright)
           +(180:2.7) coordinate                   (Ileft)
           ;
    \draw [thick] (Iright) -- 
              node[midway, above right]    {$\cal{J}^+$}
              node[midway, below, sloped]  {$r\rightarrow +\infty\quad t\rightarrow+\infty$}
          (Itop) --
              node[midway, above left] {$\mathcal{H}^+$}
              node[midway, below, sloped] {$r\rightarrow r_+\quad t\rightarrow+\infty$}
          (Ileft) -- 
              node[midway, below left] {$\mathcal{H}^-$}
              node[midway, above, sloped] {$r\rightarrow r_+\quad t\rightarrow-\infty$}
          (Ibot) --
              node[midway, below right]    {$\cal{J}^-$}
              node[midway, above, sloped]  {$r\rightarrow +\infty\quad t\rightarrow-\infty$}
          (Iright) -- cycle;
    \draw[fill=white] (Itop) circle (2.5pt);
    \draw[fill=white] (Ibot) circle (2.5pt);
    \draw[fill=white] (Iright) circle (2.5pt);
    \end{tikzpicture}
}
\caption{Penrose diagram for the exterior of a Kerr \ac{BH}.}
\label{fig:penrose_diagram}
\end{figure}
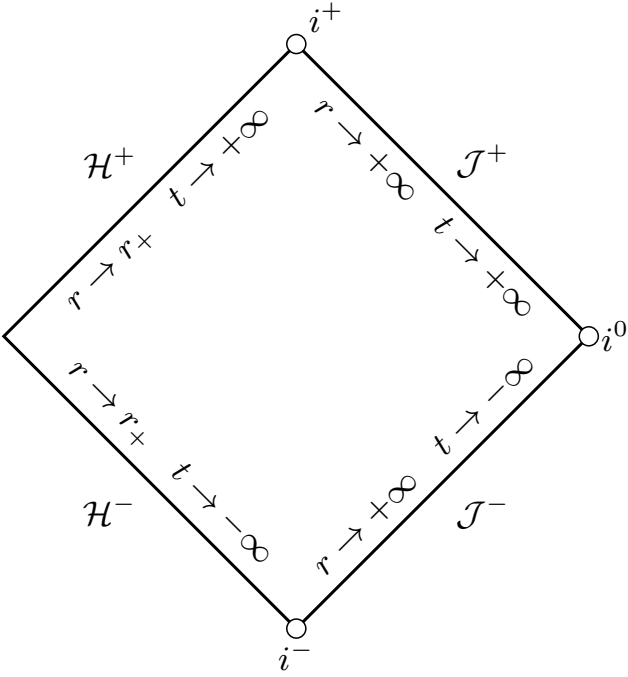

We will restrict our study to \ac{BH}s such that $M\in \mathbb R^+$. In the limit $a\rightarrow0$, the spacetime is spherically symmetric, and the line element reduces to the Schwarzschild metric. On the other hand, at $a = M$, we have the extremal limit. If we allowed $a>M$, $\Delta_r$ would have no real roots, and the metric would describe a naked singularity. Therefore, we restrict to $0<a<M$. The causal structure of the \ac{BH} exterior can be seen in figure~\ref{fig:penrose_diagram}. $\mathcal H ^+$ and $\mathcal H ^-$ are the \ac{BH} and white hole horizons, respectively, whereas $\mathcal J^{\pm}$ represents future and past null infinity. Using Kerr coordinates, the spacetime can be analytically extended to the \ac{BH} interior (see~\cite{Wald:1984rg}). At $r=r_-$ there is a Cauchy horizon ($\mathcal {CH}$). The surface gravity $\kappa$ of $\mathcal H^+$ is:
\begin{equation}
    \kappa=\frac{1}{2\,a}\Delta_r'(r_+)\,\Omega_+ = \frac{r_+-r_-}{2\l(a^2+ r_+^2\r)}\,,\label{eq:surface_gravity}
\end{equation}
where:
\begin{equation}
    \Omega_+=\frac{a}{a^2+r_+^2}\,.\label{eq:omega_definition}
\end{equation}
This parameter describes the angular velocity of $\mathcal H^+$  as measured by an observer at infinity. In the extremal limit, $r_-\rightarrow r_+$, and $\kappa\rightarrow 0$. In this limit, we no longer have a Cauchy horizon, and $\mathcal H$ becomes a doubly degenerate Killing horizon. This behavior makes calculations in the extremal limit increasingly difficult.\\

Finally, we define the dimensionless angular momentum:
\begin{equation}
    J = \frac{a}{M} \label{eq:define_J}
\end{equation}
If we replace $a$ by $J$ in the metric, a change $M\rightarrow \alpha M$ is equivalent to the coordinate transformation $r\rightarrow r/\alpha\,,t\rightarrow t/\alpha$. Consequently, all dimensionless background quantities must be independent of $M$. $J$ is the only physically relevant parameter.

\subsection{The GHP formalism \label{subsec:GHP_intro}}

The \ac{GHP} formalism~\cite{Geroch:1973am} is a more covariant version of the \ac{NP} approach~\cite{Newman:1961qr,Penrose:1985bww}. This formalism is useful to derive compact \ac{EOM} for various fields, especially when working within the context of algebraically special spacetimes.\\

As in the \ac{NP} formalism, the core idea is to project all relevant quantities and equations into a complex null tetrad basis $e^a{}_{(\alpha)}$. To define this structure, we begin by selecting two null curve congruences, $\gamma_l$ and $\gamma_n$, with tangents $l^a$ and $n^a$ satisfying $l\cdot n=-1$. Subsequently, we choose two spacelike vector fields $X^a$ and $Y^a$ orthogonal to both $l^a$ and $n^a$, such that:
\begin{equation}
\begin{aligned}
    &X_a X^a=Y_a Y^a=1/2\,,\\
    &X_a Y^a=0\,.\label{eq:defineGHP_XY}
\end{aligned}
\end{equation}
These are then used to define complex vector fields $m^a$ and $\bar{m}^a$:
\begin{equation}
\begin{aligned}
    m^a&:=X^a + i Y^a\,,\\
    \bar{m}^a&:=X^a - i Y^a\,.
\end{aligned}
\end{equation}
Throughout this paper, the bar $(\,\bar{}\,)$ operation will denote complex conjugation. As a direct consequence of equation~\eqref{eq:defineGHP_XY}, we find $m^2= \bar m^2 =0$ and $m\cdot \bar m =1$. Finally, the null tetrad will be $e^a{}_{(\alpha)}:=(l^a, n^a, m^a, \bar m^a)_{(\alpha)}$. The normalization conditions reduce to:
\begin{equation}
    g_{a b}\,e^a{}_{(\alpha)}\,e^b{}_{(\beta)}\,= \,\hat \eta_{(\alpha)(\beta)}\,,\label{eq:normalization_NP}
\end{equation}
with:
\begin{equation}
    \hat \eta_{(\alpha)(\beta)}=\begin{pmatrix}
    0&-1&0&0\\
    -1&0&0&0\\
    0&0&0&1\\
    0&0&1&0
    \end{pmatrix}_{(\alpha)(\beta)}\,.
\end{equation}

After fixing the directions of $\gamma_l$ and $\gamma_n$, there are still multiple consistent choices of $e_{(\alpha)}$, related to each other by a continuous transformation. We understand this redundancy as the gauge freedom of the formalism. On one hand, we may reparametrize $\gamma_l$ and $\gamma_n$, such that:
\begin{equation}
    l^a\rightarrow r\,l^a, \quad n^a\rightarrow \frac{1}{r}\,n^a\,, \label{eq:r_transformation_ln}
\end{equation}
where $r>0$ may vary across the spacetime. On the other hand, we can rotate the spacelike basis used to define $m^a$ by a real angle $\theta$, which, again, need not be a constant:
\begin{equation}
    m^a\rightarrow e^{i \theta}\,m^a, \quad \bar m^a\rightarrow e^{-i \theta}\,\bar m^a\,. \label{eq:theta_transformation_m}
\end{equation}

The two transformations can be encoded in a single scalar field $\lambda=\sqrt{r} e^{i \theta/2}$:
\begin{equation}
\begin{aligned}
    &l^a\rightarrow \lambda \bar\lambda\,l^a, \quad &&n^a\rightarrow \lambda^{-1} \bar\lambda^{-1}\,n^a, \\
    &m^a\rightarrow \lambda \bar\lambda^{-1}\,m^a, \quad &&\bar m^a\rightarrow \lambda^{-1} \bar\lambda\,\bar m^a\label{eq:lambda_transformation_tetrad}
\end{aligned}
\end{equation}

In general, if under~\eqref{eq:lambda_transformation_tetrad}, $\eta$ transforms as:
\begin{equation}
    \eta\rightarrow \lambda^p \bar \lambda^q \eta\,,\label{eq:lambda_transformation_general}
\end{equation}
we say that $\eta$ is \ac{GHP}-covariant, with type (or weight) ${p,q}$. We deduce that the tetrad components have type:
\begin{equation}
\begin{aligned}
    l^a:\quad&\{1,1\}\,,\\
    n^a:\quad&\{-1,-1\}\,,\\
    m^a:\quad&\{1,-1\}\,,\\
    \bar m^a:\quad&\{-1,1\}\,.\label{eq:tetrad_weights}
\end{aligned}
\end{equation}

To preserve \ac{GHP} covariance, we cannot add two quantities with different weights. Conversely, multiplying a $\{p, q\}$ scalar by a $\{p', q'\}$ scalar is allowed, yielding a $\{p+p', q+q'\}$ scalar. To further simplify the notation, we introduce the prime $(\,'\,)$ operation. This effectively interchanges instances of $l^a$ and $n^a$, and swaps instances of $m^a$ and $\bar m^a$:
\begin{equation}
    \l(l^a\r)' = n^a, \mkern4mu \l(n^a\r)' = l^a, \mkern4mu \l(m^a\r)' = \bar m^a, \mkern4mu \l(\bar m^a\r)' = m^a\,.
\end{equation}
If $\eta$ is of type $\{p,q\}$, then $\eta'$ will be of type $\{-p,-q\}$. Conversely, $\bar\eta$ will be of type $\{q,p\}$. Both operations commute, and $\l(\eta'\r)' = \bar{\bar\eta} = \eta$. It is also useful to define the \textit{spin} and \textit{boost} weights of $\eta$ as $s=\frac{1}{2}(p-q)$ and $p=\frac{1}{2}(p+q)$, respectively. Under complex conjugation, the spin weight's sign is flipped, whereas the boost weight remains unchanged. Under a $(\,'\,)$ transformation, both quantities have their signs inverted.\\

To make progress, we must define the \ac{GHP} equivalent of the \ac{NP} spin coefficients~\cite{Newman:1961qr}. These can be divided into two groups\footnote{These quantities differ from those in~\cite{Geroch:1973am} by a $(-)$ sign. This is because we follow the $(-,+,+,+)$ signature convention, whereas \cite{Geroch:1973am} use a $(+,-,-,-)$ signature. See appendix E of~\cite{Frolov:1998wf} for further details.}:
\begin{equation}
\begin{aligned}
    &\kappa := -m^b l^a\nabla_a l_b\,,&&\sigma :=-m^b m^a\nabla_a l_b\,,\\
    &\rho := -m^b \bar m^a\nabla_a l_b\,,&&\tau :=-m^b n^a\nabla_a l_b\,,
\end{aligned} \label{eq:pq_ghp_scalars}
\end{equation}
and
\begin{equation}
\begin{aligned}
    &\beta:=-1/2\l(n^b m^a \nabla_a l_b - \bar m^b m^a\nabla_a m_b\r)\,,\\
    &\varepsilon:=-1/2\l(n^b l^a \nabla_a l_b - \bar m^b l^a\nabla_a m_b\r)\,,\label{eq:not_pq_ghp_scalars}
\end{aligned}
\end{equation}
together with the corresponding primed versions. One can show that the scalars in~\eqref{eq:pq_ghp_scalars} are \ac{GHP}-covariant, with types:
\begin{equation}
    \kappa: \{3,1\}\quad \sigma: \{3,-1\} \quad \rho: \{1,1\}\quad \tau: \{1,-1\}.
\end{equation}
Conversely, because $\lambda$ is spacetime varying, the scalars in~\eqref{eq:pq_ghp_scalars} do not transform according to~\eqref{eq:lambda_transformation_general}, gaining an additional affine contribution:
\begin{equation}
\begin{aligned}
    \beta &\rightarrow \lambda \bar\lambda^{-1}\,\beta + \bar \lambda^{-1}\,m^a \nabla_a \lambda\,,\\
    \varepsilon &\rightarrow \lambda \bar\lambda\,\varepsilon + \bar \lambda\,l^a \nabla_a \lambda\,.\label{eq:affine_transf_spin_coefs}
\end{aligned}
\end{equation}

A similar problem arises when taking the covariant derivative of \ac{GHP}-covariant quantities. An affine contribution will stem from the derivatives of $\lambda$ and $\bar \lambda$. Consider $l^a\nabla_a A_{(l)}:=l^a\nabla_a \l(l^b A_b\r)$. Under~\eqref{eq:lambda_transformation_tetrad} we obtain:
\begin{equation}
    l^a\nabla_a A_{(l)} \rightarrow \lambda^2 \bar \lambda^2\,l^a \nabla_a A_{(l)} + \lambda \bar \lambda\,A_{(l)}\, l^a \nabla_a\l(\lambda \bar \lambda\r)\,. \label{eq:affine_transf_spin_scalars}
\end{equation}

Comparing equations~\eqref{eq:affine_transf_spin_coefs} and~\eqref{eq:affine_transf_spin_scalars}, we deduce that for the quantity $\l(l^a\nabla_a- \varepsilon - \bar\varepsilon\r) A_{(l)}$, the affine contributions are cancelled. Thus, this quantity is \ac{GHP}-covariant, with weight ${1+1,1+1}$. Therefore, to preserve \ac{GHP} covariance under differentiation, we need to supplement the derivative operators with appropriate combinations of the spin coefficients. This is entirely analogous to the introduction of Christoffel symbols in the definition of \textit{standard} covariant derivatives. We define four derivative operators:
\begin{alignat}{2}
    \thn\,\eta &= \l(l^a\nabla_a -  p\,\varepsilon - q\,\bar\varepsilon\r)\,\eta:\quad\quad&&\{1,1\}\,,\\
    \thn'\,\eta &= \l(n^a\nabla_a +  p\,\varepsilon' + q\,\bar\varepsilon'\r)\,\eta:\quad\quad&&\{-1,-1\}\,,\\
    \eth\,\eta &= \l(m^a\nabla_a +  p\,\beta - q\,\bar\beta'\r)\,\eta: \quad\quad&&\{1,-1\}\,,\\
    \eth'\,\eta &= \l(\bar m^a\nabla_a -  p\,\beta' + q\,\bar\beta\r)\,\eta: \quad\quad&&\{-1,1\}\,,
\end{alignat}
where the numbers on the right denote the \ac{GHP} weight. $\thn$, $\thn'$ respectively raise and lower the boost-weight, leaving the spin-weight unchanged. Conversely, $\eth$ and $\eth'$ act as raising and lowering operators for the spin-weight, leaving the boost-weight unchanged. Finally, we can define a new (doubly) covariant derivative operator that transforms covariantly under both coordinate and tetrad transformations:
\begin{equation}
    \Theta_a :=-\,n_a\,\thn\,-\,l_a \thn'\,+\,\bar m_a \eth\,+\,m_a\eth'\,.\label{eq:define_covariant_der_theta}
\end{equation}
This operator obeys a Leibniz rule. For $\alpha, \beta$ \ac{GHP}-covariant quantities:
\begin{equation}
    \Theta_a\l(\alpha \beta\r) = (\Theta_a\alpha)\beta \,+\, \alpha (\Theta_a \beta)\,.\label{eq:leibniz_theta}
\end{equation}

In 4 dimensions, the algebraic symmetries of the Weyl tensor allow for 10 degrees of freedom. Projecting it into the null tetrad, we can encode them in 5 complex scalars:
\begin{equation}
\begin{aligned}
    \psi_0 &:= C_{1313} = \psi_4':&&\{4,0\},\\
    \psi_1 &:= C_{1213} = \psi_3':&&\{2,0\},\\
    \psi_2 &:= C_{1342} = \psi_2':&&\{0,0\},\\
    \psi_3 &:= C_{2124} = \psi_1':&&\{-2,0\},\\
    \psi_4 &:= C_{2424} = \psi_0':&&\{-4,0\},
\end{aligned}\label{eq:define_NP_scalars}
\end{equation}
these are usually known as the \ac{NP} scalars. Similarly, the 6 components of the Maxwell tensor, can be encoded in 3 complex scalars:
\begin{equation}
\begin{aligned}
    \phi_0&=l^a m^b F_{ab} :&&\{2,0\}\\
    \phi_1&=\frac{1}{2}\l(l^a n^b-m^a \bar m^b\r) F_{ab}:&&\{0,0\}\\
    \phi_2&=\bar m^a n^b F_{ab}:&&\{-2,0\}
\end{aligned}\label{eq:define_maxwell_scalars}
\end{equation}
with $\phi_i'=-\phi_{2-i}$.\\

In certain cases, the algebraic structure of the Weyl tensor greatly simplifies the formalism. We say that a null vector is tangent to a principal null direction of the Weyl tensor if it satisfies the following~\cite{Wald:1984rg}:
\begin{equation}
    k^b k^ c k_{[e}C_{a]bc[d}k_{f]}=0\,.\label{eq:principal_null_direction_deffinition}
\end{equation}
In general, a spacetime has 4 principal null directions, however, in certain instances, they may overlap. When they coincide in two pairs of two, we state that the spacetime is of Petrov type-D. This is the case for Kerr \acp{BH}. By taking $l^a$ and $n^a$ to be tangent to these null directions, many of the quantities defined above vanish. In fact, for a vacuum spacetime, we can use the \textit{Goldberg-Sachs theorem} to establish that:
\begin{equation}
\begin{aligned}
    \psi_0&=\psi_1=\psi_3=\psi_4=0\,,\\
    \kappa&=\kappa'=\sigma=\sigma'=0\,.
\end{aligned}\label{eq:type_D_simplifications}
\end{equation}
Thus, the only non vanishing \ac{GHP} scalars are $\psi_2,\,\rho,\,\rho',\,\tau$ and $\tau'$. The Kinnersley tetrad is an example that satisfies this condition~\cite{Kinnersley:1969zza,Chandrasekhar:1985kt}:
\begin{equation}
\begin{aligned}
    l^\mu&=\l(\frac{a^2+r^2}{\Delta_r},1,0,\frac{a}{\Delta_r}\r)^\mu,\\
    n^\mu&=\frac{1}{2 \Sigma}\l(a^2+r^2,-\Delta ,0,a\r)^\mu,\\
    m^\mu&=\frac{i}{\sqrt{2}\l(r + i a x\r)} \l(a \sqrt{\Delta_x},0,i \sqrt{\Delta_x},\frac{1}{\sqrt{\Delta_x}}\r)^\mu,\\
    \bar m^\mu&=\frac{-i}{\sqrt{2}\l(r - i a x\r)} \l(a \sqrt{\Delta_x},0,-i \sqrt{\Delta_x},\frac{1}{\sqrt{\Delta_x}}\r)^\mu.
\end{aligned}\label{eq:kerr_null_tetrad}
\end{equation}

\subsection{Master wave equation \label{subsec:master_equation}}

Teukolsky first proved that the \ac{EOM} of most fields in a Kerr background can be described by a single wave equation~\cite{Teukolsky:1973ha}. Following the approach in~\cite{Aksteiner:2010rh}, we will express this result in a more covariant way. We start by defining a generalized d'Alembert operator~\cite{Aksteiner:2010rh}:
\begin{equation}
    \teuk_p = \l(\Theta^a+p B^a\r)\l(\Theta_a+p B_a\r)\,,
\end{equation}
where
\begin{equation}
    B^a=\rho\,n^a -  \tau\,\bar m^a\,. \label{eq:define_B}
\end{equation}
and we defined $\Theta_a$ in equation~\eqref{eq:define_covariant_der_theta}. When acting on a $\{0,0\}$ qunatity, $\Theta_a$ reduces to $\nabla_a$, thus $\teuk_0$ is the standard d'Alembert operator $\Box$. The Teukolsky master equation is a generalization of this that acts on $\psi^{(s)}$, a spin-s field with weight $\{2s,0\}$ (see table~\ref{tab:meaning_psi_s}). The master Teukolsky equation reads~\cite{Aksteiner:2010rh}:
\begin{equation}
    \mathcal O_s\,\psi^{(s)}:=\frac{1}{2}\l(\teuk_{2s}- 4 s^ 2\psi_2\r)\psi^{(s)}=0\,.\label{eq:master_teukolsky}
\end{equation}
Using the Kinnersley tetrad (equation~\eqref{eq:kerr_null_tetrad}), this equation can be written:
\begin{widetext}
\begin{equation}
    \mathcal O_s\,\psi^{(s)}=\l((\thn -2 s\rho -\bar \rho)(\thn'-\rho')- (\eth-2 s\tau-\bar \tau')(\eth'-\tau') - (2s^2-3s+1)\psi_2\r)\psi^{(s)}=0\,.\label{eq:teuk_expanded}
\end{equation}
\end{widetext}

Now, to separate the resulting \ac{PDE}, we follow the approach in~\cite{Teukolsky:1973ha}. We make the \textit{Ansatz}:
\begin{equation}
    \psi^{(s)}(t,r,x,\phi)= S(t,\phi) \, \Psi_s(r,x)\,,\label{eq:sep_t_phi}
\end{equation}
where
\begin{equation}
    S(t,\phi):=e^{-i\l(\omega t -m \phi\r)}\,, \label{eq:define_S}
\end{equation}
and $m \in \mathbb Z$, ensuring periodic \acp{BC} are met for $\phi$. Substituting this in the Teukolsky equation, we obtain:
\begin{equation}
    \mathcal O_{s, \omega}\,\Psi_s:=\frac{1}{S}\mathcal O_s \psi^{(s)}\,=0,\label{eq:teukolsky_OL}
\end{equation}
where $\mathcal O_{s, \omega}$ is an operator that only depends on $(r,x)$. Finally, taking $\Psi_s=R_s(r) \Theta_s(x)$, we can separate equation~\eqref{eq:teukolsky_OL} as the sum of a radial and an angular \ac{ODE} operator~\cite{Teukolsky:1973ha,Leaver:1985ax,Borissov:2009bj}:
\begin{equation}
    \mathcal O_{s, \omega}\,\Psi_s=\frac{1}{\Sigma(r,x)}\l(L_{r,s}+L_{x,s} \r)\l(\Theta_s(x)R_s(r)\r) = 0\,,\label{eq:teukolsky_rx}
\end{equation}
where:
\begin{widetext}
\begin{align}
    L_{x,s} \Theta_s(x) &:= \dv{x} \Delta_x \dv{x}\Theta_s(x) +\l(s-\frac{(m+sx)^2}{\Delta_x} - 2a\omega s x+a^2\omega^2x^2 +  {}_s \mathcal A_{\ell m}\r)\Theta_s(x)=0\,,\label{eq:master_teukolsky_angular}\\[1em]
    L_{r,s}  R_s(r) &:= \Delta_r^{-s}\dv{r} \Delta_r^{s+1} \dv{r}R_s(r)+ \l(\frac{K^2-i s\Delta_r'K}{\Delta_r} + 4is\omega r -{}_s \Lambda_{lm}\r)R_s(r)=0\,,\label{eq:master_teukolsky_radial}
\end{align}
\end{widetext}
with:
\begin{equation}
\begin{aligned}
    K(r)&= \l(a^2+r^2\r) \l(\omega -m \frac{a}{a^2+r^2}\r)\,,\\
    {}_s \Lambda_{lm}&= {}_s \mathcal A_{\ell m} - a^2 \omega ^2 + 2 a m \omega\,,
\end{aligned}
\end{equation}
and ${}_s \mathcal A_{\ell m}$ is a separation constant. In the Schwarzschild limit, regular solutions of equation~\eqref{eq:master_teukolsky_angular} are the spin-weighted spherical harmonics, with ${}_s \mathcal A_{\ell m} \rightarrow l(l+1)-s(s+1)$, see~\cite{Newman:1966ub}. When we turn on rotation, the equation depends on $\omega$, and we no longer know of a closed-form solution. Regular solutions in this regime are known as spin-weighted \textit{spheroidal} harmonics. \\

The singular structure of equations~\eqref{eq:master_teukolsky_angular} and~\eqref{eq:master_teukolsky_radial} is identical. They both have two regular singular points at finite values of their domain ($x=\pm1$ for~\eqref{eq:master_teukolsky_angular} and $r=r_\pm$ for~\eqref{eq:master_teukolsky_radial}), and an irregular singular point at infinity. In fact, through suitable reparametrizations, they can be both expressed as a confluent Heun equation, see~\cite{Borissov:2009bj,Fiziev:2009wn}.

\begin{table}
\begin{tabular}{r|r|l}
$s\hspace{.4em}$&$\psi^{(s)}\hspace{.4em}$& \\\hline\hline
0\hspace*{.4em}&$\Phi\hspace{.4em}$&Scalar field\\\hline
$\frac{1}{2}$\hspace*{.4em}&$\chi_0\hspace{.4em}$&\multirow{ 2}{*}{Dirac spinors}\\[.1em]
$-\frac{1}{2}$\hspace*{.4em}&$\psi_2^{-\sfrac{1}{3}}\,\chi_1\hspace{.4em}$&\\\hline
$1$\hspace*{.4em}&$\phi_0\hspace{.4em}$&\multirow{ 2}{*}{Maxwell fields}\\
$-1$\hspace*{.4em}&$\psi_2^{-\sfrac{2}{3}}\,\phi_2\hspace{.4em}$&\\\hline
$2$\hspace*{.4em}&$\psi_0{}^{(1)\hspace{.4em}}$&\multirow{ 2}{*}{\acp{GW}}\\
$-2$\hspace*{.4em}&$\psi_2^{-\sfrac{4}{3}}\,\psi_4{}^{(1)}\hspace{.4em}$&\\\hline
\end{tabular}
\caption{Physical meaning of the fields that solve equation~\eqref{eq:master_teukolsky}\label{tab:meaning_psi_s}.}
\end{table}

\subsection{Quasinormal Modes\label{subsec:QNMs_intro}}

\acp{QNM} are the natural basis to encode the late time behaviour of linear fields in a \ac{BH} spacetime. Because the system is naturally dissipative, the corresponding frequencies are complex valued. To formally define these modes, the standard approach is to introduce hyperboloidal coordinates in a conformally compactified version of spacetime. \acp{QNM} are solutions of a given wave equation, with defined frequency, that are regular at $\mathcal H^+$ and $\mathcal J^+$, see~\cite{Ansorg:2016ztf,PanossoMacedo:2019npm, Ripley:2022ypi}. Physically, this is the statement that \acp{QNM} are waves that are ingoing through $\mathcal H^+$ and outgoing at $\mathcal J^+$. For simplicity, in this paper, we will work with a constant $t$ slicing of spacetime, and derive the \acp{BC} for the radial function that encode the appropriate ingoing / outgoing behaviour of the full solution. We will specialize in \ac{QNM} solutions of the Teukolsky equation, (see~\eqref{eq:master_teukolsky}).\\

A Frobenius analysis of the radial \ac{EOM} near $r=r_+$ shows that the radial function can be decomposed as the sum of an ingoing and an outgoing mode~\cite{Teukolsky:1973ha,Leaver:1985ax}:
\begin{equation}
    R_s(r) = \mathcal A(\omega) R_{\text{out},r_+}(r) + \mathcal B(\omega) R_{\text{in},r_+}(r)\,,
\end{equation}
where:
\begin{equation}
\begin{aligned}
    R_{\text{out},r_+} &= \l( r-r_+\r)^{i\rho}\l(1+\mathcal O(r-r_+)^{-1}\r)\,,\\\label{eq:Rasymptotics_H+}
    R_{\text{in},r_+} &= \l( r-r_+\r)^{-s-i\rho}\l(1+\mathcal O(r-r_+)^{-1}\r)\,,
\end{aligned} 
\end{equation}
with:
\begin{equation}
    \rho:= \frac{\omega -m\,\Omega_+}{2\,\kappa}\,.\label{eq:definition_rho}
\end{equation}
Similarly, an asymptotic expansion near $r\rightarrow \infty$ yields a decomposition into an ingoing and an outgoing mode:
\begin{equation}
    R_s(r) =  \mathcal C(\omega) R_{\text{out},+\infty}(r) + \mathcal D(\omega) R_{\text{in},+\infty}(r)\,,
\end{equation}
where:
\begin{equation}
\begin{aligned}
    R_{\text{out},+\infty}(r)&= r^{-1 - 2 s + 2 i M \omega}e^{i \omega r}\l(1+\mathcal O\l(r^{-1}\r)\r)\,,\\
    R_{\text{in},+\infty}(r)&= r^{-1 - 2i M \omega}e^{-i \omega r}\l(1+\mathcal O\l(r^{-1}\r)\r)\,.\label{eq:Rasymptotics_I+}
\end{aligned}
\end{equation}
\acp{QNM} are defined as solutions of the Teukolsky equation with $\mathcal A(\omega) = \mathcal D(\omega)=0$. This defines a \ac{BVP} with eigenfunction $R_s$ and eigenvalue $\omega$.\\

Numerically, there is a more practical way of setting up the \ac{BVP}. Define the compact radial coordinate:
\begin{equation}
    y := 1- \frac{r_+}{r}\label{eq:define_compact_y}\,.\\
\end{equation}
At $y=0$ ($r=r_+$), and $y=1$ ($r=+\infty$), the \acp{BC} corresponding to ingoing and outgoing modes and the multiplicative factor that transforms one into the other are singular. Define:
\begin{equation}
    R_s(r(y)) = \alpha_{y,s}(y) \widetilde R_s(y)\,,\label{eq:def_tildeR}
\end{equation}
where $\alpha_{y,s}$ encodes the singular part of the appropriate \acp{BC},~\cite{Leaver:1985ax}:
\begin{widetext}
\begin{equation}
    \alpha_{y,s}(y(r)):=e^{i r \omega } r^{4 i M\omega } \Delta_r^{-s-i M \omega } \l(\frac{r-r_-}{r-r_+}\r)^{\frac{i\l(\frac{M}{r_+}\omega - m \Omega\r)}{2 \kappa}}\,.\label{eq:alfa_r_def}
\end{equation}
\end{widetext}
Substituting this in equation~\eqref{eq:master_teukolsky_radial}, and requiring smoothness of $\widetilde R_s(y)$ in $y\in[0,1]$, the appropriate \ac{QNM} \acp{BC} are automatically satisfied. As outlined in section~\ref{appendix:numerics}, pseudo-spectral methods can only encode smooth functions, thus, if we find a pseudo-spectral solution of the resulting \ac{EOM} in the interval $y\in[0,1]$, it must describe a \ac{QNM}.\\

In modified \ac{BL} coordinates, the Kerr metric has coordinate singularities at both $S^2$ poles ($x=\pm1$). This induces apparent singular behaviour on Teukolsky modes in these regions. As above, we can circumvent this by explicitly factoring out the problematic component. Take:
\begin{equation}
    \Theta_s(x) = \alpha_{x,s}(x) \widetilde \Theta_s(x)\,,\label{eq:def_tildeTheta}
\end{equation}
with:\footnote{$\alpha_{x,s}(x)$ is smooth for $x\in[-1,1]$, however, when we combine this with $e^{-i\,m \,\phi}$ and convert to Cartesian coordinates, we get an irregular function.}
\begin{equation}
    \alpha_{x,s}(x):=(1-x)^{\frac{\abs{m+s}}{2}} (1+x)^{\frac{\abs{m-s}}{2}}\,,\label{eq:alfa_x_def}
\end{equation}
\ac{QNM} solutions will have $\widetilde \Theta_s(x)$ to be a smooth function for $x \in[-1,1]$.\\

This definition of \acp{QNM} relies on the separability of the \ac{EOM} into a radial and an angular \ac{ODE}. However, in sections~\ref{sec:KG_modes} and~\ref{sec:EM_modes}, we will study \ac{EOM} that do not separate. The same set of steps is still valid. First, using a Frobenius analysis of the \ac{PDE} near the boundaries of the domain, determine the \acp{BC} that encode the correct ingoing / outgoing behaviour. Then, define a function $\alpha_s(y,x)$ that encodes the corresponding singular behaviour and factor it out from the \ac{QNM} solution. Finally, solve the resulting \ac{EOM} requiring smoothness of the solution for $[y,x]\in[0,1]\cross[-1,1]$. Concretely, for Teukolsky modes, the joint singular behaviour is:
\begin{equation}
    \alpha_s(y,x) = \alpha_{y,s}(y) \alpha_{x,s}(x)\,. \label{eq:define_alpha_s}   
\end{equation}
Define:
\begin{equation}
    \widetilde \Psi_s:=\frac{1}{\alpha_s}\Psi_s\,.
\end{equation}
Substituting this in equation~\eqref{eq:teukolsky_OL}, we get:
\begin{equation}
    0=\widetilde{\mathcal O}_{s, \omega}\,\widetilde\Psi_s(y,x):=\frac{1}{\alpha_s}\mathcal O_{s, \omega}\,\alpha_s\widetilde\Psi_s(y,x)\,.\label{eq:define_tilde_O}
\end{equation}
\acp{QNM} are solution of the full \ac{PDE} that are smooth for $[y,x]\in[0,1]\cross[-1,1]$.

\subsection{Parity transformations\label{subsec:background_parity}}

Denote $\mathbb P$ the parity transformation around $r=0$. In modified \ac{BL} coordinates, $\mathbb P$ takes $x\rightarrow -x$ and $\phi\rightarrow \pi + \phi$. The action~\eqref{eq:Maxwell_Weyl_Action} is invariant under parity, and so is the Kerr metric. For the Kinnersley tetrad,see~\eqref{eq:kerr_null_tetrad}, we can prove:
\begin{equation}
\begin{aligned}
    \mathbb P l^a\partial_a &= l^a\partial_a,\\
    \mathbb P n^a\partial_a &= n^a\partial_a,\\
    \mathbb P m^a\partial_a &= -\bar m^a\partial_a,\\
    \mathbb P \bar m^a\partial_a &= -m^a\partial_a.
\end{aligned}
\label{eq:parity_on_null_tetrad}
\end{equation}

The vector potential $A_a$,  corresponding field strength $F_{ab}$, and the spacetime metric are all real quantities. However, because of the complex nature of $m$ and $\bar m$, when we project these tensors to the null tetrad, the resulting scalars are complex-valued. Notwithstanding, this is merely an artifact of the representation. In particular, the complex conjugation operator simply maps $m\rightarrow \bar m$ and $\bar m \rightarrow m$.  This motivates the definition of the \textit{conjugate-parity-transform} operator $\mathcal P$. For some scalar $\eta$, we have:
\begin{equation}
    \mathcal P \eta := \overline{\mathbb P \eta}\,.\label{eq:define_cal_P}
\end{equation}
As an example, consider the action in the Teukolsky field $\psi^(s)$. Using equation~\eqref{eq:sep_t_phi}, we have that:
\begin{equation}
\begin{aligned}
    \mathcal P \psi^{(s)}(t,r,x,\phi) &= \mathcal P S(t,\phi)\,\Psi_s(r,x)\\
    &= (-1)^m \bar S(t,\phi) \bar\Psi_s(r,-x)\,,
\end{aligned}
\end{equation}
where the $(-1)^m$ contribution comes from taking $\phi\rightarrow \pi + \phi$.\\

The transformation rule in equation~\eqref{eq:parity_on_null_tetrad} is tensorial. Thus, if $\eta$ is constructed by taking covariant derivatives and tensor products of tetrad quantities, we have:
\begin{equation}
    \mathcal P \eta = (-1)^s \eta\,,\label{eq:cal_P_on_eta}
\end{equation}
where $s$ is the number of $m^a$ and $\bar m^a$ instances in the definition of $\eta$. As an example, take $\rho$ and $\tau$. Using equation~\eqref{eq:pq_ghp_scalars}, we have:
\begin{equation}
    \mathcal P \rho = (-1)^2 \rho \qquad \mathcal P \tau = (-1)^1 \tau\,.
\end{equation}
By definition, $\eta$ is a \ac{GHP} covariant quantity. Thus, $s$ is simply its spin weight: $s=(p-q)/2$.\\

Now, consider $\mathcal G$, a differential operator constructed with tensor products and covariant derivatives of tetrad quantities, acting on some scalar $f$. Following the same reasoning, we have:
\begin{equation}
    \mathcal P\,\l(\mathcal G\,f\r) = (-1)^s\mathcal G\,\mathcal P\,f\,.\label{eq:GHP_operators_under_parity}
\end{equation}
where $s$ is the spin-weight of $\mathcal G$. Because Kerr is invariant under parity, this rule also extends to the case where $\mathcal G$ contains pure gravitational scalars, e.g. $\psi_2$.\\

In particular, the Teukolsky operator $\mathcal O_s$ has spin-weight $0$, thus:
\begin{equation}
    \l[\mathcal P, \mathcal O_s\r]=0,
\end{equation}
where $[\cdot, \cdot]$ denotes the commutator operator. Standard results tell us that if $\psi^{(s)}$ is in the kernel of $\mathcal O_s$, then so is $\mathcal P \psi^{(s)}$. Using equation~\eqref{eq:sep_t_phi}, we have that:
\begin{equation}
\begin{aligned}
    &\mathcal O_s\,S\,\Psi_s = 0 \implies\\
    &\mathcal O_s\,(-1)^m\bar S\,\mathbb P \bar \Psi_s = 0\,.
\end{aligned}
\end{equation}

This proves that if $\l(m, \omega\r)$ is in the \ac{QNM} spectrum of $\mathcal O_s$, then so is $\l(-m, -\bar \omega\r)$. The parity invariance of the Kerr metric leads to a \textit{degeneracy} between these two families of \acp{QNM}.\\

The action of $\mathbb P$ on the Kinnersley tetrad, see~\eqref{eq:parity_on_null_tetrad}, does not generalize to all the equivalent tetrads. In general, if we generate a new null tetrad with a \ac{GHP} transformation, see~\eqref{eq:lambda_transformation_tetrad}, the identities in~\eqref{eq:parity_on_null_tetrad} are no longer respected. To circumvent this, we can restrict the allowed \ac{GHP} transformations, such that the resulting tetrads obey~\eqref{eq:parity_on_null_tetrad}. Under a generic \ac{GHP} transformation, we have:
\begin{equation}
\begin{aligned}
    \mathbb P l^a\partial_a &\rightarrow \mathbb P\l(\lambda \bar \lambda\r)\mathbb P l^a\partial_a,\\
    \mathbb P n^a\partial_a &\rightarrow \mathbb P\l(\frac{1}{\lambda \bar \lambda}\r) \mathbb P n^a\partial_a,\\
    \mathbb P m^a\partial_a &\rightarrow \mathbb P \l(\frac{\lambda}{\bar \lambda}\r) \mathbb P m^a\partial_a,\\
    \mathbb P \bar m^a\partial_a &\rightarrow \mathbb P \l(\frac{ \bar \lambda}{\lambda}\r) \mathbb P \bar m^a\partial_a.
\end{aligned}
\end{equation}
Thus, to preserve~\eqref{eq:parity_on_null_tetrad}, we must have:
\begin{equation}
\begin{aligned}
    \mathbb P \lambda \bar \lambda &= \lambda \bar \lambda\,,\\
    \mathbb P\l(\frac{\lambda}{\bar \lambda}\r) &=\frac{ \bar \lambda}{\lambda}\,.
\end{aligned}\label{eq:allowed_GHP_transformations_parity}
\end{equation}
The absolute value and phase of $\lambda$ must be parity even and odd respectively. Take $\eta$ to be a \ac{GHP} scalar with boost weight $r$ and spin weight $s$. Restricting to \ac{GHP} transformations that respect~\eqref{eq:allowed_GHP_transformations_parity}, we get:
\begin{equation}
\begin{aligned}
    \mathbb P \eta &\rightarrow \l(\lambda \bar \lambda\r)^r \l(\frac{\lambda}{\bar \lambda} \r)^{-s} \mathbb P \eta\,,\\
    \mathcal P \eta &\rightarrow \l(\lambda \bar \lambda\r)^r \l(\frac{\lambda}{\bar \lambda} \r)^s \mathcal P \eta\,.
\end{aligned}
\end{equation}
The action $\mathbb P$ is akin to complex conjugation. It takes a $\{p,q\}$ \ac{GHP} scalar to a $\{q,p\}$ scalar. Consequently, $\mathcal P$ preserves the \ac{GHP} weight. This fact is particularly useful, as it allows us to take linear combinations of $\eta$ and $\mathcal P \eta$ without breaking \ac{GHP} covariance.

\subsection{Hodge duals and p forms\label{subsec:background_hodge}}
In this section, we will review some key concepts regarding $p$ forms, and their hodge duals. The hodge dual operator, $\star$, is a linear map from the space $p$ forms to the space of $d-p$ forms. We will follow the convention:
\begin{equation}
    \l(\star \omega\r)_{a_1\cdots a_{d-p}} = \frac{1}{p!}\varepsilon_{a_1\cdots a_{d-p}b_{1}\cdots b_p} \omega^{b_1\cdots b_p}\,,
\end{equation}
where $\varepsilon$ is the Levi-Civita tensor. We have $\varepsilon_{a_1\cdots a_d}= \varepsilon_{[a_1\cdots a_d]}$ and $\varepsilon_{123\cdots d} = \sqrt{-g}$. For a Lorentzian metric, we have:
\begin{equation}
    \star\star \omega = -(-1)^{p\l(d-p\r)}\omega \,.
\end{equation}
Finally, we can define the differential and co-differential operators:
\begin{equation}
\begin{aligned}
    \l(\dd \omega\r)_{a_1\cdots a_{p+1}} &:= (p+1)\nabla_{[a_1}\omega_{a_2\cdots a_{p+1}]}\,,\\
    \l(\dd^\dagger \omega\r)_{a_1\cdots a_{p-1}}&:=\l(\star \dd \star \omega\r)_{a_1\cdots a_{p-1}}\\
     &= - (-1)^{p\l(d-p\r)} \nabla^b \omega_{a_1\cdots a_{p-1}b}\,.
\end{aligned}
\end{equation}

The Weyl tensor is not fully antisymmetric, and thus cannot be regarded as a $4$ form. However, it is antisymmetric in the first two and the last two indices, and so, we can take their dual. In 4-dimensions, we define the left and right dual of $\mathcal C$:
\begin{equation}
\begin{aligned}
    \l(\star \mathcal C\r)_{abcd} = \frac{1}{2!}\varepsilon_{abef}\mathcal C^{ef}{}_{cd}\,,\\
    \l(\mathcal C \star\r)_{abcd} = \frac{1}{2!}\mathcal C_{ab}{}^{ef}\varepsilon_{efcd}\,.
\end{aligned}
\end{equation}
Now, taking the left and right dual at the same time, we get:
\begin{equation}
\begin{aligned}
    \l(\star \mathcal C\star\r)^{ab}{}_{cd} &=\frac{1}{4}\varepsilon^{abef}\varepsilon_{cdgh}\mathcal C_{ef}{}^{gh}\\
    &=-\frac{4!}{4}\delta^a_{[c}\delta^b_d\delta^e_g\delta^f_{h]}\,\mathcal C_{ef}{}^{gh}\\
    &=-\frac{4!}{4\cdot3!}\delta^e_{[c}\delta^f_{d]}\delta^a_{[g}\delta^b_{h]}\,\mathcal C_{ef}{}^{gh}\\
    &=-\mathcal C^{ab}{}_{cd}\,.
\end{aligned}
\end{equation}
In the second step we used the standard result for the contraction of two instances of $\varepsilon$, and in the the third step, we used the fact that $\mathcal C$ is traceless, i.e. $\mathcal C^c{}_{acb} =0$. Using $\star^2 = -1$ (for a 2 form in 4 dimensions), we get:
\begin{equation}
    \star \mathcal C = \mathcal C \star\,, \label{eq:left_equals_right_hodge_weyl}
\end{equation}
The left and right duals are equivalent.\\

There are multiple ways to fix the degeneracy between the two electromagnetic / gravitational degrees of freedom. Usually, it is helpful to decompose the relevant quantities in a basis of eigenstates of a given operator. If we pick that operator to be $\star$, we decompose the electromagnetic and Weyl tensor into their \textit{self-dual} and \textit{anti-self-dual} parts:
\begin{equation}
\begin{aligned}
    F^\pm &= \frac{1}{2}\l(F \mp i \star F\r)\,,\\
    \mathcal C^\pm &= \frac{1}{2}\l(\mathcal C \mp i \star \mathcal C\r)\,,
\end{aligned}\label{eq:define_F_C_+-}
\end{equation}
As intended, $F^\pm$ and $\mathcal C^\pm$ are $\star$ eigenstates:
\begin{equation}
    \star F^\pm = \pm i F^\pm\,, \qquad \star \mathcal C^\pm = \pm i \mathcal C^\pm\,.
\end{equation}

When projected to the Kinnersley tetrad~\eqref{eq:kerr_null_tetrad}, the Levi-Civita tensor takes a simple form:
\begin{equation}
\begin{aligned}
    \varepsilon_{(\alpha)(\beta)(\gamma)(\delta)} &= \varepsilon_{abcd} e^a_{(\alpha)}e^b_{(\beta)}e^c_{(\gamma)}e^d_{(\delta)}\\
    &=-i\,E_{(\alpha)(\beta)(\gamma)(\delta)}
\end{aligned}
\end{equation}
where $E$ is fully antisymmetric and $E_{1234}$. Thus, we get $\varepsilon_{(1)(2)(3)(4)}=-i$. Dualized quantities, carry an extra factor of $i$, and thus, complex conjugation will add an extra $-$ contribution. Concretely, consider the \textit{dualized \ac{GB} term}:
\begin{equation}
\begin{aligned}
    \widetilde{\mathcal G} &= \l(\star\mathcal C\r)^{abcd}\mathcal C_{abcd}\\
    &=24i \l(\bar \psi_2^2 - \psi_2^2\r)\,.
\end{aligned}
\end{equation}
$\widetilde{\mathcal G}$ can be understood as a product of $i$ and a spin weight $0$ \ac{GHP} scalar that obeys the transformation rules detailed in section~\ref{subsec:background_parity}. Thus, under the \textit{conjugate-parity-transform}, we have:
\begin{equation}
    \mathcal P \widetilde{\mathcal G} = -\widetilde{\mathcal G}\,.\label{eq:transformation_G_tilde}
\end{equation}
As a rule of thumb, quantities that are dualized an odd amount of times are parity odd.\\

\subsection{The vacuum Maxwell equations\label{subsec:hertz_potential}}

In the \ac{GHP} formalism, we represent the electromagnetic field with 3 complex scalar fields. However, we know that electromagnetism is fully specified by two independent polarizations. Thus, two of the \ac{GHP} scalars must be redundant. In~\cite{Wald:1978vm}, Wald derived  an approach to fully reconstruct the electromagnetic tensor $F_{ab}$ from a single complex scalar. This is commonly known in the literature as the Hertz potential approach. In this section, we will start by deriving the Teukolsky equation for $\phi_0$, and then explain how to generate $F_{ab}$ from it.\\

The vacuum Maxwell equations take the form:
\begin{equation}
\begin{aligned}
    &\nabla^a F_{a b} = 0,\\
    &\dd F=0.
\end{aligned}\label{eq:vacuum_maxwell_eqs}
\end{equation}
Here, the second equation, commonly known as the Bianchi identity, implies that $F_{ab}$ must be a closed $2$ form. Because $F$ is a real field, we can use the results in section~\ref{subsec:background_hodge}, to combine the two equations into a single complex equation:
\begin{equation}
    M_a:=i\l(\star\dd F^-\r)_a=0\,,\label{eq:vacuum_maxwell_eqs_complex}
\end{equation}
where $F^-$ was defined in~\eqref{eq:define_F_C_+-}. The, first and second equations in~\eqref{eq:vacuum_maxwell_eqs}, are encoded by the real and imaginary parts of~\eqref{eq:vacuum_maxwell_eqs_complex} respectively. In the tetrad basis, using the \ac{GHP} formalism, $M$ takes a simple form~\cite{Geroch:1973am}:
\begin{equation}
    M_{(\alpha)}=
    \begin{pmatrix}
        \l(\eth ' -\tau'\r)\phi_0-\l(\thn -2 \rho\r) \phi_1\\
        \l(\thn'-2 \rho '\r) \phi_1-\l(\eth - \tau\r) \phi_2\\
        \l(\thn'- \rho'\r)\phi_0-\l(\eth - 2 \tau\r)  \phi_1\\
        \l(\eth ' - 2 \tau'\r)\phi_1-\l(\thn-\rho\r)\phi_2
    \end{pmatrix}_{(\alpha)}\,.\label{eq:maxwell_GHP_vacuum}
\end{equation}
The corresponding \ac{GHP} weights are:
\begin{equation}
    \{1,1\},\quad\{-1,-1\},\quad\{1,-1\},\quad\{-1,1\} \,.
\end{equation}

If we increase the spin-weight of $M_{l}$ and the boost-weight of $M_{m}$ by 1, the two components will have the same \ac{GHP} weights and can be combined. That is precisely the role of the operator $\zeta$ introduced by Teukolsky in~\cite{Teukolsky:1973ha}:
\begin{equation}
    \zeta^a := \l(\eth- 2 \tau + \bar \tau'\r) l^a - \l(\thn - 2 \rho + \bar \rho\r) m^a\,.\label{eq:define_zeta}
\end{equation}
Using the \ac{GHP} relations in~\cite{Geroch:1973am}, we can obtain the Teukolsky equation for $s=1$ (see equation~\eqref{eq:teuk_expanded}):
\begin{equation}
    \zeta^a M_a =\mathcal O_1 \phi_0 = 0\,.
\end{equation}

$F$ is a closed form, thus, locally, there is a form $A$, such that $F= \dd A$. This is known as the vector potential. We can express the components of $F$ directly in terms of the $A$. Using the definition of $\phi_i$ (equation~\eqref{eq:define_maxwell_scalars}), and treating $A^a$ as a $\{0,0\}$ \ac{GHP} covariant vector field, we have:
\begin{equation}
    \begin{aligned}
    \phi_0 &=\tau_0{}^a A_a:=\l(m^a\thn -l^a \eth\r)A_a\,,\\
    \phi_1 &=\tau_1{}^a A_a:= \frac{1}{2} \l(n^a\thn-l^a\thn'+m^a\eth'  -\bar m^a\eth\r)A_a\,,\\
    \phi_2 &=\tau_2{}^a A_a:=\l(n^a \eth'-\bar m^a\thn'\r)A_a\,.\\
    \end{aligned}
\end{equation}\label{eq:define_tau_i}
Thus, in terms of $A$, the Teukolsky equation is:
\begin{equation}
    \mathcal O_1\,\tau_0\,A=0\,.
\end{equation}

On the other hand, working directly in an abstract basis, we can obtain $M_a$ in terms of $A_a$. We have:
\begin{equation}
    M_a = \l(\mathcal E A\r)_a:=\nabla_b \nabla^b A_a -\nabla_b \nabla_a A^b\\
\end{equation}
Acting with $\zeta^a$ on the left, we must recover the Teukolsky equation:
\begin{equation}
    \mathcal O_1 \phi_0 =\zeta\,\mathcal E\,A\,.
\end{equation}

Thus, we have the operator identity:
\begin{equation}
    \mathcal O_1 \tau_0 = \zeta \mathcal E\,.
\end{equation}

The key insight comes from taking the transpose\footnote{Here the transpose refers to the classical transpose of a differential operator, no complex conjugation is needed. See~\cite{Wald:1978vm}for details.} of this identity. We have that:
\begin{equation}
    \tau_0{}^T \mathcal O_1{}^T = \mathcal E^T \zeta^T\,.
\end{equation}
Now, we can prove that $\mathcal E^T = \mathcal E$, thus, for a mode $\psi_H$ such that:
\begin{equation}
    \mathcal O_1{}^T \psi_H = 0\,,\label{eq:define_hertz_potential}
\end{equation}
we have:
\begin{equation}
    \mathcal E\l(\zeta^T \psi_H\r) =0\,.
\end{equation}

We proved that $\hat A_a = \zeta_a{}^T \psi_H$, solves the vacuum Maxwell equations. Thus, by solving equation~\eqref{eq:define_hertz_potential}, we can generate all the components of $F_{ab}$. This is not yet the complete story. In general, $\psi_H$ is a complex scalar, thus, $\hat A_a$ is a complex vector. Nevertheless, the reality of $A_a$ and consequently $F_{ab}$ is a crucial assumption in the derivation of equation~\eqref{eq:vacuum_maxwell_eqs_complex}. $\mathcal E$ is a real linear operator, thus if $\hat A$ is in its kernel, then so must be $\overline{\hat A}$. Hence, for any mode $\psi_H$ that solves~\eqref{eq:define_hertz_potential}, we can construct a real solution to Maxwell's equation given by:
\begin{equation}
    A_a :=\mathcal Re(\zeta_a{}^T \psi_H). \label{eq:hert_potential_A}
\end{equation}
In the literature, $\psi_H$ is usually known as the Hertz potential.\\

From equation~\eqref{eq:master_teukolsky}, we can deduce that $\mathcal O_s{}^T=\mathcal O_{-s}$. Thus, $\psi_H$ is a solution to the master Teukolsky equation with $s=-1$. In fact, this is a re-parametrization of the \ac{EOM} for $\phi_2$~\cite{Aksteiner:2010rh}:
\begin{equation}
    \mathcal O_{-1} \l(\psi_2^{-2/3} \phi_2 \r)= 0
\end{equation}
Thus, the Hertz potential is proportional to the $\phi_2$ component of a solution to the Maxwell equations, and we can deduce $\psi_H$ has GHP weight $\{-2,0\}$. We can prove that any \ac{QNM} of $\phi_2$ will generate a \ac{QNM} for $A^a$ with the same frequency, and consequently, all \acp{QNM} of $\phi_2$ must be \acp{QNM} of $\phi_0$. The Hertz potential derivation can be reversed, and we can generate a vector potential $A^a$ from any $\phi_0$. Thus, all $\phi_0$ \acp{QNM} must also be \acp{QNM} of $\phi_2$. Hence, $\phi_0$ and $\phi_2$ must be isospectral, as seen in~\cite{Chandrasekhar:1985kt}.\\

Notwithstanding, the $\phi_2$ proportional to the Hertz potential, is not the same as the component of the electromagnetic tensor generated by~\eqref{eq:hert_potential_A}. Instead, we have to use the operators $\tau_i$ defined in~\eqref{eq:define_tau_i}. Applying the \ac{GHP} identities in~\cite{Geroch:1973am} we get~\cite{Wald:1978vm}:
\begin{equation}
\begin{aligned}
    \phi_0 &=  \tau_0{}^a A_a = \frac{1}{2} \thn \thn\bar \psi_H\\
    \phi_1 &=  \tau_1{}^a A_a = \frac{1}{2} \l(\thn\,\eth'+\tau'\,\thn\r)\bar\psi_H\\
    \phi_2 &=  \tau_2{}^a A_a = \frac{1}{2} \eth' \eth'\bar \psi_H
\end{aligned}\label{eq:phi_i_hertz}
\end{equation}

Because we required $A$ to be a real vector, $\bar\phi_i$ can be obtained by simply taking the complex conjugate of these expressions. If we worked with complex $A$, $\phi_i$ and $\bar \phi_i$ would be independent variables. Conveniently, $\phi_i$ depends exclusively on $\bar \psi_H$. \footnote{This coincidence is merely a consequence of the definition of $\phi_i$. If the projection of $F$ into the tetrad basis was done differently, we would likely have a mixing of $\psi_H$ and $\bar \psi_H$.} Thus, we can encode equation~\eqref{eq:phi_i_hertz}, in operators $\mathcal T_i$ such that:
\begin{equation}
    \mathcal T_i \bar \psi_H:=\phi_i \label{eq:define_Ti}
\end{equation}

\section{Corrections to scalar modes\label{sec:KG_modes}}

In this section, we will compute the \ac{EFT} corrections to \ac{QNM} frequencies of a massless scalar, when considering a Kerr background. We will mostly focus in the parity even corrections, as seen in the action~\eqref{eq:scalar_EFT_action}, however, in section~\ref{subsec:scalar_paraity_odd}, we will briefly outline how to adapt our methods to consider parity odd corrections.

\subsection{Parity even equations of motion \label{subsec:KG_EOM}}

The \ac{EOM} considering parity even \ac{EFT} corrections, were obtained in~\eqref{eq:EOM_quadraticEsGB}. For a Kerr background, using the results in section~\ref{subsec:master_equation}, they reduce to:
\begin{equation}
    \l(\mathcal O_0-L^2\lambda\,\mathcal G\r)\Phi=0\,\label{eq:scalar_EOM_teukolsky},
\end{equation}
where $\mathcal O_0$ is the $s=0$ Teukolsky operator and $\mathcal G$ is the \ac{GB} term, defined in~\eqref{eq:define_GB}. Because Kerr is a vacuum, type-D spacetime, we get:
\begin{equation}
    \mathcal G = 24\l(\psi_2^2 + \bar\psi_2^2\r)\,.
\end{equation}
Following the discussion in section~\ref{subsec:background_parity}, it is clear that the \textit{conjugate-parity-transform} operator, $\mathcal P$, commutes with all terms in \eqref{eq:scalar_EOM_teukolsky}:
\begin{equation}
    [\mathcal P , \mathcal O_0] = [\mathcal P , \mathcal G] =0\,,
\end{equation}
thus, if $\Phi$ is a solution to the \ac{EOM}, then so is $\mathcal P \Phi$. As argued in the end of section~\ref{subsec:background_parity}, Assuming $\Phi$ is proportional to $S(t,\phi)=\exp(-i\omega t + i m \phi)$, we get that if $(m,\omega)$ define a \ac{QNM}, then so does $(-m,-\bar\omega)$. This relates \acp{QNM} whose frequency have positive real part with the ons with negative real part, and thus, we can restrict to \ac{QNM} frequencies with positive real part.\\

In the Kinnersley tetrad basis, 
\begin{equation}
    \psi_2= -\frac{M}{\l(r- i\,a x\r)^3} \label{eq:psi2_definition}
\end{equation}
Thus, the \ac{EOM} are, in general, not separable. Notwithstanding, we can still follow the procedure in section~\ref{subsec:QNMs_intro} to define the \acp{QNM}. First, as seen in~\eqref{eq:sep_t_phi}, we factor out the $(t,\phi)$ dependence, by making the \textit{Ansatz}:
\begin{equation}
    \Phi(t,r,x,\phi) = S(t,\phi)\,\Psi(r,x) \label{eq:sep_t_phi_PhiEq}
\end{equation}
We get:
\begin{equation}
    \mathcal O_{0,\omega}\,\Psi(r,x) = \lambda L^2 \,\mathcal G \Psi(r,x) \label{eq:EOM_scalar_WeylSub_rx}
\end{equation}
where $\mathcal O_{0,\omega}$ was defined in equation~\eqref{eq:teukolsky_OL}. Now, we must enforce \acp{BC} corresponding to outgoing behavior at $\mathcal J^+$ and ingoing at $\mathcal H^+$. To do so, we will replace $r$ with the compactified radial coordinate $y= 1-r/r_+$. In the case $\lambda=0$, as argued in section~\ref{subsec:QNMs_intro}, we can set the \acp{BC} by factoring out $\alpha_0(r,x)$ (see equation~\eqref{eq:define_alpha_s}) and requiring the remainder to be smooth for $(y,x)\in[-1,1]\cross[0,1]$. $\mathcal G$ is regular at $r_+$ and at $x= \pm1$. Furthermore, it is subleading with respect to $\mathcal O_{0,\omega}$ at $r\rightarrow\infty$. This implies that the leading order of a Frobenius expansion at the domain boundaries must be independent of $\lambda$. Thus, \acp{QNM} of the corrected equation obey the same \acp{BC} as the \acp{QNM} in the uncorrected case. Define:
\begin{equation}
    \widetilde \Psi(y,x):=\frac{1}{\alpha_0(y,x)}\Psi(r(y),x)\,.
\end{equation} 
Replacing this in equation~\eqref{eq:EOM_scalar_WeylSub_rx}, we get:
\begin{equation}
    \widetilde{\mathcal O}_{0, \omega}\,\widetilde\Psi(y,x)=L^2 \lambda\,\mathcal G\,\widetilde\Psi(y,x)\,.\label{eq:EOM_scalar_WeylSub_rx_non_sing}
\end{equation}
\acp{QNM} are solutions of this equation, that are smooth for $(y,x)\in[-1,1]\cross[0,1]$.\\

\begin{table*}[ht]
\begin{tabular}{c|c|r||r|r|r|r|r|r|r|r|r}
\multicolumn{3}{c||}{Parameters}&\multicolumn{9}{c}{$a/M$}\\\hline\hline
$\ell$&$m$&Component&$0$&$0.25$&$0.5$&$0.75$&$0.95$&$0.99$&$0.999$&$0.9999$&$0.999975$\\\hline\hline
\multirow{4}{*}{$0$}&\multirow{4}{*}{$0$}&$M\mathcal Re(\omega_0)$&$0.110455$&$0.110943$&$0.112381$&$0.114322$&$0.111992$&$0.110444$&$0.110265$&$0.110247$&$0.110246$\\
&&$M\mathcal Im(\omega_0)$&$-0.104896$&$-0.104289$&$-0.102183$&$-0.097298$&$-0.089559$&$-0.089493$&$-0.089439$&$-0.089434$&$-0.089433$\\
&&$100\,M\mathcal Re(\omega_1)$&$14.056932$&$13.842350$&$13.018046$&$10.639102$&$5.154504$&$6.752369$&$7.113310$&$7.148748$&$7.151690$\\
&&$100\,M\mathcal Im(\omega_1)$&$-2.446792$&$-2.302637$&$-1.884293$&$-1.484410$&$-5.668677$&$-8.548625$&$-8.774436$&$-8.796021$&$-8.797808$\\\hline
\multirow{4}{*}{$1$}&\multirow{4}{*}{$-1$}&$M\mathcal Re(\omega_0)$&$0.292936$&$0.275671$&$0.261572$&$0.249698$&$0.241372$&$0.239810$&$0.239462$&$0.239428$&$0.239425$\\
&&$M\mathcal Im(\omega_0)$&$-0.097660$&$-0.097362$&$-0.096505$&$-0.095278$&$-0.094124$&$-0.093882$&$-0.093828$&$-0.093822$&$-0.093822$\\
&&$100\,M\mathcal Re(\omega_1)$&$4.631840$&$3.835932$&$3.130117$&$2.487689$&$2.055239$&$1.982260$&$1.966471$&$1.964905$&$1.964775$\\
&&$100\,M\mathcal Im(\omega_1)$&$0.066211$&$0.362974$&$0.571957$&$0.660057$&$0.625281$&$0.610799$&$0.607378$&$0.607033$&$0.607005$\\\hline
\multirow{4}{*}{$1$}&\multirow{4}{*}{$1$}&$M\mathcal Re(\omega_0)$&$0.292936$&$0.314934$&$0.344753$&$0.390378$&$0.462261$&$0.493423$&$0.503344$&$0.500896$&$0.500198$\\
&&$M\mathcal Im(\omega_0)$&$-0.097660$&$-0.097005$&$-0.094395$&$-0.086394$&$-0.060091$&$-0.036712$&$-0.017519$&$-0.007904$&$-0.004088$\\
&&$100\,M\mathcal Re(\omega_1)$&$4.631840$&$5.590809$&$6.833001$&$8.512134$&$9.133060$&$6.466215$&$2.063483$&$-0.540870$&$-0.129697$\\
&&$100\,M\mathcal Im(\omega_1)$&$0.066211$&$-0.312890$&$-0.793476$&$-1.424866$&$-2.149343$&$-2.470718$&$-2.627015$&$-0.405923$&$-0.009214$\\\hline
\multirow{4}{*}{$2$}&\multirow{4}{*}{$-2$}&$M\mathcal Re(\omega_0)$&$0.483644$&$0.450060$&$0.422751$&$0.399883$&$0.384002$&$0.381041$&$0.380384$&$0.380318$&$0.380313$\\
&&$M\mathcal Im(\omega_0)$&$-0.096759$&$-0.096432$&$-0.095616$&$-0.094525$&$-0.093536$&$-0.093330$&$-0.093283$&$-0.093279$&$-0.093278$\\
&&$100\,M\mathcal Re(\omega_1)$&$2.545392$&$1.945246$&$1.494968$&$1.151127$&$0.937867$&$0.900983$&$0.892928$&$0.892128$&$0.892061$\\
&&$100\,M\mathcal Im(\omega_1)$&$0.160232$&$0.266663$&$0.315865$&$0.321760$&$0.304842$&$0.300007$&$0.298872$&$0.298758$&$0.298748$\\\hline
\multirow{4}{*}{$2$}&\multirow{4}{*}{$2$}&$M\mathcal Re(\omega_0)$&$0.483644$&$0.526740$&$0.585990$&$0.679504$&$0.840982$&$0.928028$&$0.978023$&$0.993234$&$0.996634$\\
&&$M\mathcal Im(\omega_0)$&$-0.096759$&$-0.096148$&$-0.093494$&$-0.085026$&$-0.056471$&$-0.031063$&$-0.010980$&$-0.003525$&$-0.001761$\\
&&$100\,M\mathcal Re(\omega_1)$&$2.545392$&$3.384658$&$4.653470$&$6.805746$&$9.331381$&$7.636031$&$3.618486$&$1.324989$&$0.675931$\\
&&$100\,M\mathcal Im(\omega_1)$&$0.160232$&$-0.026160$&$-0.338287$&$-0.872204$&$-1.415846$&$-1.032269$&$-0.322180$&$-0.038936$&$-0.000052$\\\hline
\end{tabular}
\caption{Numerical values for the \ac{QNM} frequencies and corresponding \ac{EFT} correction of scalar modes. All the included digits are significant.\label{tab:QNM_frequencies_scalar}}
\end{table*}

\begin{figure*}[ht]
    \subfloat[\label{subfig:convergence_plot_angular}]{
    \includegraphics[width=0.457\textwidth]{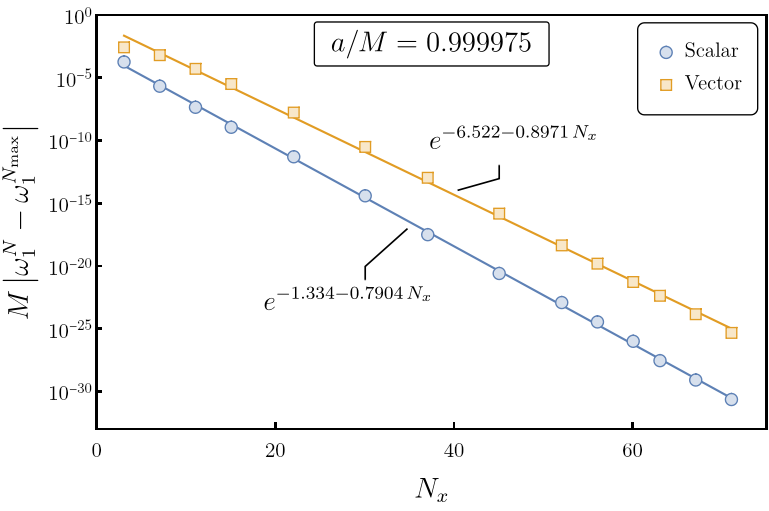}}\hskip 2em
    \subfloat[\label{subfig:convergence_plot_radial}]{
    \includegraphics[width=0.457\textwidth]{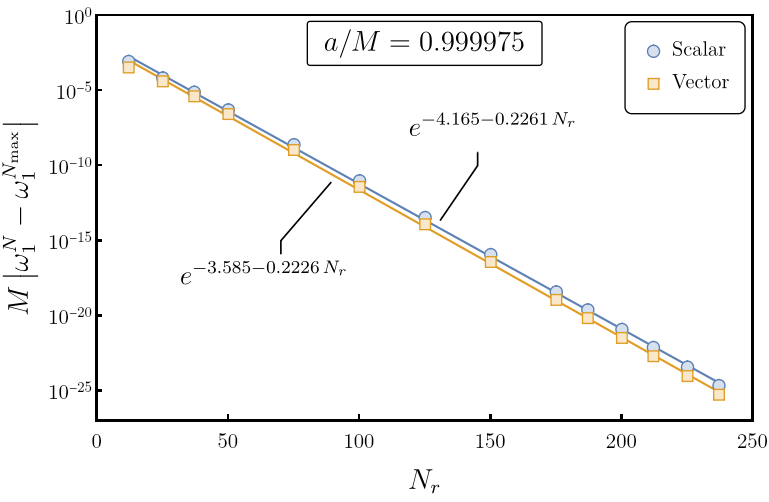}}

\caption{In this plot we show that our numerical approach converges exponentially with increasing grid size. Here, $N_x$, $N_r$ denote the sizes of the angular and radial grids respectively. We computed the \ac{EFT} corrections to scalar and electromagnetic \acp{QNM} of \acp{BH} with $a/M = 0.999975$, and $\ell=m=1$, for multiple values of $N_x$ and $N_r$. Then, we compared them with a reference $\omega_1$ obtained with the largest grid, plotting the difference. On the left panel, we fixed $N_r=250$ and varied $N_x$, whereas on the right panel we fixed $N_x = 75$, and varied $N_r$. The electromagnetic case was particularly difficult to compute, to achieve accurate results we had to obtain $\omega_0$, with $70$ digits of precision.\label{fig:convergence_plot}}
\end{figure*}

\begin{figure*}[ht]
    \subfloat[\label{subfig:scalar_QNM_complex_plot_background}]{
    \includegraphics[width=0.457\textwidth]{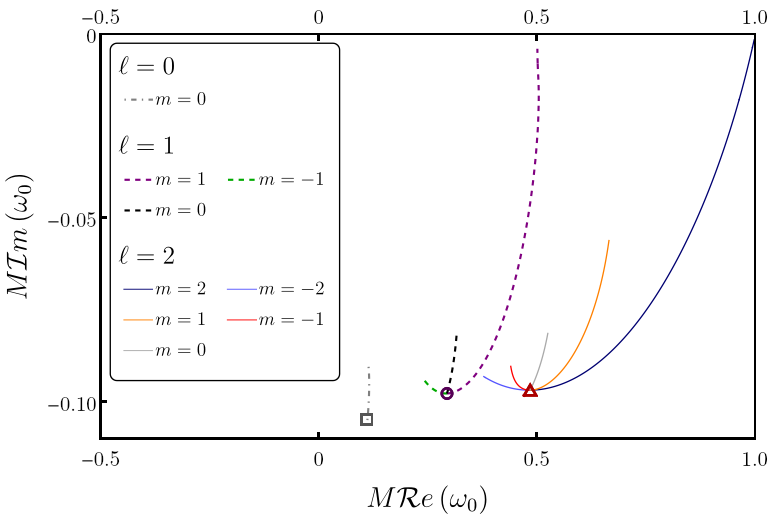}}\hskip 2em
    \subfloat[\label{subfig:scalar_QNM_complex_plot_correction}]{
    \includegraphics[width=0.457\textwidth]{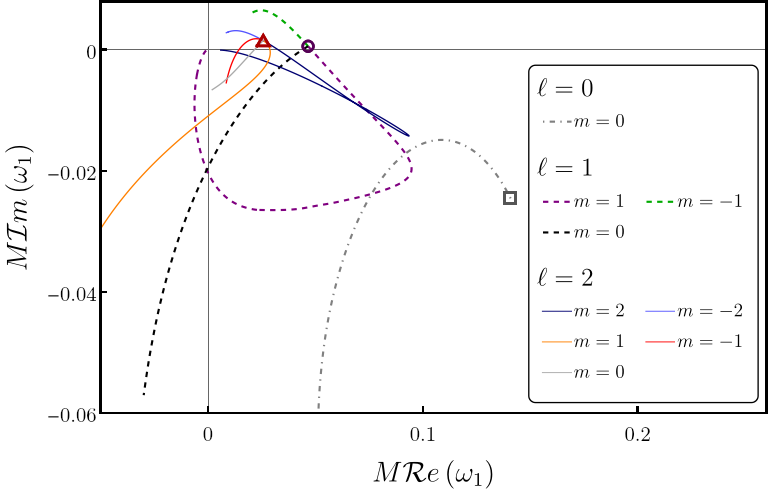}}

\caption{Here, we show a parametric plot of the real and imaginary parts of the scalar \ac{QNM} frequencies as $a$ is varied. On the left, we have the background \ac{QNM} frequencies, and on the right the \ac{EFT} correction. To produce this image, we computed the \ac{QNM} frequencies of \acp{BH} with $a/M\in (0, 0.999975)$, by solving equation~\eqref{eq:EOM_scalar_final}. The plot markers represent the \ac{QNM} frequencies obtained for the Schwarzschild limit using an independent non-perturbative method. We find perfect agreement between the two results. $\ell=m\neq0$ modes vary very quickly near extremality, with $M\omega \rightarrow \ell / 2$ as $a\rightarrow M$. If we truncated the spin at $a/M\sim 0.99$, we would lose roughly the last third of the parametric curves.\label{fig:scalar_QNM_complex_plot}}
\end{figure*}

\begin{figure*}[ht]
    \subfloat[\label{subfig:slowly_damped_near_extremal_scalar_QNM_background}]{
    \includegraphics[width=0.457\textwidth]{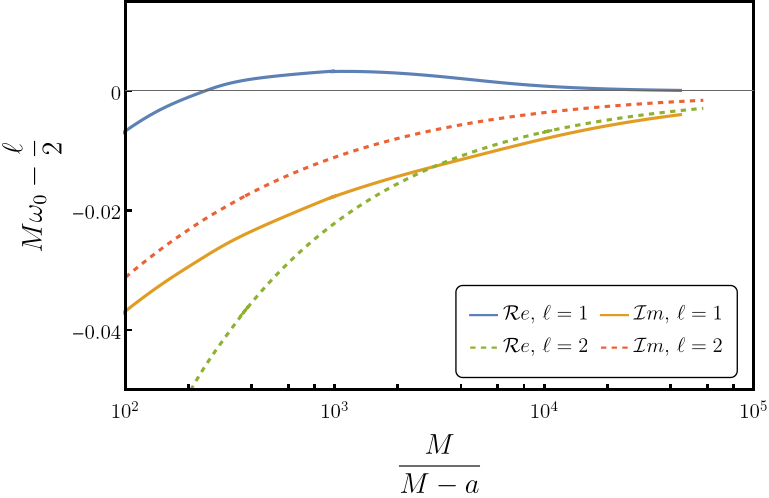}}\hskip 2em
    \subfloat[\label{subfig:slowly_damped_near_extremal_scalar_QNM_correction}]{
    \includegraphics[width=0.457\textwidth]{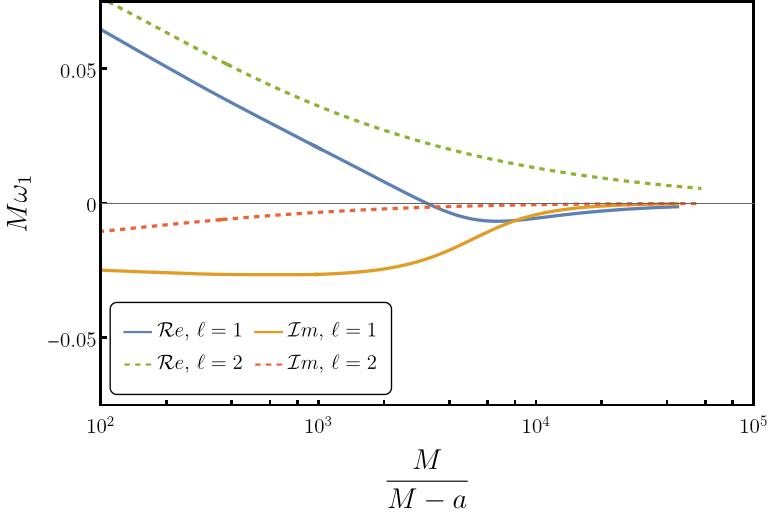}}

\caption{In this image we explore the scalar \ac{QNM} frequencies of $\ell=m\neq0$ modes of \acp{BH} very close to extremality. On the left panel, we show the background frequency, whereas the right has the \ac{EFT} correction. We establish that, as $a \rightarrow M$, $M\omega_0\rightarrow \ell/2$, agreeing with the prediction in~\cite{Hod:2008zz}. Similarly, in this limit, $M\omega_1$ converges to $0$. Notice that the near-extremal behaviour of these modes has some structure. In particular, $M\mathcal Re(\omega_0^{\ell=1})$ crosses $\ell/2$ at $a/M\approx 0.992$, $M\mathcal Re(\omega_1^{\ell=1})$ crosses $0$ at $a/M\approx 0.9993$, and $M\mathcal Im(\omega_1^{\ell=2})$ crosses $0$ at $a/M\approx 0.99994$. \label{fig:slowly_damped_near_extremal_scalar_QNM}}
\end{figure*}

\begin{figure}
\centering
    \includegraphics[width=.95\columnwidth]{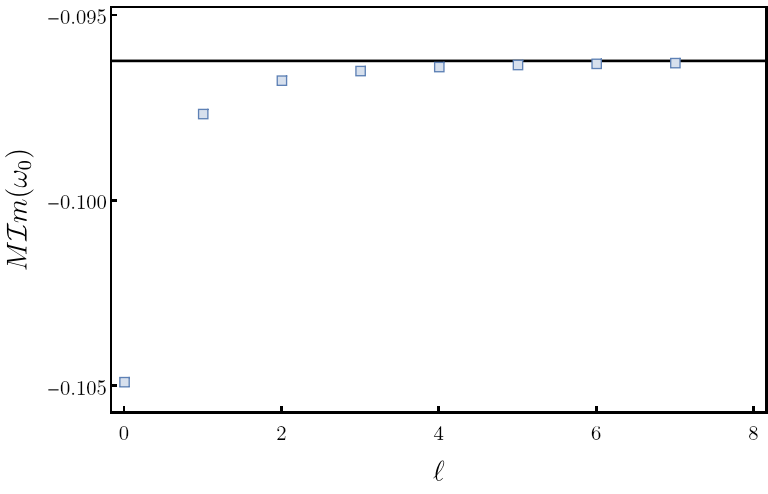}

\caption{Here, we show $M \mathcal Im(\omega_0)$ for scalar modes in the Schwarzschild limit as a function of $\ell$. The dark line represents the Eikonal limit prediction. Contrary to gravitational \acp{QNM}, $\mathcal Im(\omega_1)$ increases with increasing $\ell$, approaching the Eikonal limit frequency from below. This indicates that the longest-lived modes will have $\ell\rightarrow \infty$. \label{fig:Schwarzchild_fundamental_modes}}
\end{figure}

In the Schwarzschild limit, the \ac{RHS} reduces to $\l(48 L^2 M^2 \lambda /(r_+(1-y))^6\r) \widetilde\Psi(y,x)$, thus the \ac{EOM} separate into an angular and a radial \acp{ODE}. This is expected due to the spherical symmetry of the problem. The angular part can be solved analytically, yielding the familiar spherical harmonics $Y_{\ell m}$. The radial part is an eigenvalue problem for $\omega$, with $(L/M)^2\lambda$ and $\ell$ as free parameters. This can be solved directly with the pseudospectral methods discussed in the appendix~\ref{appendix:numerics}. The discussion is more involved when we consider the spinning case ($a\neq 0$). While the \ac{LHS} still separates into an angular and a radial operator, this is no longer the case for the \ac{RHS}. To make progress, we will use the validity condition for the \ac{EFT} approximation. The action~\eqref{eq:scalar_EFT_action} is at most quadratic in $L/M$, thus our results are only valid to that order. Assuming $\lambda= \mathcal O(1)$, the relevant perturbation parameter is:
\begin{equation}
    \widetilde \lambda := L^2/M^2 \lambda\,.
\end{equation}
Our results are only valid to leading order in $\widetilde \lambda$.\\

We can solve the \ac{EOM} using standard quantum mechanical perturbation theory. We make the following \textit{Ansatz}:\\
\begin{align}
    \widetilde\Psi(y,x) &= \widetilde\Psi_{(0)}(y,x) +  \widetilde \lambda\,\widetilde\Psi_{(1)}(y,x) +  \mathcal O(\widetilde \lambda^2)\,,\label{eq:define_phi_expansion}\\
    \omega &= \omega_0 + \widetilde \lambda\,\omega_1 + \mathcal O(\widetilde \lambda^2)\,.\label{eq:define_omega_expansion}
\end{align}
Grouping together same powers of $\widetilde\lambda$, and neglecting higher order terms the \ac{EOM} reduce to:
\begin{subnumcases}{\hskip -1cm\label{eq:EOM_scalar_final}}
    \widetilde{\mathcal O}_{0, \omega_0}\,\widetilde\Psi_{(0)}=0\,,\label{eq:EOM_background_scalar}\\
    \widetilde{\mathcal O}_{0, \omega_0}\,\widetilde\Psi_{(1)}=-\omega_1\partial_{\omega_0}\widetilde{\mathcal O}_{0, \omega_0}\widetilde\Psi_{(0)}+M^2\mathcal G\,\widetilde\Psi_{(0)}\,.\label{eq:EOM_perturbation_scalar}
\end{subnumcases}

Equation~\eqref{eq:EOM_background_scalar} is simply the $s=0$ of~\eqref{eq:define_tilde_O}. Consequently, its solutions will be the \acp{QNM} and frequencies of a massless scalar field. Henceforth, we will denote these, the \textit{background modes}. Substituting these in equation~\eqref{eq:EOM_perturbation_scalar}, the \ac{EOM} are fully specified, and we get an eigenvalue problem for $(\widetilde\Psi_{(1)}, \omega_1)$.\\

Apart from $(\widetilde\Psi_{(1)}, \omega_1)$, equation~\eqref{eq:EOM_perturbation_scalar} is fully specified by the background quantities. Thus, after solving the background equation, the \ac{EOM} for the correction reduce to an eigenvalue problem.

\subsection{On parity odd EFT corrections\label{subsec:scalar_paraity_odd}}

To achieve full generality, we must consider parity odd \ac{EFT} corrections. For a vacuum spacetime, under the assumptions we made when deriving the parity even correction, the most general odd correction can be obtained by replacing $\mathcal G$ with:
\begin{equation}
    \widetilde {\mathcal G} := (\star \mathcal C)_{abcd} \mathcal C^{abcd}\,.
\end{equation}
As argued at the end of section~\ref{subsec:background_hodge}, $\mathcal P$ anticommutes with $\widetilde {\mathcal G}$:
\begin{equation}
    \mathcal P \widetilde {\mathcal G} + \widetilde {\mathcal G} \mathcal P\,=0,
\end{equation}
thus, the action of $\mathcal P$ in the \ac{EOM} will be:
\begin{equation}
    \mathcal P\l(\mathcal O_0-L^2\lambda\widetilde{\mathcal G}\r)\Phi = \l(\mathcal O_0+L^2\lambda\widetilde{\mathcal G}\r)\mathcal P \Phi\,.
\end{equation}
$\mathcal P$ flips the sign of $\lambda$. Thus, again, we can relate \acp{QNM} whose frequency has positive real part with the ones with negative real part. If $(m,\omega(\lambda))$ defines a \ac{QNM} frequency, then so does $(-m,-\bar \omega(-\lambda))$. Bearing this in mind, and following the approach in section~\ref{subsec:KG_EOM}, we could obtain the parity odd \ac{EFT} corrections to scalar \acp{QNM} numerically.

\subsection{Results\label{subsec:KG_results}}

Using the pseudo-spectral methods outlined in section~\ref{subsec:numerics_perturbation}, we solved the background and perturbation \ac{EOM}. In figure~\ref{fig:convergence_plot} we show that, as expected, these methods converge exponentially with increasing grid size. We fixed the size of the radial/angular grid and computed $\omega_1$ for varying angular/radial grid sizes. By comparing the result with a more accurate value obtained for a large grid, we see that the accuracy of $\omega_1$ increases exponentially.\\

The main results are presented in figure~\ref{fig:scalar_QNM_line_plot} and table~\ref{tab:QNM_frequencies_scalar}. For the background \acp{QNM} we obtain slightly unusual results. The imaginary part of the frequency controls the decay rate of \acp{QNM}. The larger it is, the slower the decay rate. Thus, the longest-lived \ac{QNM} has the largest imaginary part of the frequency. For gravitational and, as we will see below, electromagnetic \acp{QNM} of most \acp{BH}, the longest-lived \acp{QNM} have $\ell=m=\abs{s}$. Usually, the ringdown spectrum is mostly driven by these modes, with the relative intensity of modes with smaller imaginary parts decaying exponentially. However, in figure~\ref{fig:Schwarzchild_fundamental_modes}, we show that in the Schwarzschild limit the slowest decaying modes have $\ell \rightarrow \infty$. This is consistent with the literature, (see table 1 of~\cite{Mamani:2022akq}). By inspecting figure~\ref{fig:scalar_QNM_line_plot}, we show that this feature generalizes for \acp{BH} with arbitrary spin. Note that this does not guarantee that the ringdown does not include small $\ell$ modes. The non-linear dynamics of the \ac{BH} collision control the initial intensity of each \ac{QNM}, and that could be heavily skewed towards modes with small $\ell$. An in-depth discussion of this lies beyond the scope of this paper.\\

In the Schwarzschild limit, the spacetime is spherically symmetric, and the \ac{EOM} can be solved non-perturbatively in $\widetilde \lambda$, using the direct methods in section~\ref{subsec:numerics_background}. We obtained the \ac{QNM} frequency $\omega$ for several values of $\widetilde \lambda$. Taking $\widetilde \lambda$ to be arbitrarily small, we disentangled $\omega$ into a background component $\omega_0$ and the \ac{EFT} correction $\omega_1$. The results can be seen in the plot markers of figure~\ref{fig:scalar_QNM_complex_plot}. For $a\neq0$, the \ac{EOM} are no longer separable, thus we proceed perturbatively, as outlined in section~\ref{subsec:KG_EOM}. The results are shown in the plot lines of figures \ref{fig:scalar_QNM_line_plot} and~\ref{fig:scalar_QNM_complex_plot}, and in table~\ref{tab:QNM_frequencies_scalar}. We found excellent agreement between the direct results in the Schwarzschild limit and the ones for slowly rotating Kerr \acp{BH}. This happens both for the background and \ac{EFT} correction of the \ac{QNM} frequencies.\\

Except very close to extremality, the real and imaginary parts of $\omega_1$ have maximal absolute value for $\ell=m$ modes. For $\ell\neq0$, the correction is largest at $a \sim 0.9 M$, quickly approaching $0$ afterwards. For $a \gtrsim 0.9 M$, the correction to modes with $\ell\neq m >0$, or $\ell=m=0$ increases quickly, before approaching finite values at $a\rightarrow M$. This can more easily be seen in a parametric form in figure~\ref{fig:scalar_QNM_complex_plot}. $\ell=m\neq0$ modes, also known as \textit{slowly damped \acp{QNM}}, are particularly interesting in the near extremal limit. Thus, in figure~\ref{fig:slowly_damped_near_extremal_scalar_QNM}, we show the \ac{QNM} frequencies in the region $0.99 < a/M \lesssim 0.999975$. As predicted in~\cite{Hod:2008zz}, we find that $\omega_0$ approaches $\ell /2$. Furthermore, for $\ell=m=2$, the convergence rate of $M \omega_0-1$ seems to be consistent with $\sqrt{1-a/M}$, as predicted in~\cite{Hod:2008zz}. However, for $\ell=m=1$, $M \omega_0$ seems to approach to $1/2$ slightly faster than expected. The author is convinced that the behaviour should agree better with the prediction in~\cite{Hod:2008zz} if we studied \acp{BH} even closer to extremality. However, extracting accurate results in this regime involves using very large collocation grids, increasing dramatically the size of the linear systems to solve. Thus, we did not cross $a \sim 0.999975M$. On the other hand, we found that, for slowly damped modes, $M\omega_1$ approaches $0$ in the same limit. As outlined in the caption of figure~\ref{fig:slowly_damped_near_extremal_scalar_QNM}, the near-extremal region of parameter space has an intricate structure. Both the real and imaginary parts of $\omega_1$ cross $0$ very close to extremality, making it very difficult to accurately estimate the convergence rate. Again, this might be overcome by studying \acp{BH} that are even closer to extremality.\\

\section{Corrections to electromagnetic modes \label{sec:EM_modes}}

Drawing from the methods and intuition developed in section~\ref{sec:KG_modes}, we will obtain the corrections to electromagnetic \acp{QNM}, under the parity even / odd terms introduced in the action~\eqref{eq:Maxwell_Weyl_Action}. Contrary to a scalar field, electromagnetism has two degrees of freedom, encoded in the two polarizations of light. In a vacuum, both follow null geodesics of the background spacetime, propagating at the speed of light. However, when passing through certain media, the propagation speed can be polarization dependent, leading to birefringent behaviour. In~\cite{Balakin:2017eur}, considering the parity even part of~\ref{eq:Maxwell_Weyl_Action}, the authors proved that the \ac{EFT} correction, effectively turns vacuum into a birefringent medium. For a type-D spacetime, in the Eikonal limit, they found that light propagates along the null geodesics of two different effective metrics, depending on its polarization. There is a known correspondence between null geodesics and the large $\ell$ limit of \ac{QNM} frequencies, see~\cite{Mashhoon:1985cya,Dolan:2010wr,Yang:2012he}. Hence, the birefringent behaviour of light should lead to a splitting of the \ac{QNM} frequencies into two families. In section~\ref{subsec:EM_even}, we will set $\lambda_{(o)} =0$ in the action~\eqref{eq:Maxwell_Weyl_Action}, and obtain the corrections to the two families under parity even \ac{EFT} corrections. Then, in section~\ref{subsec:EM_mixed}, we will keep $\lambda_{(o)}\neq 0$ and generalize the result to mixed parity \ac{EFT} corrections. Surprisingly, we find that the correction to \ac{QNM} frequencies due to parity odd \ac{EFT} terms coincides with the parity even case. We find that, for each background \ac{QNM} frequency there are two possible corresponding corrections with opposing signs:
\begin{equation}
    \omega^{\pm} = \omega_0 +\l(\frac{L}{M}\r)^2\abs{\lambda}\,\omega_1^{\pm}\,.\label{eq:schematic_correction_vector_QNFs}
\end{equation}
where:
\begin{equation}
    \omega_1^\pm=\pm (-1)^{\ell + m +1}\,\omega_1\label{eq:correction_splitting_QNFs}
\end{equation}
and:
\begin{equation}
    \abs{\lambda} = \sqrt{\lambda_{(e)}^2 + \lambda_{(o)}^2}\,.
\end{equation}
The factor of $(-1)^{\ell + m +1}$ was introduced to ensure that, in all cases studied, the Schwarzschild limit of $\omega_1$ has positive real and imaginary parts.\footnote{In section~\ref{subsubsec:EM_eikonal}, we will show that, for parity even corrections, in the Schwarzschild limit, the imaginary part of $\omega_1$ vanishes in the limit $\ell\rightarrow\infty$. As will be clear in section~\ref{subsubsec:EM_Schwarzschild_results}, this feature does not generalize for finite $\ell$, thus we can, in general, enforce the positivity of $\mathcal Im\,\omega_1$.} Because $\omega_1^+=-\omega_1^-$, this sign carries no physical meaning. This is in close analogy to the results for the gravitational case, see~\cite{Cano:2021myl,Cano:2023tmv,Cano:2023jbk}. Finally, in section~\ref{subsec:EM_results}, we will use the pseudo-spectral methods in the appendix~\ref{appendix:numerics} to solve the \ac{EOM}, obtaining the corrections to the \ac{QNM} frequencies in all cases.

\subsection{Equations of motion \label{subsec:EM_EOM}}

In this section, we will derive the corrected \ac{EOM} for electromagnetic waves in a Kerr background. Using $p$ forms and hodge duals, we will cast them into a form that can be easily tackled in later sections. Varying action~\eqref{eq:Maxwell_Weyl_Action} with respect to the vector potential, and using equation~\eqref{eq:left_equals_right_hodge_weyl}, we get:
\begin{equation}
    \nabla^b\l(F_{ab} - \frac{1}{3}L^2\l(\lambda_{(e)}\mathcal C + \lambda_{(o)}\star\mathcal C\r)_{ab}{}^{cd}F_{cd}\r)=0\,.\label{eq:EOM_maxwell_tensor}
\end{equation}
We want to solve this equation directly in terms of the components of $F_{ab}$, thus, we must also enforce the electromagnetic Bianchi identity, $\dd F=0$, guaranteeing $F$ is a closed form. Following the approach in section~\ref{subsec:hertz_potential}, the two conditions can be encoded by the real and imaginary part of a single complex equation. We get:
\begin{equation}
    i\star\dd\l(F^- + i \frac{1}{3}L^2\l(\lambda_{(e)}\star \mathcal C - \lambda_{(o)} \mathcal C\r)\,: F\r)=0\,,\label{eq:EOM_maxwell_forms}
\end{equation}
where, $F^-$ was defined in equation~\eqref{eq:define_F_C_+-}, and we define ``$:$'' as a double contraction. For some 4 tensor $\mathcal A_{ab}{}^{cd}$ and 2 tensor $\mathcal B_{ab}$ we have:
\begin{equation}
    \l(\mathcal A : \mathcal B\r)_{ab} := \mathcal A_{ab}{}^{cd}\mathcal B_{cd}\,.
\end{equation}
Assuming $\mathcal{C}$ and $F$ are real, the real part of~\eqref{eq:EOM_maxwell_forms} reduces to equation~\eqref{eq:EOM_maxwell_tensor}, whereas the imaginary part encodes the Bianchi identity. By solving the equation for a general $F$, we guarantee that both conditions are satisfied. This equation takes a simpler form, if we aggregate $\lambda_{(e)}$ and $\lambda_{(o)}$ into a single complex coupling constant:
\begin{equation}
    \lambda = \lambda_{(e)} + i \lambda_{(o)}\label{eq:define_complex_lambda}\,.
\end{equation}
Using the identities in section~\ref{subsec:background_hodge}, we get:
\begin{equation}
    i\star \dd\l(F^- - L^2 \frac{\bar\lambda}{6}\,\mathcal C^-:F^- + L^2\frac{\lambda}{6}\,\mathcal C^+:F^+\r)=0\,.\label{eq:EOM_maxwell_forms_simple}
\end{equation}
By definition, $F^+ = \bar F^-$, thus the \ac{EOM} depend on both, $F^-$ and its complex conjugate. Consequently, the equation is not holomorphic in $F^\pm$, and we must independently solve its real and imaginary parts.\\

To make progress, define $\mathcal F^\pm$, such that:
\begin{equation}
    \mathcal F^\pm = F^\pm - \l(\frac{L}{M}\r)^2 \l(\lambda_{(e)}\pm i  \lambda_{(o)}\r)M^2\mathcal C^\pm: F^\pm\,.\label{eq:define_CalF}
\end{equation}
Note that $\mathcal F^\pm$ obeys the same properties as $F^\pm$. We have:
\begin{equation}
\begin{aligned}
    \bar{\mathcal F}^\pm &= \mathcal F^\mp\,,\\
    \star \mathcal F^\pm &= \pm i  \mathcal F^\pm\,.
\end{aligned}
\end{equation}
Thus, we can define:
\begin{equation}
    \mathcal F = \mathcal F^+ + \mathcal F^-\,,
\end{equation}
as a $2$ form that reduces to $F$ in the $L\rightarrow 0$ limit. Working perturbatively in $L/M$, in terms of $\mathcal F^\pm$, the \ac{EOM} are:
\begin{equation}
    \star\dd\l(\mathcal F^- + L^2 \frac{\lambda}{6}\,\mathcal C^+ : \mathcal F^+\r)=\mathcal O\l(\frac{L}{M}\r)^4\,.\label{eq:EOM_maxwell_calF}
\end{equation}

To project this into a null tetrad basis, we must first define \ac{NP} scalars to encode $\mathcal F$. Following the convention in equation~\eqref{eq:define_maxwell_scalars}, we get:
\begin{equation}
\begin{aligned}
    \Phi_0&:=l^a m^b \mathcal F_{ab} = \l(1- \frac{1}{3} \bar \lambda\,L^2\,\psi_2\r)\phi_0 \,,\\
    \Phi_1&:=\frac{1}{2}\l(l^a n^b-m^a \bar m^b\r) \mathcal F_{ab}=\l(1+ \frac{2}{3}\bar \lambda\,L^2\,\psi_2\r)\phi_1\,,\\
    \Phi_2&:=\bar m^a n^b \mathcal F_{ab}=\l(1- \frac{1}{3} \bar \lambda\,L^2\,\psi_2\r)\phi_2 \,.
\end{aligned}\label{eq:define_calF_maxwell_scalars}
\end{equation}
In the limit $L\rightarrow 0$, $\Phi_i\rightarrow \phi_i$. The Weyl tensor also obeys a Bianchi identity $\nabla_{[e}\mathcal C_{ab]cd} = 0$. In a vacuum, this implies that $\nabla^a \mathcal C_{abcd} =0$. Putting the two identities together, we can prove:
\begin{equation}
    \l(\star \dd\,\mathcal C^+ : \mathcal F^+\r)_a = \mathcal C^+_{ab}{}^{cd} \nabla^b \mathcal F_{cd}^+\,.
\end{equation}
Using this, we can prove that in the null tetrad basis, \eqref{eq:EOM_maxwell_calF} reduces to:
\begin{equation}
    \mathcal M =\frac{1}{3} \bar \psi_2 L^2 \lambda\,J + \mathcal O\l(\frac{L}{M}\r)^4\,, \label{eq:maxwell_GHP_corrected}
\end{equation}
where:
\begin{equation}
    \mathcal M_{(\alpha)}:=
    \begin{pmatrix}
        \l(\eth ' -\tau'\r)\Phi_0-\l(\thn -2 \rho\r) \Phi_1\\
        \l(\thn'-2 \rho '\r) \Phi_1-\l(\eth - \tau\r) \Phi_2\\
        \l(\thn'- \rho'\r)\Phi_0-\l(\eth - 2 \tau\r)  \Phi_1\\
        \l(\eth ' - 2 \tau'\r)\Phi_1-\l(\thn-\rho\r)\Phi_2
    \end{pmatrix}_{(\alpha)}\,,\label{eq:maxwell_GHP_LHS}
\end{equation}
and:
\begin{equation}
J_{(\alpha)}:=
\begin{pmatrix}
    \l(\eth + 2\bar \tau ' \r)\bar\Phi_0+2 \l(\thn + \bar\rho \r)\bar\Phi_1\\
    -\l(\eth' + 2\bar \tau\r)\bar\Phi_2-2 \l(\thn' + \bar\rho' \r)\bar\Phi_1\\
    -\l(\thn + 2\bar \rho\r) \bar \Phi_2 - 2\l(\eth+ \bar \tau'\r)\bar \Phi_1\\
    \l(\thn' + 2\bar \rho'\r) \bar \Phi_0+2\l(\eth'+ \bar \tau\r)\bar \Phi_1 
\end{pmatrix}_{(\alpha)}\,.\label{eq:maxwell_GHP_RHS}
\end{equation}
As expected from the identities $n^a = \l(l^a\r)'$ and $\bar m^a = \l(m^a\r)'$, the $2^{nd}$ and $4^{th}$ line can be obtained from the $1^{st}$ and $3^{rd}$ by using the prime operation.\\

While $M_{(\alpha)}$ depends exclusively on $\Phi_i$, $J_{(\alpha)}$, is a function of $\Bar \Phi_i$. Thus, if we try to factor out $\exp(- i \omega t)$ from each $\Phi_i$, we will get a term proportional to $\exp(- i \omega t)$ on the \ac{LHS} and a term proportional to $\exp(i \bar \omega t)$ on the \ac{RHS}. The standard approach to define \acp{QNM} cannot be applied directly. As remarked before, this is a consequence of the non holomorphic character of equation~\eqref{eq:EOM_maxwell_forms_simple}. Physically, this happens because the \ac{EFT} correction breaks the degeneracy between the two polarizations of light. To tackle this issue, we must combine~\eqref{eq:maxwell_GHP_corrected} with its complex conjugate, and choose a basis for $\Phi_i$ and $\bar \Phi_i$, that leads to decoupled \ac{EOM}. This is analogous to finding a basis for the polarizations of light that propagate along the two effective light cones of the corrected theory. If we consider exclusively parity even corrections, we expect that this basis will be related with parity eigenstates. In section~\ref{subsec:EM_even}, we will make this statement more concrete. When considering mixed parity corrections, the answer is not as obvious. We will explore this case in section~\ref{subsec:EM_mixed}.

\subsection{Parity even corrections\label{subsec:EM_even}}

In this section, we set $\lambda_{(o)} =0$ and consider only parity even \ac{EFT} corrections. In section~\ref{subsubsec:EM_eikonal}, Using the effective metrics derived in~\cite{Baibhav:2023clw}, we will obtain an analytic expression for the \acp{QNM} in the large $\ell$ limit. Then, in section~\ref{subsubsec:EM_even_Schwarzschild}, we specialize the discussion to the Schwarzschild limit. Working non-perturbatively in $L/M$, we obtain decoupled \acp{ODE} for the \acp{QNM} that can readily be solved with the direct methods in the appendix~\ref{appendix:numerics}. We compare these equations with the ones derived in~\cite{Chen:2013ysa}, and show that they are equivalent. Finally, in section~\ref{subsubsec:EM_even_Kerr}, we generalize the discussion to Kerr \acp{BH} of any spin, obtaining decoupled \acp{PDE} for the \acp{QNM}, that can be solved with the pseudo-spectral approach outlined in~\ref{subsec:numerics_perturbation}.

\subsubsection{Eikonal limit \label{subsubsec:EM_eikonal}}

The Eikonal approximation, also known as the geometric optics limit, is valid when the wavelength of light, dubbed $\varepsilon$, is much smaller than the characteristic length scale of the background spacetime. For a Kerr background, this is the statement that $\alpha \ll 1$, with:
\begin{equation}
    \alpha = \frac{\varepsilon}{M}\,.
\end{equation}

In~\cite{Baibhav:2023clw}, the authors explored this limit for the parity even part of~\eqref{eq:Maxwell_Weyl_Action}, considering several algebraically special \ac{BH} spacetimes. In the beginning of this section, we will outline their approach, and obtain the effective metrics that encode the propagation of light on a Kerr background. Then, we will exploit the connection with the large $\ell$ limit of \acp{QNM} to derive analytic expressions for the frequencies in this limit.\\

We start with a plane wave \textit{Ansatz}:
\begin{equation}
    A^a = \mathcal A^a e^{i\frac{\Theta}{\alpha}}\,.\label{eq:Eikonal_Limit_Ansatz}
\end{equation}
We assume that the characteristic length scale of all variables is $M$, with the factor of $1/\alpha$ in the exponential ensuring that the signal has a short wavelength. Plugging this into equation~\eqref{eq:EOM_maxwell_tensor}, and setting $\lambda_{(o)}$ we get:
\begin{equation}
    C_{a b c d} K^b K^c \mathcal A ^d + \mathcal O\l(\frac{\varepsilon}{M}\r)=0\,.\label{eq:geometric_optics_MaxwellEqs}
\end{equation}
where:
\begin{equation}
    C_{ab}{}^{cd}= \delta_a^{[c}\,\delta_b^{d]} - \frac{1}{3}\, \lambda_{(e)} L^2\,\mathcal C_{ab}{}^{cd}\,,\label{eq:define_eikonal_C}
\end{equation}
$K_a:=\nabla_a \Theta$ is the wavevector of~\eqref{eq:Eikonal_Limit_Ansatz}, and parametrizes the propagation direction.\\

As pointed out in~\cite{Reall:2021voz}, there is an issue with this approach. By dimensional analysis, we have $\mathcal C \sim 1/M^2$, thud, the size of the correction in~\eqref{eq:define_eikonal_C} is:
\begin{equation}
    \l(\frac{L}{M}\r)^2 \lambda_{(e)} = \l(\frac{L}{\varepsilon}\r)^2 \l(\frac{\varepsilon}{M}\r)^2 \lambda_{(e)}\,.
\end{equation}
Now, the \ac{EFT} approximation is only valid, if $L$ is much smaller than any other length scale in the theory. In particular, $L/ \varepsilon \ll 1$, or else, higher order derivative corrections become relevant. Thus:
\begin{equation}
    \l(\frac{L}{M}\r)^2 \ll \l(\frac{\varepsilon}{M}\r)^2
\end{equation}
In the Eikonal approximation, we neglected all terms that are $\mathcal O(\varepsilon/M)$, thus we must also neglect the contribution of the \ac{EFT} correction. The \ac{EFT} correction is invisible in the Eikonal limit.\\

The same argument can be made on the \ac{QNM} side. Bellow, we will obtain an asymptotic expansion of the \ac{QNM} frequencies in the large $\ell$ limit. Roughly, we have:
\begin{equation}
    \omega = \ell\,\omega_{(1)} + \omega_{(0)} + \frac{1}{\ell} \omega_{(-1)} + \cdots
\end{equation}
The angular part of a \ac{QNM} with azimuthal number $\ell$ has $\ell$ nodes, thus, the length-scale of this mode is at least $M/\ell$. Hence, the validity of the \ac{EFT} approximation requires $L/M \ll 1/\ell$. We expect the correction to the \ac{QNM} frequency to be $\mathcal O(L/M)^2$ smaller than the background, thus it should only be present at $\mathcal O(1/\ell^2)$ above. We conclude that we cannot extract information from the Eikonal limit while being consistent with the \ac{EFT} approximation. In spite of these facts, throughout the remainder of this section, we compute the \ac{EFT} corrections to large $\ell$ \acp{QNM}, ignoring the requirement $L \ll \varepsilon$. We will do so exclusively to derive an analytical check to test our numerics.\\

$C_{ab}{}^{cd}$, usually known as the susceptibility tensor, is a rank-4 tensor with the same algebraic symmetries as the Weyl tensor. Thus, equation~\eqref{eq:geometric_optics_MaxwellEqs} is trivially solved if $\mathcal A^a$ is in the same direction as $K^a$, however, this is a pure gauge solution ($A_a \sim \nabla_a e^{i \Theta}$), and thus, unphysical. To obtain physical solutions, we must pick the direction of $K$ that the matrix $U_{ad} = C_{abcd} K^b K^c$ has at most rank 2. This requirement leads to the derivation of the Fresnel equation. This is a quartic equation on $K$, that takes the form:
\begin{equation}
    \mathcal G_{abcd} K^a K^b K^c K^d =0\,.\label{eq:Fresnel}
\end{equation}
where $\mathcal G^{abcd}$ is a cubic contraction of $C_{abcd}$ defined in equation 15 of~\cite{Balakin:2017eur}.\\

For a type-D spacetime, choosing a frame aligned with the two principal null directions, The Fresnel equation factorizes as the product of two equations, quadratic in $K$. We get \cite{Balakin:2017eur}:
\begin{equation}
    \l(\mathfrak g^{(+)}_{ab} K^a K^b\r)\l(\mathfrak g^{(-)}_{cd} K^c K^d\r)=0\,\label{eq:Fresnel_typeD}
\end{equation}
where:
\begin{equation}
    \mathfrak g^{(\pm)}_{ab} = g_{ab} + \sigma^{\pm}\l(m_a \bar m_b+ m_b \bar m_a\r)\,.\label{eq:define_effective_metrics}
\end{equation}
Here, $\mathfrak g^{(\pm)}_{ab}$ is the background Kerr metric, and:
\begin{equation}
    \sigma^{\pm}= \frac{\pm 2 \lambda_{(e)} L^2 \l|\psi_2\r|}{\l|1+\frac{1}{3}\lambda_{(e)} L^2\l(2\psi_2 - \bar \psi_2\r)\r| \mp\lambda_{(e)} L^2\l|\psi_2\r|}\,.\label{eq:define_sigma_pm}
\end{equation}
For $L\ll M$, $\mathfrak g^{(\pm)}_{ab}$ are non-degenerate symmetric 2-tensors, with Lorentzian signature. Thus, they can be understood as effective metrics, with the corresponding solutions $K^a$ following their null curves. In fact, because $K_a$ is the gradient of a scalar, we can prove that these curves must be geodesic. Thus, in the Eikonal limit, depending on the polarization, light will propagate along either of two distinct effective light cones. The \ac{EFT} correction effectively turns spacetime into a birefringent medium.\\

To draw a parallel between the analytic results obtained here and the numerical results obtained in section~\ref{subsubsec:EM_Kerr_results}, we will work perturbatively in $L/M$. At leading order, we have:
\begin{equation}
    \sigma^{\pm} = \pm2 \lambda_{(e)} L^2 \abs{\psi_2} + \mathcal O\l(\frac{L}{M}\r)^4\,\label{eq:define_sigma_perturbative}.
\end{equation}
Substituting this in equation~\eqref{eq:define_effective_metrics}, we get:
\begin{equation}
    \mathfrak g^{(\pm)}_{ab} = g_{ab} + 2\widetilde \lambda_{(\pm)} M^2 \abs{\psi_2} \,\l(m_a \bar m_b+ m_b \bar m_a\r)\,,\label{eq:define_effective_metrics_perturbative}
\end{equation}
where:
\begin{equation}
    \widetilde \lambda_{(\pm)} := \pm \l(\frac{L}{M}\r)^2 \lambda_{(e)}\,.
\end{equation}

For a Kerr spacetime, the large $\ell$ limit of \acp{QNM} is fully specified by circular photon orbits~\cite{Mashhoon:1985cya,Dolan:2010wr,Yang:2012he}. In fact, $\ell =\pm m$ \acp{QNM} are related to the two sets of unstable equatorial null geodesics. The following relation holds:
\begin{equation}
    \omega_{\ell =\pm m} = \ell\,\Omega +\mathcal O\l(\ell^{0}\r) + i\l(\l(\frac{1}{2} + n\r) \gamma^ L +\mathcal O\l(\ell^{-2}\r)\r)\,.\label{eq:QNM_Eikonal_correspondence}
\end{equation}
Here, $\Omega = \frac{d\phi}{dt}$ represents the angular frequency of the orbit, and $\gamma^ L$ is the corresponding Lyapunov exponent, which indicates the instability time scale with respect to the coordinate $t$. The order of magnitude of the corrections was derived in~\cite{Dolan:2010wr}, by comparing the Eikonal limit with a WKB expansion of the Teukolsky equation. We will make the logical leap that the results generalize to photons propagating in the effective metrics~\eqref{eq:define_effective_metrics}. Intuitively, this should be the case because of the similarities between the geometric optics approximation and a WKB expansion, however, without a more formal treatment, we cannot make a statement regarding the quality of the approximation. $\ell = m$ modes relate to null-geodesics that \textit{co-rotate} with the \ac{BH}, while $\ell=-m$ modes are specified by \textit{counter-rotating} photons.\\

\begin{figure*}[ht]
    \subfloat[\label{subfig:eikonal_limit_plot_background}]{\includegraphics[width=0.457\textwidth]{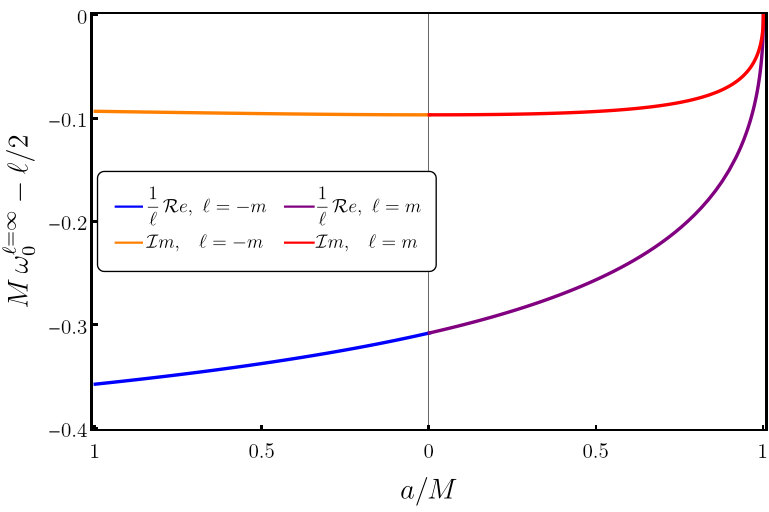}}\hskip 2em
    \subfloat[\label{subfig:eikonal_limit_plot_correction}]{\includegraphics[width=0.457\textwidth]{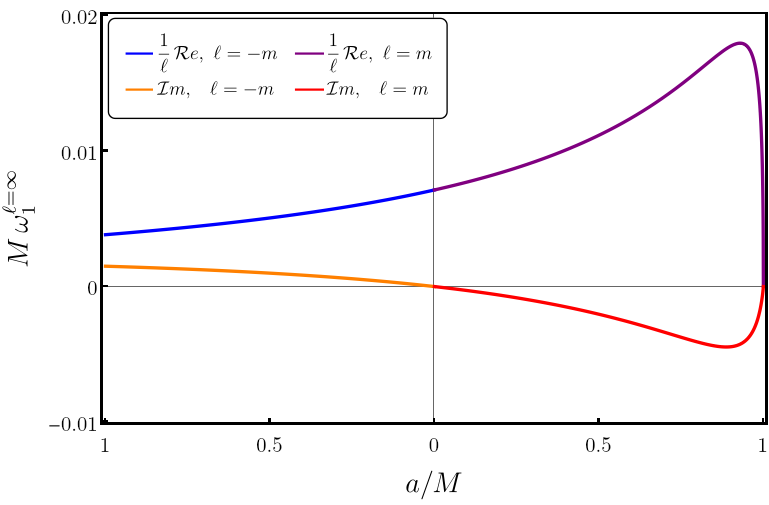}}

\caption{Here, we show the prediction for the large $\ell$ limit of electromagnetic \ac{QNM} frequencies. On the left, we have the background component, and on the right, we present the corresponding correction, as defined in equation~\eqref{eq:schematic_correction_vector_QNFs}. For each plot, the left side showcases the $\ell=-m$ modes, with $a/M \in[0,1]$, running from the middle to the left. Conversely, the right side illustrates the $\ell=m$ modes, with $a/M\in[0,1]$, from the middle to the right. We arrange the plots in this manner to underscore the equivalence between the $m\rightarrow -m$ operation and $a\rightarrow -a$. In the near extremal limit, for $\ell=m$ modes, we find that $M\omega_0^{\ell=m}\rightarrow \ell/2$, and $M\omega_1^{\ell=m}\rightarrow 0$. Finally, note that in the Schwarzschild limit, $\mathcal Im (\omega_1)\rightarrow 0$, as can be deduced from equation~\eqref{eq:powers_of_gamma_L}\label{fig:eikonal_limit_plot}}
\end{figure*}

Perturbatively, the two effective metrics are related by reversing the sign of $\lambda$. Thus, in the Eikonal limit, the corresponding \acp{QNM} must be related by reversing the sign of $\lambda$. Thus, following the convention in equation~\eqref{eq:schematic_correction_vector_QNFs}, we can conclude that:
\begin{equation}
    \omega^{\abs{\ell}=\abs{m}\rightarrow\infty}_{\pm} = \omega_0 \mp \l(\frac{L}{M}\r)^2 \lambda_{(e)}\,\omega_1 + \mathcal O\l(\frac{L}{M}\r)^4\,.\label{eq:QNM_frequencies_splitting_Eikonal}
\end{equation}

Null geodesics extremize the Lagrangian,
\begin{equation}
    \mathcal L = \frac{1}{2}\mathfrak g^{(\pm)}_{ab} \dv{x^a}{\chi}\dv{x^b}{\chi}\,.
\end{equation}
where $\chi$ is some affine parameter. By definition $\mathcal L =0$, furthermore due to the symmetries of $\mathfrak g$, we have that:
\begin{equation}
\begin{aligned}
    \mathcal E =\pdv{\mathcal L}{\l(\dd t/ \dd \chi\r)},\\
    L =\pdv{\mathcal L}{\l(\dd \phi/ \dd \chi\r)}\,,
\end{aligned}
\label{eq:define_energy_vars}
\end{equation}
are conserved quantities. Inverting this relation, we can express $(\dd t / \dd \chi, \dd \phi / \dd \chi)$ in terms of $(\mathcal E, L)$. Because we are looking for equatorial geodesics, $x=\dd x/ \dd \chi = 0$. Substituting all of this in $\mathcal L =0$, and solving with respect to $\dd r/ \dd \chi$, we get:
\begin{equation}
    \l(\frac{1}{\mathcal E}\dv{r}{\chi}\r)^2 = V(r,b):= V_0(r, b) + \widetilde \lambda_{(\pm)} V_1(r, b)\,,\label{eq:define_V}
\end{equation}
where:
\begin{equation}
\begin{aligned}
    V_0(r, b) &=\frac{2 (a+b)^2\,M+(a^2-b^2)\,r+r^3}{4 r^3}\,,\\
    V_1(r, b) &=\frac{(a+b)^2 \Delta_r}{2 r^7}M^2\,,
\end{aligned}
\end{equation}
and:
\begin{equation}
    b= \frac{L}{\mathcal E}\,
\end{equation}
is the \textit{impact parameter} of the null geodesic. Taking the derivative of~\eqref{eq:define_V} with respect to $\chi$, we see that circular geodesics happen if and only if:
\begin{equation}
\begin{cases}
    V(r, b)=0\,,\\
    \pdv{}{r}V(r, b)=0\,.
\end{cases}\label{eq:algebraic_EOM_geometric_optics}
\end{equation}

To solve this, we will make the \textit{Ansatz}:
\begin{equation}
\begin{aligned}
    r&= r_0 + \widetilde \lambda_{(\pm)} r_1 + \mathcal O\l(\widetilde \lambda_{(\pm)}\r)^2\,,\\
    b&= b_0 + \widetilde \lambda_{(\pm)} b_1 + \mathcal O\l(\widetilde \lambda_{(\pm)}\r)^2\,.
\end{aligned}\label{eq:radius_b_epansion}
\end{equation}
At $\mathcal O\l(\widetilde \lambda_{(\pm)}{}^0\r)$, the equations reduce to:
\begin{equation}
\begin{cases}
    r_0\,(r_0 -3 M)^2=4 a^2 M\,,\\
    b_0=a\frac{r_0 + 3M}{r_0 -  3M}\,.
\end{cases}\label{eq:0_order_solution_geometric_optics}
\end{equation}
The cubic equation has two roots for $r> r_+$, given by~\cite{Dolan:2010wr}:
\begin{equation}
    r_0^{\l(\pm\r)} = 2 M\l(1 + \cos\l(\frac{2}{3}\cos^{-1}\l(\mp \frac{a}{M}\r)\r)\r)
\end{equation}
Notice that $r_0^{+} \leq 3M \leq r_0^{-}$, with equality only for $a=0$. The $+$ case describes photons that \textit{co-rotate} with the \ac{BH}, being dual to \acp{QNM} with positive real part, whereas the $-$ case describes \textit{counter-rotating} photons, dual to \acp{QNM} with negative real part. This $\pm$ is independent from the one in $\widetilde \lambda_{(\pm)}$, thus, for clarity, we will omit it below.\\

The first order is a linear equation for $(r_1, b_1)$, thus we can solve it directly. Using equation~\eqref{eq:0_order_solution_geometric_optics} to simplify the resulting expression, we get:
\begin{equation}
\begin{aligned}
    b_1&=M\,\frac{(r_0-3M) (r_0-M)}{4\,a\,r_0}\,,\\
    r_1&=M^2\,\frac{5 a^2\,+\,r_0\,(4\,r_0-9\,M)}{3\,r_0^3}\,.
\end{aligned}
\end{equation}

We can now compute the relevant parameters for equation~\eqref{eq:QNM_Eikonal_correspondence}. The Lyapunov exponent is defined as:
\begin{equation}
    \gamma^L:= \sqrt{\frac{\mathcal E^2\,\partial^2V(r,b)/ \partial r ^2}{(\dd t/ \dd \chi)^2}}\,.
\end{equation}
Expanding in powers of $\widetilde \lambda_{(\pm)}$, we get:
\begin{equation}
    \gamma^L = \gamma^L_0 + \widetilde \lambda_{(\pm)}\gamma^L_1 + \mathcal O\l(\widetilde \lambda_{(\pm)}\r)^2\label{eq:define_gamma_L}
\end{equation}
with:
\begin{equation}
\begin{aligned}
    \gamma^L_0&=\frac{\sqrt{3}}{r_0} \frac{r_0-M}{r_0+3\,M}\,,\\
    \gamma^L_1&=a\,\gamma^L_0 M \frac{(r_0-M) (5\,r_0+9\,M)}{12\,b_0 r_0^3}\,.
\end{aligned}\label{eq:powers_of_gamma_L}
\end{equation}

Similarly, the orbital frequency is given by:
\begin{equation}
    \Omega := \frac{\dd \phi / \dd \chi}{\dd t / \dd \chi} = -\frac{1}{b_0}\,+\,\widetilde \lambda_{(\pm)}\frac{\gamma^L_0}{\sqrt{3}}{\,b_0\,r_0}+ \mathcal O\l(\widetilde \lambda_{(\pm)}\r)^2
\end{equation}
Plugging these results into equation~\eqref{eq:QNM_Eikonal_correspondence}, and using the convention in equation~\eqref{eq:QNM_frequencies_splitting_Eikonal}, we get an analytic estimate for the correction to \ac{QNM} frequencies in the large $\ell$ limit. The results are summarized in figure~\ref{fig:eikonal_limit_plot}.

\subsubsection{Schwarzschild limit \label{subsubsec:EM_even_Schwarzschild}}

In the Schwarzschild limit, considering parity even \ac{EFT} corrections, decoupled \ac{EOM} for the \acp{QNM} were first obtained in~\cite{Chen:2013ysa}. There, the authors worked directly with the vector potential $A_a$ and obtaining the \ac{EOM} using vector spherical harmonics. In this section, we will arrive at the same equations independently, by working with the gauge invariant components of $F_{ab}$, using the \ac{GHP} formalism. This way, it will be easier to relate our results with the ones obtained later for general Kerr \acp{BH} in terms of the same variables. For non-spinning \acp{BH}, the spacetime is spherically symmetric, thus, the \ac{EOM} are much simpler. This allows us to proceed non-perturbatively in $L/M$, ignoring the \ac{EFT} validity conditions. We will do so to relate our results with the ones in~\cite{Chen:2013ysa}, but we must bear in mind that only solutions with $L\ll M$ are relevant from an \ac{EFT} perspective. Spherical symmetry simplifies the \ac{GHP} formalism. In addition to the \ac{GHP} identities that are satisfied for any type-D spacetime (see equation~\eqref{eq:type_D_simplifications}), we have:
\begin{equation}
\begin{alignedat}{2}
    &\rho = \bar \rho, \quad\quad&&\rho'= \bar \rho'\\
    &\psi_2 = \bar \psi_2, &&\tau= \tau' = 0
\end{alignedat}\label{eq:GHP_identies_Schwarzschild}
\end{equation}
Furthermore, all scalars depend exclusively on the radial coordinate $r$. Projecting equation~\eqref{eq:EOM_maxwell_forms_simple} into a tetrad basis, and setting $\lambda_{o}=0$, we get:
\begin{widetext}
\begin{equation}
\begin{pmatrix}
    \eth '\phi_0-\l(\thn- 2 \rho\r)\phi_1\\
    \eth \phi_2-\l(\thn' -2 \rho'\r)\phi_1\\
    \l(\thn'-\rho '\r)\phi_0-\eth \phi_1\\
    \l(\thn- \rho\r)\phi_2-\eth '\phi_1
\end{pmatrix}
=\frac{1}{3}\lambda_{(e)} L^2\psi_2
\begin{pmatrix}
\eth '\phi_0+\eth \bar \phi_0+2 \l(\thn+ \rho\r)\l(\phi_1+\bar \phi_1\r)\\
\eth \phi_2+\eth' \bar \phi_2+2 \l(\thn'+ \rho'\r)\l(\phi_1+\bar\phi_1\r)\\
\l(\thn' + 2 \rho'\r)\phi_0-\l(\thn +2 \rho\r)  \bar \phi_2+2\,\eth \l(\phi_1-\bar \phi_1\r)\\
\l(\thn + 2 \rho\r)\phi_2-\l(\thn' +2 \rho'\r)  \bar \phi_0+2\,\eth' \l(\phi_1-\bar \phi_1\r)
\end{pmatrix}
\label{eq:maxwell_eqs_GHP_Schwarzschild}
\end{equation}
\end{widetext}

To simplify these equations, our strategy is to express them as a system of real equations of real variables. $\phi_1$ has spin weight $0$ (see section~\ref{subsec:GHP_intro}), thus its \ac{GHP} weights are invariant under complex conjugation. That is why, in the \ac{RHS} of~\eqref{eq:maxwell_eqs_GHP_Schwarzschild}, we see terms that explicitly depend on the real and imaginary part of $\phi_1$, without breaking \ac{GHP} covariance. This would not be possible for $\phi_0$ and $\phi_2$, as they have spin weight $1$ and $-1$ respectively. To represent them with real, \ac{GHP}-covariant variables, we must first raise / lower their spin weight by acting with $\eth$ and $\eth'$ respectively. Define the polarized Maxwell \ac{GHP} scalars as:
\begin{equation}
\hskip -.35cm
\begin{alignedat}{3}
&P_0^+=\mathcal Re\l(\eth' \phi_0\r),\, &&P_1^+=\mathcal Re\l(\phi_1\r),\, &&P_2^+=\mathcal Re\l(\eth\phi_2\r)\,,\\
&P_0^-=\mathcal Im\l(\eth' \phi_0\r),\, &&P_1^-=\mathcal Im\l(\phi_1\r),\, &&P_2^-=\mathcal Im\l(\eth\phi_2\r)\,.\\
\end{alignedat}
\end{equation}

By definition, for any \ac{GHP} gauge choice, these variables are real. On a similar note, the first two equations in~\eqref{eq:maxwell_eqs_GHP_Schwarzschild} have spin-weight $0$, whereas the others have spin-weight $1$ and $-1$ respectively. Thus, after applying $\eth'$ and $\eth$ to the $3^{rd}$ and $4^{th}$ equations, the real and imaginary parts of the resulting system are \ac{GHP}-covariant. Remarkably, after extensive use of \ac{GHP} identities~\cite{Geroch:1973am}, we can prove that the real part of equation~\eqref{eq:maxwell_eqs_GHP_Schwarzschild} depends exclusively on $P_i^+$. Similarly, the imaginary part depends exclusively on $P_i^-$. Thus, the \ac{EOM} for the two polarizations decouple:
\begin{widetext}
\begin{equation}
\begin{pmatrix}
    \l(\thn -2 \rho\r)P_1^+ -P_0^+\\
    \l(\thn' -2 \rho'\r)P_1^+ -P_2^+\\
    \eth\,\eth'\,P_1^+ - \l(\thn'-2\rho '\r)  P_0^+ \\
    \eth\,\eth'\,P_1^+ - \l(\thn-2\rho \r)  P_2^+
\end{pmatrix}
=\frac{1}{3} \lambda_{(e)} L^2 \psi_2
\begin{pmatrix}
    4\l(\thn+ \rho\r)P_1^+  + 2 P_0^ +\\
    4\l(\thn'+ \rho'\r)P_1^+  + 2 P_2^ +\\
    \l(\thn' + \rho '\r)P_0^+ - \l(\thn+ \rho\r)P_2^+\\
    \l(\thn + \rho \r)P_2^+ - \l(\thn'+ \rho'\r)P_0^+
\end{pmatrix}\,,
\label{eq:EOM_Schwarzschild_polarization_plus}
\end{equation}
and:
\begin{equation}
\begin{pmatrix}
    \l(\thn - 2 \rho\r)P_1^- -P_0^-\\
    \l(\thn' - 2 \rho'\r)P_1^- -P_2^-\\
    \eth\,\eth'\,P_1^- - \l(\thn'-2\rho '\r)  P_0^- \\
    \eth\,\eth'\,P_1^- - \l(\thn-2\rho \r)  P_2^-
\end{pmatrix}
= \frac{1}{3} \lambda_{(e)} L^2 \psi_2
\begin{pmatrix}
    0 \\
    0 \\
    4 \eth\,\eth'\,P_1^- +\l(\thn' + \rho '\r)P_0^- + \l(\thn+ \rho\r)P_2^-\\
    4 \eth\,\eth'\,P_1^- +\l(\thn' + \rho '\r)P_0^- + \l(\thn+ \rho\r)P_2^-\\
\end{pmatrix}\,.
\label{eq:EOM_Schwarzschild_polarization_minus}
\end{equation}
\end{widetext}

In the Schwarzschild limit, $\thn$ and $\thn'$ reduce to $(t,r)$ derivative operators, while $\eth$ and $\eth'$ act exclusively on the angle variables. Thus, the only angular dependence in equations~\eqref{eq:EOM_Schwarzschild_polarization_plus} and~\eqref{eq:EOM_Schwarzschild_polarization_minus}, is encoded in the $(\eth\,\eth')$ operator. There is a known relationship between $(\eth\,\eth')$ and the spherical harmonics $Y_{\ell m}$~\cite{Penrose:1985bww}. We have:
\begin{equation}
    \eth\,\eth'Y_{\ell m} = - \frac{\ell (\ell+1)}{2 r^ 2}Y_{\ell m}\,\label{eq:eigenstates_Ylm}.
\end{equation}

Given $\thn Y_{\ell m} = \thn' Y_{\ell m}=0$, a decomposition of the scalars in spherical harmonics removes all angular dependence. Similarly, we can factor out time dependence by multiplying out $\exp(-i \omega t)$.\footnote{Both $Y_{\ell m}$ and $\exp(-i \omega t)$ are complex valued, however, we argued that $P_i^{\pm}$ must be real valued functions. Because the \ac{EOM} are now real, we can work with a complex-valued \textit{Ansatz} and in the end simply extract its real part.} $P_0^{\pm}$ and $P_2^{\pm}$ have boost weight $1$ and $-1$ respectively. Thus, to work with $\{0,0\}$ scalars, we will explicitly factor out $\rho$ and $\rho'$ respectively. The final \textit{Ansatz} will be:
\begin{equation}
\begin{aligned}
    P_0^{\pm}(t,r,x,\phi) &= e^{- i \omega t} Y_{\ell m}(x, \phi)\,\rho\,R_0^{\pm}(r)\,,\\
    P_1^{\pm}(t,r,x,\phi) &= e^{- i \omega t} Y_{\ell m}(x, \phi)\,R_1^{\pm}(r)\,,\\
    P_2^{\pm}(t,r,x,\phi) &= e^{- i \omega t} Y_{\ell m}(x, \phi)\,\rho'\,R_2^{\pm}(r)\,.
\end{aligned}
\end{equation}

Equations~\eqref{eq:EOM_Schwarzschild_polarization_plus} and~\eqref{eq:EOM_Schwarzschild_polarization_minus} reduce to two systems of four first-order \acp{ODE} for the three radial variables, an overdetermined description. For both polarizations, we can find linear combinations of the first two equations, that eliminate $R_1^{\pm}(r)$ and $\dd R_1^{\pm}(r)/ \dd r$ from the $3^{rd}$ and $4^{th}$ equations, reducing the problem to a system of two \acp{ODE} with two variables. Finally, we define $u^\pm(r)$ and $v^\pm(r)$, such that:
\begin{equation}
\begin{aligned}
    R_0^\pm(r)&=\frac{1}{r\,f\sqrt{\Xi}}\l(\sqrt{\Xi}^{\pm1}u^\pm(r)\,+\,\sqrt{\Xi}^{\mp1}\,v^\pm(r) \r)\,,\\
    R_2^\pm(r)&=\frac{1}{r\,f\sqrt{\Xi}}\l(\sqrt{\Xi}^{\pm1}u^\pm(r) - \sqrt{\Xi}^{\mp1}\,v^\pm(r) \r)\,,
\end{aligned}
\end{equation}
with
\begin{equation}
\begin{aligned}
    f &:= 1 - \frac{2 M}{r}\,,\\
    \Xi(r) &:=1 + \,\frac{2}{3} \,\widetilde \lambda_{(e)}\,\l(\frac{M}{r}\r)^3\,,
\end{aligned}    
\end{equation}
where we defined:
\begin{equation}
    \widetilde \lambda_{(e)} :=\l(\frac{L}{M}\r)^2 \lambda_{(e)}\label{eq:define_widetilde_lambda_e}
\end{equation}

The \ac{EOM} reduce to:
\begin{equation}
\begin{cases}
\dv{}{r}v^\pm +\Xi^{\pm1}\frac{i \omega}{f}\,u^\pm=0\\
\dv{}{r}u^\pm +\l(\Xi^ {\mp 1} \frac{i \omega }{f}+\frac{1}{i \omega}V_\pm\r) v^\pm=0
\end{cases}\,,\label{eq:decoupled_system_maxwell_Sch_even}
\end{equation}
where:
\begin{equation}
\begin{aligned}
    &V_\pm:=\frac{\ell(\ell+1)}{r^2}\zeta^{\mp1}\,,\\
    &\zeta:=1 - \frac{4}{3}\widetilde \lambda_{(e)}\l(\frac{M}{r}\r)^3\,.
\end{aligned}
\end{equation}
These equations can be combined into a single decoupled second-order \ac{ODE} for $v^\pm$. We get:
\begin{equation}
\dv{r}\Xi^{\mp1}\,f\,\dv{r} v^\pm\,+\,\l(\Xi^{\mp1}\,\frac{\omega^2}{f} - V_\pm\r)v^\pm=0\,.\label{eq:decoupled_maxwell_Sch_even}
\end{equation}

After suitable variable changes, we can prove that this equation agrees with the effective potentials derived in~\cite{Chen:2013ysa} (equations 16 and 17).\footnote{To match the conventions in that paper, we must take $\lambda \rightarrow 12\,\alpha/L^2$.} In the $\widetilde \lambda_{(e)}\rightarrow0$ limit, $\Xi\rightarrow 1$, $\zeta\rightarrow 1$, and the \ac{EOM} reduce to the wave equation for electromagnetic fields in a Schwarzschild \ac{BH} background, see~\cite{Motl:2003cd}.\\

As outlined in section~\ref{subsec:QNMs_intro}, to find the \ac{QNM} spectrum of~\eqref{eq:decoupled_maxwell_Sch_even}, we must set \acp{BC} that are ingoing at $r\rightarrow r_+$, and outgoing at $r\rightarrow\infty$. $\Xi$ and $\zeta$ are smooth at $r=r_+$, and asymptote to $1 +\mathcal O (1/r^3)$ at $r\rightarrow \infty$. Thus, at leading order, a Frobenius expansion near the boundaries is independent of $\widetilde \lambda_{(e)}$. Hence, to set the appropriate \acp{BC}, we can proceed exactly as in section~\ref{subsec:QNMs_intro}. Define $\widetilde v^\pm(r)$:
\begin{equation}
    \widetilde v^\pm(r) = \frac{1}{\alpha^{a=0}_r(r)}v^\pm(r)\,
\end{equation}
with:
\begin{equation}
    \alpha^{a=0}_r(r)=f^{-2 i \omega M}\,\l(\frac{r}{M}\r)^{2 i \omega M}\,e^{i r \omega}
\end{equation}

We change variables to the compactified radial coordinates $y=1-r/r_+$, and plug this in equation~\eqref{eq:decoupled_maxwell_Sch_even}. \acp{QNM} are solutions to the resulting equation that are smooth for $y\in[0,1]$. Treating $\widetilde \lambda_{(e)}$ as a free parameter. this equation can be solved numerically, using pseudospectral methods, as outlined in the appendix~\ref{appendix:numerics}. The \ac{EOM} are not perturbative in $\widetilde \lambda_{(e)}$, and can, in theory, be solved for $\widetilde \lambda_{(e)} \in(-12,6)$.\footnote{This is the range such that $\zeta$ and $\Xi$ have no roots for $r>2M$.} However, the \ac{EFT} approximation entails $\widetilde \lambda_{(e)} \ll 1$, thus, we should disregard the remaining of the parameter space. Expanding equation~\eqref{eq:decoupled_maxwell_Sch_even} in powers of $\widetilde \lambda_{(e)}$, we see that at leading order, the \ac{EOM} for the $\pm$ polarizations can be mapped into each other by changing $\widetilde \lambda_{(e)} \rightarrow - \widetilde \lambda_{(e)}$. Thus, at leading order, the two corrections to the \ac{QNM} frequencies are equal and opposite. This is in agreement with what we postulated in equation~\eqref{eq:schematic_correction_vector_QNFs}.\footnote{The convention we picked in this section actually differs from the one in~\eqref{eq:schematic_correction_vector_QNFs} by a factor of $(-1)^{(\ell + m)}$.}

\subsubsection{Generalization to Kerr \label{subsubsec:EM_even_Kerr}}

In this section, we will solve equation~\eqref{eq:maxwell_GHP_corrected}, considering only parity even \ac{EFT} corrections (i.e. $\lambda_{(o)} =0$), for a Kerr \ac{BH} background of any spin. To do so, we will make explicit use of the \ac{EFT} validity condition, and work perturbatively in $L/M$, or equivalently, in terms of $\widetilde \lambda_{(e)}$, defined in~\eqref{eq:define_widetilde_lambda_e}.\\

Take a generic differential equation:
\begin{equation}
    Q_{(0)} + \widetilde \lambda_{(e)} \l(\mathcal G Q_{(0)} + Q_{(1)}\r) + \mathcal O(\widetilde \lambda_{(e)}^2) =0,
\end{equation}
where $Q_{(0)}$, $\mathcal G$ and $Q_{(1)}$ are arbitrary. At $\mathcal O(\widetilde \lambda_{(e)}^0)$, we have $Q_{(0)}= \mathcal O(\widetilde \lambda_{(e)}^1)$. Substituting this on the $\mathcal O(\widetilde \lambda_{(e)}^1)$ part of the equation, we get:
\begin{equation}
    Q_{(0)} + \widetilde \lambda_{(e)} Q_{(1)} + \mathcal O(\widetilde \lambda_{(e)}^2) =0,
\end{equation}

Thus, any identity that is satisfied at $\mathcal O(\widetilde \lambda_{(e)}^0)$ can be used to simplify the $\mathcal O(\widetilde \lambda_{(e)}^1)$ terms. Following this idea, we equate $\mathcal M_{(\alpha)}=0$ and solve with respect to the \ac{GHP} derivatives of $\phi_1$. The result can be used to eliminate all derivatives of $\bar \phi_1$ from $J_{(\alpha)}$. Then, we apply the Teukolsky $\zeta^a$ operator, defined in equation~\eqref{eq:define_zeta}, to both sides of equation~\eqref{eq:maxwell_GHP_corrected}. Using the \ac{GHP} identities in~\cite{Geroch:1973am} we get:
\begin{equation}
    \mathcal O_1\Phi_0 =\lambda_{(e)} L^2 \bar\psi_2 \mathcal J_i \bar \Phi_i + \mathcal O(\widetilde \lambda_{(e)}^2)\,,\label{eq:corrected_Teulkosky}
\end{equation}
where the summation in $i$ is implicit, $\mathcal J_i$ are linear operators:
\begin{equation}
\begin{aligned}
    \mathcal J_0 &= \l(2 \l(\tau -2 \bar{\tau }'\r)-\eth\r)\eth +2\bar\tau'^2\,,\\
    \mathcal J_1 &= 6\l(\rho  \bar\tau'+\bar\rho \l(\tau -4 \bar\tau'\r)\r)\,,\\
    \mathcal J_2 &= \l(2 \l(\rho -2 \bar{\rho }\r)-\thn\r) \thn+2 \bar\rho ^2\,,
\end{aligned}
\end{equation}
and $\mathcal O_s$ is the master Teukolsky operator defined in equation~\eqref{eq:teuk_expanded}.\\

From section~\ref{subsec:hertz_potential}, we know that there is a Hertz potential $\psi_H$ such that:
\begin{equation}
\begin{aligned}
    &\mathcal O_{-1} \psi_H =0\,,\\
    &\phi_i =  \mathcal T_i \bar \psi_H + \mathcal O(\widetilde \lambda_{(e)}^1)\,,
\end{aligned}\label{eq:phi_i_to_order_1}
\end{equation}
where $\mathcal T_i$ was defined in equation~\eqref{eq:define_Ti}. Substituting this in equation~\eqref{eq:corrected_Teulkosky}, we get:
\begin{equation}
\begin{cases}
    \mathcal O_1 \Phi_0 =\widetilde \lambda_{(e)}\mathcal D \psi_H + \mathcal O(\widetilde \lambda_{(e)}^2)\\[.5em]
    \Phi_0= \mathcal T_0 \bar\psi_H + \mathcal O(\widetilde \lambda_{(e)}^1)
\end{cases}\,,\label{eq:corrected_Teulkosky_compact}
\end{equation}
with:
\begin{equation}
    \mathcal D:=M^2 \bar \psi_2\,\mathcal J_i\,\bar{\mathcal T_i}\,.\label{eq:define_D}
\end{equation}

Equation~\eqref{eq:corrected_Teulkosky_compact} mixes modes proportional to $S$ with modes proportional to $\bar S$. If we set $\Phi_0$ and $\psi_H$ proportional to $S=\exp(-i \omega t + i m \phi)$, $S$ factors out nicely from both sides of the first equation in~\eqref{eq:corrected_Teulkosky_compact}. However, for the second equation, we get $S$ on the \ac{LHS} and $\bar S$ on the \ac{RHS}. The reverse problem occurs if we choose $\psi_H$ to be proportional to $\bar S$ instead. In section~\ref{subsec:background_parity}, we studied the properties of the \textit{conjugate-parity-transform} operator, $\mathcal P$, defined in equation~\eqref{eq:define_cal_P}. We proved, that due to the parity invariance of the Kerr metric, the operator commutes with the master Teukolsky operator $\mathcal O_s$. Consequently, we found that if $\psi^{(s)}=S(t,\phi) \Psi_s(r,x)$ is a \ac{QNM} of $\mathcal O_s$ then so is $\mathcal P\psi^{(s)} = (-1)^m\bar S(t,\phi) \Psi_s(r,-x)$. Taking linear combinations of these states, we can express the kernel of $\mathcal O_s$ in a basis of $\mathcal P$ eigenstates. We restricted to parity invariant \ac{EFT} corrections, thus, we should be able to express solutions to equation~\eqref{eq:corrected_Teulkosky_compact} as eigenstates of $\mathcal P$. Take:
\begin{equation}
\begin{aligned}
    \Phi_0^{\pm} &:=\Phi_0 \pm \mathcal P \Phi_0\,,\\
    \psi_H^{\pm} &:=\psi_H \pm \mathcal P \phi_H\,.
\end{aligned}\label{define_cal_P_eigenstates}
\end{equation}
By definition, $\mathcal P \Phi_0^{\pm} = \pm \Phi_0^{\pm}$ and $\mathcal P \psi_H^{\pm} = \pm \psi_H^{\pm}$, thus we associate these with the \textit{polar} and \textit{axial} polarizations of light respectively.\footnote{The \textit{polar} polarization is parity even, whereas its \textit{axial} counterpart must be parity odd, hence this classification.} Because, $\mathcal O_s$, $\mathcal D$ and $\mathcal T_0$ have even spin-weight, they all commute with $\mathcal P$:
\begin{equation}
    [\mathcal P,\mathcal O_s] = [\mathcal P,\mathcal D] = [\mathcal P,\mathcal T_0]=0\,.
\end{equation}
Thus, applying $\mathcal P$ to both sides of equation~\eqref{eq:corrected_Teulkosky_compact}, we get: 
\begin{equation}
\begin{cases}
    \mathcal O_1\,\mathcal P\Phi_0 =\widetilde \lambda_{(e)}\mathcal D \,\mathcal P\psi_H + \mathcal O(\widetilde \lambda_{(e)}^2)\\[.5em]
    \mathcal P\Phi_0= \mathcal T_0\,\mathbb P\psi_H + \mathcal O(\widetilde \lambda_{(e)}^1)
\end{cases}\,,\label{eq:corrected_Teulkosky_compact_parity}
\end{equation}
where $\mathbb P$ is the \textit{parity transform} operator, (see section~\ref{subsec:background_parity}). Writing equations~\eqref{eq:corrected_Teulkosky_compact} and~\eqref{eq:corrected_Teulkosky_compact_parity} in the $\mathcal P$ eigenstates basis, we get:
\begin{equation}
\begin{cases}
    \mathcal O_1\Phi_0^\pm = \widetilde \lambda_{(e)} \mathcal D \psi_H^\pm+\mathcal O(\widetilde \lambda_{(e)}^2)\\[.5em]
    \Phi_0^\pm= \pm\mathcal T_0\,\mathbb P\psi_H^\pm + \mathcal O(\widetilde \lambda_{(e)}^1)
\end{cases}\,.\label{eq:corrected_Teulkosky_parity_eigenstates}
\end{equation}

Because $[\mathcal P, \mathcal O_{-1}]=0$, we have $\mathcal O_{-1} \psi_H^\pm=0$. However, in a Kerr background, \acp{QNM} of the Teukolsky equation are not, in general, eigenstates of $\mathcal P$, consequently $\psi_H^\pm$ cannot be a \ac{QNM} solution. The same can be said about the $\mathcal O(\widetilde \lambda_{(e)}^0)$ component of $\Phi_0^\pm$. Nevertheless, we can choose $\psi_H^\pm$ and $\Phi_0^{\pm}$ to be a linear combination of modes proportional to $S$ and $\bar S$. Extending this to first order in $\widetilde \lambda_{(e)}$, we take:
\begin{equation}
\begin{aligned}
    \Phi_0^\pm &= \Phi \pm \mathcal P \Phi\,,\\
    \psi_H^\pm &= \Psi \pm \mathcal P \Psi\,,
\end{aligned}\label{eq:define_Phi_Psi}
\end{equation}
with $\Psi$ and $\Phi$ proportional to $S$:\footnote{It may seem that $\Phi=\Phi_0$ and $\Psi=\psi_H$, however that is not the case. As argued below equation~\eqref{eq:define_D}, $(\Phi_0, \psi^H)$ cannot be proportional to $S$ or $\bar S$, thus, necessarily $\Phi$ and $\Psi$ are different variables.}
\begin{equation}
\begin{aligned}
    \Psi(t,r,x,\phi)&=S(t,\phi)\,\Psi_{r,x}(r,x)\,,\\
    \Phi(t,r,x,\phi)&=S(t,\phi)\,\Phi_{r,x}(r,x)\,.\\
\end{aligned}
\end{equation}
$\mathcal P\Psi$ and $\mathcal P\Phi$ will be proportional to $\bar S$, thus substituting this in equation~\eqref{eq:corrected_Teulkosky_parity_eigenstates} we get:
\begin{equation}
\hskip -.4cm
\begin{cases}
    S(t,\phi)\,\mathcal F^\pm(r,x)= \mp (-1)^m\bar S(t,\phi)\,\mathcal P \mathcal F^\pm(r,x) \\[.5em]
    S(t,\phi)\,\mathcal G^\pm(r,x)= \mp (-1)^m\bar S(t,\phi)\,\mathcal P \mathcal G^\pm(r,x)
\end{cases}\,,\label{eq:corrected_Teukolsky_ansatz}
\end{equation}
where:
\begin{equation}
\begin{aligned}
    \mathcal F^\pm(r,x) &:=\mathcal O_{1,\omega} \Phi_{r,x}(r,x) - \widetilde \lambda_{(e)} \mathcal D_\omega\, \Psi_{r,x}(r,x) + \mathcal O(\widetilde \lambda_{(e)}^2)\,,\\[.5em]
    \mathcal G^\pm(r,x) &:=\Phi_{r,x} \mp(-1)^m\,\mathcal T_{0,\omega}\,\mathbb P \Psi_{r,x} + \mathcal O(\widetilde \lambda_{(e)}^1)\,,
\end{aligned}
\end{equation}
with:
\begin{equation}
\begin{aligned}
    \mathcal D_\omega &= \frac{1}{S} \mathcal D S\,,\\
    \mathcal T_{0,\omega} &= \frac{1}{S} \mathcal T_0 S\,,\\
    \mathcal O_{1,\omega} &= \frac{1}{S} \mathcal O_1 S\,.\\
\end{aligned}
\end{equation}
$S$ and $\bar S$ are linearly independent, thus, their pre-factors must vanish. We get:
\begin{subnumcases}{\hskip -.5cm\label{eq:corrected_Teukolsky_rx}}
    \mathcal F^\pm = 0\,,\label{eq:corrected_Teukolsky_rx_equation}\\
    \mathcal G^\pm =0\,.\label{eq:corrected_Teukolsky_rx_normalization}
\end{subnumcases}

Equation~\eqref{eq:corrected_Teukolsky_rx_normalization} relates the background of $\Phi_{r,x}$ with the background of $\Psi_{r,x}$. We choose the latter to be a \ac{QNM} of the $s=-1$ Teukolsky equation, and consequently a Hertz potential for the background \ac{QNM} solutions. This implies that, in the limit $\widetilde \lambda_{(e)}\rightarrow0$,  $\Phi$ must be a \ac{QNM} of the $s=1$ Teukolsky equation. We substitute this in equation~\eqref{eq:corrected_Teukolsky_rx_equation}, and work perturbatively in $\widetilde \lambda_{(e)}$ to find the \ac{QNM} spectrum of the full theory. Concretely, we will solve~\eqref{eq:corrected_Teukolsky_rx_equation} in two steps. First, treating $\omega$ as a parameter independent of $\widetilde \lambda_{(e)}$, we set the appropriate \ac{QNM} \acp{BC}. As outlined in section~\ref{subsec:QNMs_intro}, this is done by factoring out the appropriate corresponding singular behaviour. Then, we expand $\omega$ in powers of $\widetilde \lambda_{(e)}$, and solve equation~\eqref{eq:corrected_Teukolsky_rx} perturbatively. If we don't follow this sequence, we cannot ensure that the \acp{BC} for \acp{QNM} are properly enforced, and we might get logarithmic singularities plaguing the numerics.\\

The singular behaviour of $\Psi_{r,x}$ is well understood. Working with the compactified radial coordinate, we have:
\begin{equation}
    \Psi_{r,x}(r(y),x) = \alpha_{-1}(y,x) \widetilde\Psi_{y,x}(y,x)\,, \label{eq:expand_psi_all_vars}
\end{equation}
where $\alpha_{-1}$, defined in~\eqref{eq:define_alpha_s}, is singular at the domain border, and $\widetilde\Psi_{y,x}$ is smooth for $(y,x)\in[0,1]\cross[-1,1]$. Working perturbatively in $\widetilde \lambda_{(e)}$, $\Phi_{r,x}$ takes the form:
\begin{widetext}
\begin{equation}
    \Phi_{r,x}(r(y),x)=\l(\alpha_1(y,x) + \widetilde \lambda_{(e)}\,\delta\alpha(y,x)\r)\l(\widetilde\Phi_{y,x}^{(0)}(y,x)+\widetilde \lambda_{(e)} \Phi_{y,x}^{(1)}(y,x)\r) + \mathcal O(\widetilde \lambda_{(e)}^2)\,.
\end{equation}
\end{widetext}

By definition, $\widetilde\Phi_{y,x}^{(0)}$ is smooth for $(y,x)\in[0,1]\cross[-1,1]$. For $\Phi$ to be a \ac{QNM}, we require $\widetilde\Phi_{y,x}^{(1)}$ to be smooth in the same interval. $\delta \alpha$ is included to ensure that this is the case. It acts as a counter-term to cancel non-smooth behaviour. With a Frobenius analysis of equation~\eqref{eq:corrected_Teukolsky_final}, we can actually prove that $\delta \alpha = 0$.\footnote{This is expected if we take the view of \acp{QNM} as modes that are smooth at $\mathcal H^+$ and through a conformal compactification of $\mathcal J^+$. The \ac{EFT} correction does not affect the spacetime in these regions, thus we do not expect it to change the \acp{BC} for \acp{QNM}.} Because this term, simply overloads notation, we will omit it in the remainder of the discussion. Inserting this in equation~\eqref{eq:corrected_Teukolsky_rx}, we get:
\begin{equation}
\begin{cases}
    \widetilde{\mathcal O}_{1,\omega} \widetilde \Phi_{y,x}^{(0)} = \widetilde \lambda_{(e)}\l(- \widetilde{\mathcal O}_{1,\omega} \widetilde\Phi_{y,x}^{(1)}+ \frac{\alpha_{-1}}{\alpha_1}\widetilde{\mathcal D}_\omega \widetilde \Psi_{y,x}\r)+ \mathcal O (\widetilde \lambda_{(e)}^2)\\[.5em]
    \widetilde \Phi_{y,x}^{(0)}=\pm\,(-1)^m\widetilde{\mathcal T}_{0,\omega}\,\mathbb P\widetilde\Psi_{y,x} + \mathcal O (\widetilde \lambda_{(e)}^1)
\end{cases}\label{eq:corrected_Teukolsky_rx_non_singular}
\end{equation}
\vskip 1em
\noindent where $\,\widetilde{\,}\,\,$ identifies $\alpha_{\pm1}$, accordingly, was commuted past the corresponding operator, and we used $\mathbb P\mathcal \alpha_{-1} = \mathcal \alpha_{1}$. Finally, we expand $\omega$ in powers of $\widetilde \lambda_{(e)}$:
\begin{equation}
    \omega =  \omega_0 + \widetilde \lambda_{(e)}\,\omega_1 +  \mathcal O (\widetilde \lambda_{(e)}^2)\,,
\end{equation}
where $\omega_0$ is the background \ac{QNM} frequency. We substitute this in~\eqref{eq:corrected_Teukolsky_rx_non_singular} and equate each order of $\widetilde \lambda_{(e)}$ to $0$. The $0^{th}$ order is satisfied by definition, and the $1^{st}$ order yields:
\begin{equation}
    L^{(0)}\,\widetilde \Phi_{y,x}^{(1)} = \l(\mp (-1)^m\omega_1\,\delta L\, + L^{(1)} \r)\widetilde \Psi_{y,x}\,,\label{eq:corrected_Teukolsky_final}
\end{equation}
where:
\begin{equation}
\begin{aligned}
    L^{(0)} &= \widetilde{\mathcal O}_{1,\omega_0}\,,\\
    L^{(1)} &=\frac{\alpha_{-1}}{\alpha_1}\widetilde{\mathcal D}_{\omega_0}\,,\\
    \delta L &=\partial_{\omega_0}\widetilde{\mathcal O}_{1,\omega_0}\widetilde{\mathcal T}_0\,\mathbb P\,.
\end{aligned}
\end{equation}
Mapping $\omega_1 \rightarrow - \omega_1$ takes the $+$ into the $-$ polarization. Thus, for each background frequency $\omega_0$, the two possible corrections have opposing sign. This is in agreement with what was postulated in equation~\eqref{eq:schematic_correction_vector_QNFs} for $\lambda_{(o)} =0$. Consequently, we find agreement with the results in the Eikonal limit, (section~\ref{subsubsec:EM_eikonal}) and the results derived for the Schwarzschild limit (section~\ref{subsubsec:EM_even_Schwarzschild}). $\widetilde \Psi_{y,x}$ separates as the product of a radial and an angular function. Thus, we can use the corresponding \ac{EOM} to iteratively reduce the order of the derivatives of $\widetilde \Psi_{y,x}$ in equation~\eqref{eq:corrected_Teukolsky_final} into at most first order. The resulting equation is in the form of equation~\eqref{eq:EOM_general}, and can be solved with the pseudo-spectral methods outlined the in appendix~\ref{appendix:numerics}.\\

\subsection{Mixed parity corrections\label{subsec:EM_mixed}}

In this section, we will generalize the discussion of~\ref{subsec:EM_even} to mixed parity \ac{EFT} corrections, i.e. modes where both $\lambda_{(e)}$ and $\lambda_{(o)}$ are non-zero. In section~\ref{subsec:EM_even}, the main non-trivial step was to find a basis such that the \ac{EOM} for the two possible electromagnetic polarizations decouple. In the Schwarzschild limit, we found that this could be done by taking the real and imaginary parts of suitably transformed \ac{EOM}. These components are invariant under complex conjugation, thus, by taking the real and imaginary part of the \ac{EOM}, we are effectively projecting them into a basis of \textit{complex conjugation eigenstates}. To generalize the result to Kerr \acp{BH}, we followed a similar approach, projecting the \ac{EOM} into a basis of eigenstates of the \textit{parity complex conjugation operator} $\mathcal P$. In both cases, we expected the approach to work because of the parity invariance of the equations of motion.\footnote{In the Schwarzschild limit, due to spherical symmetry, when we take the real and imaginary part of the \ac{EOM}, we are also indirectly projecting into a basis of eigenstates of $\mathcal P$. We did not make this more explicit because the \ac{EOM} look slightly more complicated.} In this section, we are no longer dealing with parity invariant \ac{EFT} corrections, thus the approach should break down. Notwithstanding, it may be possible to choose a different basis that shares the same properties as the ones in section~\ref{subsec:EM_even}, where the equations decouple.\\

Define $\mathcal A_\delta$, with $\delta \in [-\pi,\pi)$, as an operator that takes $\alpha\in \mathbb C$ and projects perpendicularly onto the line in the complex plane defined by $r e^{i \delta}, r\in \mathbb R$. We get:
\begin{equation}
    \mathcal A_\delta\,\alpha := \frac{1}{2}\l(e^{-i \delta} \alpha + e^{i \delta} \bar \alpha\r)\,.\label{eq:define_calA_delta}
\end{equation}
As expected, $\mathcal A_0\,\alpha = \mathcal Re(\alpha)$ and $\mathcal A_{\frac{\pi}{2}}\,\alpha = \mathcal Im(\alpha)$. Furthermore, note that the output is always a real number, and thus invariant under complex conjugation. Crucially, if we know the projection of $\alpha$ at two different angles, $\delta_1, \delta_2$, as long as they do not differ by an integer multiple of $\pi$, we can always recover $\alpha$ and $\bar \alpha$:
\begin{equation}
\begin{aligned}
    \alpha&=2\,\frac{e^{-i\delta_1} \mathcal A_{\delta_1}\,\alpha -e^{-i\delta_2} \mathcal A_{\delta_2}\,\alpha}{e^{-2i\delta_1} - e^{-2i\delta_2}}\,,\\[.5em]
    \bar\alpha&=2\,\frac{e^{i\delta_1} \mathcal A_{\delta_1}\,\alpha -e^{i\delta_2} \mathcal A_{\delta_2}\,\alpha}{e^{2i\delta_1} - e^{2i\delta_2}}\,.
\end{aligned}\label{eq:inverse_calA_delta}
\end{equation}
$\mathcal A_\delta$, allows for a systematic way to find a basis to express $(\alpha, \bar \alpha)$ whose components are real valued. Note that $\mathcal A_\delta$ is periodic in $\delta$. For any integer $k$, we have:
\begin{equation}
    \mathcal A_{\delta + k \pi}\,\alpha= e^{i k \pi} \mathcal A_\delta\,\alpha\,.\label{eq:periodicity_calA}
\end{equation}
Furthermore, flipping the sign of $\delta$ is equivalent to taking the complex conjugate of the argument:
\begin{equation}
    \mathcal A_{-\delta}\,\alpha= \mathcal A_\delta\,\bar \alpha\,.
\end{equation}
Finally, note that rotating $\alpha$ by some phase $\gamma \in [-\pi, \pi)$, is the same as rotating the $\delta$ line in the opposite direction, i.e. $\delta \rightarrow \delta - \gamma$:
\begin{equation}
    \mathcal A_\delta e^{i \gamma}\alpha = \mathcal A_{\delta - \gamma} \alpha
\end{equation}

Now, lets employ this approach to a toy problem, similar to equation~\eqref{eq:EOM_maxwell_calF}. Imagine we want to solve the following complex valued differential equation:
\begin{equation}
    A \psi + \lambda\,B \bar \psi=0\,.\label{eq:toy_equation_mixed_parity}
\end{equation}
Here, $A,B$ are real valued differential operators, $\psi$ is a complex valued function, and $\lambda \in \mathbb C$ has phase $\lambda_\theta \in [-\pi, \pi)$:
\begin{equation}
    \lambda:= \abs{\lambda} e^{i \lambda_\theta}
\end{equation}
Because the equation depends on both $\psi$ and $\bar \psi$, it is not holomorphic in $\psi$. Thus, we need to explicitly solve the real and imaginary parts of the equation, and this is in general non trivial. However, there might be a lines in the complex plane where the equations decouple. Applying $\mathcal A_\delta$ to both sides, and using the fact that $A$ and $B$ \textit{commute with complex conjugation}, we get:
\begin{equation}
    A\,\mathcal A_\delta\,\psi +\abs{\lambda} B\,\mathcal A_{\lambda_\theta-\delta}\,\psi=0\label{eq:calA_transformed_toy}
\end{equation}
Now, we want to pick two distinct values of $\delta$ such that~\eqref{eq:calA_transformed_toy} depends on a single real function, i.e. $\mathcal A_\delta \psi \sim \mathcal A_{\lambda_\theta-\delta}$. Using equation~\eqref{eq:periodicity_calA}, we get two distinct classes of values for $\delta$, labelled as $\delta_\pm$:
\begin{equation}
\begin{aligned}
    \delta_+ &= - \delta_+ + \lambda_\theta +  2 k_+ \pi\\
    \delta_- &= - \delta_- + \lambda_\theta +  (2 k_-+1) \pi
\end{aligned}
\end{equation}
where $k_\pm \in \mathbb Z$. Solving this, we get:
\begin{equation}
\begin{aligned}
    \delta_+ &= \frac{\lambda_\theta}{2}+ k_+ \pi\,,\\
    \delta_- &= \frac{\lambda_\theta}{2}+ \frac{\pi}{2} + k_-\pi\,.
\end{aligned}
\end{equation}
Plugging this into~\eqref{eq:calA_transformed_toy} we get two decoupled equations:
\begin{equation}
\begin{cases}
    (A+ \abs{\lambda} B)\,\mathcal A_{\delta_+}\,\psi=0\\
    (A- \abs{\lambda} B)\,\mathcal A_{\delta_-}\,\psi=0
\end{cases}\label{eq:decoupled_toy}
\end{equation}
We can solve this in terms of $\mathcal A_{\delta_+}\,\psi$ and $\mathcal A_{\delta_-}\,\psi$\footnote{All values of $k_+,k_-$ yield the same \ac{EOM}.}, and then, using equation~\eqref{eq:inverse_calA_delta}, recover the value of $\psi$ and $\bar \psi$. Note that the second equation can be obtained from the first by mapping $\abs{\lambda}\rightarrow-\abs{\lambda}$.\\

Working perturbatively in $L/M$, in section~\ref{subsubsec:EM_mixed_Schwarzschild}, we will use this approach to correct the \acp{QNM} of Schwarzschild \acp{BH}. Then, in section~\ref{subsubsec:EM_mixed_Kerr}, we will generalize the approach to find the corrections to \acp{QNM} of Kerr \acp{BH} of any spin. To do so, we will define an analogue of $\mathcal A_\delta$ that depends on $\mathcal P$ instead of the complex conjugation operator. In both cases, we will show that \ac{QNM} frequencies reduce to the results of section~\ref{subsec:EM_even} with $\lambda_{(e)}$ replaced by $\abs{\lambda} = \sqrt{\lambda_{(e)}^2 + \lambda_{(o)}^2}$, in agreement with equation~\eqref{eq:schematic_correction_vector_QNFs}.

\subsubsection{Schwarzschild limit\label{subsubsec:EM_mixed_Schwarzschild}}

In this section, we will derive the \ac{EOM} for the corrected \acp{QNM} of a Schwarzschild \acp{BH}, when considering mixed parity \ac{EFT} corrections. In section~\ref{subsubsec:EM_even_Schwarzschild}, we computed the corrected \ac{EOM} for $\lambda_{(o)}=0$, non perturbatively in $L/M$. It is much harder to proceed non-perturbatively for mixed parity corrections, as there isn't an obvious way to decouple the \ac{EOM}. Nevertheless, validity of the \ac{EFT} approximation requires $L\ll M$. Thus, only the leading order correction is physically relevant and we can safely proceed perturbatively.\\

Our starting point is equation~\eqref{eq:maxwell_GHP_corrected}. Making use of the \ac{GHP} identities that are specific to spherically symmetric backgrounds (equation~\eqref{eq:GHP_identies_Schwarzschild}), the \ac{EOM} simplify. We get:
\begin{equation}
    \mathcal M =\frac{1}{3} \psi_2 L^2 \lambda\,J + \mathcal O\l(\frac{L}{M}\r)^4\,. \label{eq:maxwell_GHP_corrected_Sch}
\end{equation}
where:
\begin{equation}
    \mathcal M_{(\alpha)}=
    \begin{pmatrix}
        \eth'\,\Phi_0-\l(\thn -2 \rho\r) \Phi_1\\
        \l(\thn'-2 \rho '\r)\,\Phi_1-\eth\,\Phi_2\\
        \l(\thn'- \rho'\r)\,\Phi_0- \eth\,\Phi_1\\
        \eth '\,\Phi_1-\l(\thn-\rho\r)\,\Phi_2
    \end{pmatrix}_{(\alpha)}\,,\label{eq:maxwell_GHP_LHS_Sch}
\end{equation}
and:
\begin{equation}
J_{(\alpha)}=
\begin{pmatrix}
    \eth\,\bar\Phi_0+2 \l(\thn + \bar\rho \r)\bar\Phi_1\\
    -\eth'\,\bar\Phi_2-2 \l(\thn' + \bar\rho' \r)\bar\Phi_1\\
    -\l(\thn + 2\bar \rho\r) \bar \Phi_2 - 2 \eth\,\bar \Phi_1\\
    \l(\thn' + 2\bar \rho'\r) \bar \Phi_0+2\eth'\bar \Phi_1 
\end{pmatrix}_{(\alpha)}\,.\label{eq:maxwell_GHP_RHS_Sch}
\end{equation}

The \ac{EOM} are non-holomorphic, as they depend on both $\Phi_i$, and $\bar \Phi_i$. To tackle this, we will project the equations along suitably chosen directions of the complex plane, as outlined in the beginning section~\ref{subsec:EM_mixed}. However, the action of $\mathcal A_\delta$ breaks \ac{GHP} covariance. When we take $\mathcal A_\delta \Phi_0$, we are adding $\Phi_0$ and $\bar \Phi_0$, \ac{GHP} scalars with different spin weights. Thus, as was done in section~\ref{subsubsec:EM_even_Schwarzschild}, we must reduce / increase the spin weight of $\Phi_0 / \Phi_2$ respectively, to work with \textit{spinless} scalars. We define the following set of quantities with $0$ spin weight:
\begin{equation}
    P_0 :=\eth' \Phi_0\,,\quad P_1 :=\Phi_1\,, \quad P_2 :=\eth \Phi_2\,.
\end{equation}
Similarly, we must work with \ac{EOM} of vanishing spin. To set the spin weights of the $m$ and $\bar m$ components of~\eqref{eq:maxwell_GHP_corrected_Sch} to $0$, we act with $\eth'$ and $\eth$ respectively. The resulting \ac{EOM} will contain factors of $\eth \eth' P_1$ and $\eth \eth' \bar P_1$. As argued in section~\ref{subsubsec:EM_even_Schwarzschild}, there is a relationship between the $\eth \eth'$ operator and the spherical harmonics $Y_{\ell m}$, see~\eqref{eq:eigenstates_Ylm}. Using the fact that in the Schwarzschild limit, $\thn, \thn'$, are exclusively $(t, r)$ derivatives and $\eth, \eth'$ act exclusively in $(x,\phi)$, we can generalize this relation. For $\eta$, a \ac{GHP} scalar with spin weight $0$, that can be decomposed as a linear combination of spherical harmonics with angular quantum number $\ell$:
\begin{equation}
    \eta(t,r,x,\phi) := \sum_{m=-\ell}^{m=\ell} \eta_m(t,r) Y_{\ell m}(x,\phi)\,, \label{eq:decompose_spherical_harmonics}
\end{equation}
we have:
\begin{equation}
    \eth \eth' \eta=- \frac{\ell (\ell+1)}{2 r^ 2} \eta\,.
\end{equation}
Complex conjugation preserves the angular quantum number, thus it is safe to assume that $P_1$ and $\bar P_1$ obey this condition. In terms of $P_i$ the 0 spin-weight \ac{EOM} are:\footnote{Note that $\widetilde{\mathcal M}$ and $\widetilde{J}$ are no longer the tetrad components of some tensor.}
\begin{equation}
    \widetilde{\mathcal M}_i\,P_i =\frac{1}{3} \psi_2 L^2 \lambda\,\widetilde J_i \bar P_i + \mathcal O\l(\frac{L}{M}\r)^4\,. \label{eq:maxwell_GHP_corrected_Sch_spinless}
\end{equation}
where the summation in $i$ is implicit and:
\begin{equation}
\begin{aligned}
    \widetilde{\mathcal M}_i P_i&:=
    \begin{pmatrix}
        -\l(\thn- 2 \rho\r) \,P_1+P_0\\[.3em]
        \l(\thn'-2\rho'\r) \,P_1-P_2\\[.3em]
        \l(\thn'-2\rho'\r)\,P_0+\frac{\ell(\ell+1)}{2 r^2}\,P_1 \\[.3em]
        -\l(\thn- 2 \rho\r)\,P_2 -\frac{\ell(\ell+1)}{2 r^2}\,P_1
    \end{pmatrix}\,,\\[1em]
\widetilde J_i \bar P_i&:=
\begin{pmatrix}
   2\l(\thn+\rho\r)\,\bar P_1\,+\,\bar P_0\\[.3em]
   -2\l(\thn'+\rho'\r)\,\bar P_1-\bar P_2\\[.3em]
   -\l(\thn+\rho\r)\,\bar P_2+\frac{\ell(\ell+1)}{r^2}\,\bar P_1\\[.3em]
   \l(\thn'+\rho'\r)\,\bar P_0-\frac{\ell(\ell+1)}{r^2}\,\bar P_1 
\end{pmatrix}\,.
\end{aligned}\label{eq:maxwell_GHP_Sch_spinless_components}
\end{equation}

Note that the \ac{LHS} depends exclusively on $P_i$, whereas the \ac{RHS} is a function of $\bar P_i$. Furthermore, for all $i$, $\widetilde{\mathcal M}_i$ and $\widetilde J_i$ are real operators, the only complex parameter in~\eqref{eq:maxwell_GHP_corrected_Sch_spinless} is $\lambda$. Thus, the \ac{EOM} have the same form as~\eqref{eq:toy_equation_mixed_parity}, and we can decouple them into two polarizations following the same procedure.\footnote{Note that in~\eqref{eq:toy_equation_mixed_parity} we work with a single scalar $\psi$, whereas here we have a vector $\psi_i$. Notwithstanding, because all $\widetilde{J}_i$ commute with complex conjugation, the same approach works.} We find two phases, $\delta_\pm$, that decouple the \ac{EOM}. Taking $\lambda_\theta$ to be the phase of $\lambda$, we get:
\begin{equation}
\begin{aligned}
    \delta_+ &=\frac{\lambda_\theta}{2}\,,\\
    \delta_- &=\frac{\lambda_\theta}{2}+ \frac{\pi}{2}\,.
\end{aligned}
\end{equation}
The decoupled variables are:
\begin{equation}
    \mathcal P_i^\pm := \mathcal A_{\delta^\pm}P_i\,
\end{equation}
And the \ac{EOM} separate as:
\begin{equation}
    \l(\widetilde{\mathcal M}_i \mp \frac{1}{3} \psi_2 L^2 \abs{\lambda}\,\widetilde J_i\r)\,\mathcal P_i^\pm =\mathcal O\l(\frac{L}{M}\r)^4\,. \label{eq:maxwell_GHP_corrected_Sch_spinless_decoupled}
\end{equation}

Now, to derive radial decoupled \ac{EOM}, we follow an identical procedure to the one in section~\ref{subsubsec:EM_even_Schwarzschild}. First, we factorize the angular and radial dependence:\footnote{$\mathcal P_i^{\pm}$ are real variables, but we chose a complex valued Ansatz. Because the \ac{ODE} operators are real valued, we can fix this by taking the real value of the solution in the end.}
\begin{equation}
\begin{aligned}
    \mathcal P_0^{\pm}(t,r,x,\phi) &= e^{- i \omega t} Y_{\ell m}(x, \phi)\,\rho(r)\,R_0^{\pm}(r)\,,\\
    \mathcal P_1^{\pm}(t,r,x,\phi) &= e^{- i \omega t} Y_{\ell m}(x, \phi)\,R_1^{\pm}(r)\,,\\
    \mathcal P_2^{\pm}(t,r,x,\phi) &= e^{- i \omega t} Y_{\ell m}(x, \phi)\,\rho'(r)\,R_2^{\pm}(r)\,.
\end{aligned}
\end{equation}
The spherical harmonics factor out, as there are no angular derivatives in~\eqref{eq:maxwell_GHP_corrected_Sch_spinless_decoupled}. For each polarization, we get a system of four first-order \acp{ODE} of three radial variables. This is over-determined, thus, as was done in~\ref{subsubsec:EM_even_Schwarzschild}, we can find linear combinations of the first two equations that eliminate $R_1^{\pm}(r)$ and $\dd R_1^{\pm}(r)/ \dd r$ from the $3^{rd}$ and $4^{th}$ equations. The remaining two equations can be decoupled by setting:
\begin{equation}
\begin{aligned}
    R_0^\pm(r)&=\frac{1}{r\,f(r)}\l(\xi_\pm\,u^\pm(r)\,+\,\xi_\mp\,\,v^\pm(r) \r)\,,\\
    R_2^\pm(r)&=\frac{1}{r\,f(r)}\l(\xi_\pm\,u^\pm(r)\,-\,\xi_\mp\,\,v^\pm(r) \r)\,,\\
\end{aligned}
\end{equation}
where:
\begin{equation}
\begin{aligned}
    \xi_\pm &:=1\pm\frac{1}{3}\abs{\lambda}\l(\frac{L}{M}\r)^2\l(\frac{M}{r}\r)^3 \,,\\[.3em]
    f(r)&:=1- \frac{2M}{r}\,.
\end{aligned}
\end{equation}
The \ac{EOM} reduce to:
\begin{equation}
\begin{cases}
\dv{}{r}v^\pm +\Xi_\pm\,\frac{i \omega}{f}\,u^\pm=\mathcal O\l(\frac{L}{M}\r)^4\\
\dv{}{r}u^\pm +\l(\Xi_\mp \frac{i \omega }{f}+\frac{1}{i \omega}V_\pm\r) v^\pm=\mathcal O\l(\frac{L}{M}\r)^4
\end{cases}\,,\label{eq:decoupled_maxwell_Sch_mixed}
\end{equation}
where:
\begin{equation}
\begin{aligned}
    V_\pm&:=\frac{\ell(\ell+1)}{r^2}\l(1\pm \frac{4}{3}\abs{\lambda}\l(\frac{L}{M}\r)^2\l(\frac{M}{r}\r)^3\r)\,,\\
    \Xi_\pm&:=1\pm\abs{\lambda}\frac{2}{3}\l(\frac{L}{M}\r)^2\l(\frac{M}{r}\r)^3 \,.
\end{aligned}
\end{equation}

Expanding equation~\eqref{eq:decoupled_maxwell_Sch_even} to leading order in $\widetilde \lambda_{(e)}$, and replacing $\widetilde \lambda_{(e)}$ with $\l(L/M\r)^2\abs{\lambda}$, we recover~\eqref{eq:decoupled_maxwell_Sch_mixed}. Thus, to leading order in $L/M$, we find agreement between the \acp{QNM} obtained directly for parity even \ac{EFT} corrections, and the more general mixed parity case. Both cases are captured by equation~\eqref{eq:schematic_correction_vector_QNFs}.

\subsubsection{Generalization to Kerr\label{subsubsec:EM_mixed_Kerr}}

We can now generalize our conclusions to Kerr \acp{BH} of any spin. As before, we will prove that the corrections to \ac{QNM} frequencies can be obtained from the parity even results, section~\ref{subsubsec:EM_even_Kerr}, by simply replacing $\lambda_{(e)}$ with $\abs{\lambda} =\sqrt{\lambda_{(e)}^2 + \lambda_{(o)}^2}$. As in the previous section, the starting point will be equation~\eqref{eq:maxwell_GHP_corrected}. Following the procedure outlined in section~\ref{subsec:EM_even} until equation~\eqref{eq:corrected_Teulkosky_compact}, but keeping $\lambda_{(o)}\neq0$, we get:
\begin{equation}
\begin{cases}
    \mathcal O_1 \Phi_0 -\l(\frac{L}{M}\r)^2\,\lambda\,\mathcal D \psi_H = \mathcal O\l(\frac{L}{M}\r)^4\\[.5em]
    \Phi_0- \mathcal T_0 \mathbb P\,\mathcal P\psi_H = \mathcal O\l(\frac{L}{M}\r)^2
\end{cases}\,,\label{eq:corrected_Teulkosky_compact_mixed}
\end{equation}
where we remind the reader that $\Phi_0$ is the component of the corrected electromagnetic tensor we want to determine, $\psi_H$ is a Hertz potential for the background, $\lambda = \lambda_{(e)} + i \lambda_{(o)}$ is a complex $\mathcal O(1)$ constant, $\mathbb P$ is the \textit{parity} operator, and $\mathcal P$ the \textit{conjugate-parity-transform} operator (see section~\ref{subsec:background_parity}). $\mathcal D$ and $\mathcal T_0$, defined in section~\ref{subsubsec:EM_even_Kerr}, are \ac{GHP} operators with spin-weight $s=0$. In the parity even section, we have proved that $\mathcal P$ commutes with all operators in~\eqref{eq:corrected_Teulkosky_compact_mixed}:
\begin{equation}
    [\mathcal P,\mathcal O_s] = [\mathcal P,\mathcal D] = [\mathcal P,\mathcal T_0]=[\mathcal P,\mathbb P]=0\,.
\end{equation}
Equation~\eqref{eq:corrected_Teulkosky_compact_mixed} is almost in the same form as equation~\eqref{eq:toy_equation_mixed_parity}. However, in the place of $A, B$, differential operators that commute with complex conjugation, we have operators that commute with $\mathcal P$. This motivates the definition of a new projection operator $\mathcal B_\delta$, analogous to $\mathcal A_\delta$ defined in~\eqref{eq:calA_transformed_toy}. For some complex function $\eta$, we have:
\begin{equation}
    \mathcal B_\delta\,\eta := \frac{1}{2}\l(e^{-i \delta} \eta + e^{i \delta} \mathcal P \eta\r)\,.\label{eq:define_calB_delta}
\end{equation}
As argued at the end of section~\ref{subsec:background_parity}, $\mathcal P$ preserves \ac{GHP} weight, thus this is also true for $\mathcal B_\delta$. Furthermore, note that all the properties listed for $\mathcal A_\delta$ at the beginning of section~\ref{subsec:EM_mixed}, generalize to $\mathcal B_\delta$, by replacing \textit{complex conjugation} with $\mathcal P$.\\

Because $\mathcal P$ commutes with all the \ac{GHP} operators in equation~\eqref{eq:corrected_Teulkosky_compact_mixed}, so does $\mathcal B_\delta$. Acting with it on both equations, we get:
\begin{equation}
\begin{cases}
    \mathcal O_1\,\mathcal B_\delta \Phi_0 -\l(\frac{L}{M}\r)^2\,\abs{\lambda}\,\mathcal D\,\mathcal B_{\delta- \lambda_\theta}\psi_H = \mathcal O\l(\frac{L}{M}\r)^4\\[.5em]
    \mathcal B_\delta \Phi_0- \mathcal T_0 \mathbb P\,\mathcal B_{-\delta}\psi_H = \mathcal O\l(\frac{L}{M}\r)^2
\end{cases}\,,\label{eq:corrected_Teulkosky_compact_mixed_calB}
\end{equation}
where $\lambda_\theta$ is the phase of $\lambda$. In both equations we see $\mathcal B_\delta \Phi_0$, however, in the top equation it is related to $\mathcal B_{\delta- \lambda_\theta}\psi_H$, whereas in the bottom equation we relate it to $\mathcal B_{-\delta}\psi_H$. To get decoupled equations we must have $\mathcal B_{-\delta}\psi_H \sim \mathcal B_{\delta- \lambda_\theta}\psi_H$. This condition is entirely analogous to what we explored in the toy model at the beginning of section~\ref{subsec:EM_mixed}. We find two decoupled equations, analogous to~\eqref{eq:decoupled_toy}:
\begin{equation}
\begin{cases}
    \mathcal O_1\,\mathcal B_{\delta^\pm} \Phi_0 \mp\l(\frac{L}{M}\r)^2\,\abs{\lambda}\,\mathcal D\,\mathcal B_{-\delta^\pm}\psi_H = \mathcal O\l(\frac{L}{M}\r)^4\\[.7em]
    \mathcal B_{\delta^\pm} \Phi_0 - \mathcal T_0 \mathbb P\,\mathcal B_{-\delta^\pm}\psi_H = \mathcal O\l(\frac{L}{M}\r)^2
\end{cases}\label{eq:corrected_Teulkosky_compact_mixed_decoupled}
\end{equation}
with:
\begin{equation}
\begin{aligned}
    \delta_+ &= \frac{\lambda_\theta}{2}+ k_+ \pi\,,\\
    \delta_- &= \frac{\lambda_\theta}{2}+ \frac{\pi}{2} + k_-\pi\,.
\end{aligned}
\end{equation}
This reduces to the \ac{EOM} for the parity even case, equation~\eqref{eq:corrected_Teulkosky_parity_eigenstates}, under the transformations:\footnote{The factor of $i^{\frac{1\mp1}{2}}$ is included because $\Phi_0^\pm$ and $\psi_H^\pm$ are parity even / odd scalars, whereas $\mathcal B_{\delta^\pm} \Phi_0$ and $\mathcal B_{-\delta^\pm}\psi_H$ are both parity even. This is purely technical and has no bearing on the final result.}
\begin{equation}
\begin{aligned}
    \mathcal B_{\delta^\pm} \Phi_0&\rightarrow i^{\frac{1\mp1}{2}}\Phi_0^\pm\,,\\
    \mathcal B_{-\delta^\pm}\psi_H&\rightarrow-i^{\frac{1\mp1}{2}}\psi_H^\pm\,,\\
    \abs{\lambda}&\rightarrow\lambda_{(e)} \,.
\end{aligned}
\end{equation}
Thus, in agreement with the findings of section~\ref{subsubsec:EM_mixed_Schwarzschild}, we prove that the corrections to Kerr \ac{QNM} frequencies under mixed parity \ac{EFT} corrections, can be obtained from the parity even case by taking $\lambda_{(e)}\rightarrow \sqrt{\lambda_{(e)}^2 + \lambda_{(o)}^2}$. We have proved that for any Kerr \ac{BH} background, \ac{QNM} frequencies are described by equation~\eqref{eq:schematic_correction_vector_QNFs}.

\subsection{Results\label{subsec:EM_results}}

In this section, we will solve the \ac{EOM} numerically, to find the corrections to the \ac{QNM} frequencies. In section~\eqref{subsubsec:EM_Schwarzschild_results} we will study the Schwarzschild limit of parity even \ac{EFT} corrections, and then, in~\ref{subsubsec:EM_Kerr_results} we generalize the result to Kerr \acp{BH} of arbitrary spin. As argued in~\ref{subsec:EM_mixed}, at leading order in $L/M$, the results can be readily generalized to mixed parity corrections by replacing $\lambda_{(e)}$ with $\sqrt{\lambda_{(e)}^2 + \lambda_{(o)}^2}$

\subsubsection{Schwarzschild limit \label{subsubsec:EM_Schwarzschild_results}}

\begin{figure}[H]
\centering
    \includegraphics[width=.95\columnwidth]{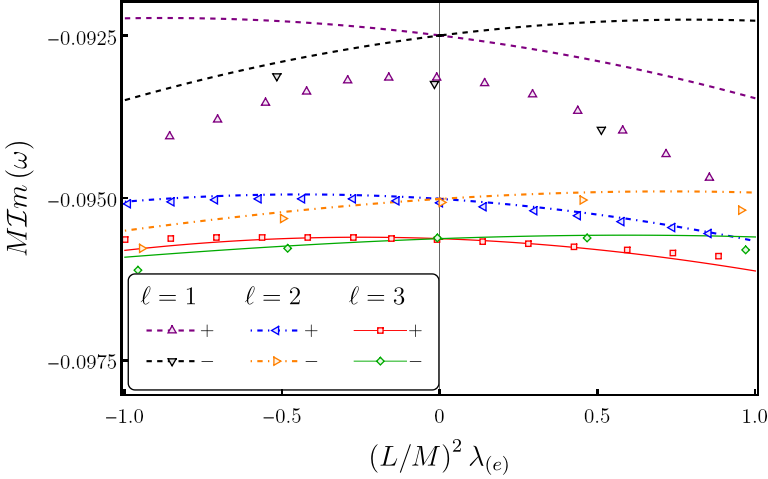}

\caption{In this plot, setting $\lambda_{(o)}=0$, we show the corrected electromagnetic \ac{QNM} frequency, for Schwarzschild \acp{BH} as a function of $\lambda_{(e)}$. The plot lines show the solutions to equation~\eqref{eq:decoupled_maxwell_Sch_even} obtained non-perturbatively in $L/M$. The plot markers were extracted from figure 4 of ~\cite{Chen:2013ysa}. In that paper, the authors computed the \acp{QNM} using a third-order WKB approximation, valid in the limit $\ell\gg1$. $\ell=1$ is outside this regime, thus, as expected, there is very poor agreement between our results for $\ell = 1$. The situation improves for $\ell = 2$ and $\ell=3$, but for large values of $\widetilde \lambda_{(e)}$ the difference is still noticeable. As argued before, the \ac{EFT} approximation is only valid in the limit $L\ll M$, thus most of this plot is non-physical. \label{fig:sch_limit_non_pertubative_imaginary_part}}
\end{figure}

Using the direct pseudospectral methods outlined in the appendix~\ref{appendix:numerics}, we solved equation~\eqref{eq:decoupled_maxwell_Sch_even} for $\ell = 1,2$,\ $m \in \{-\ell,\cdots, \ell\}$, by working non-perturbatively in $L/M$. We took $\sim 50$ values of $\widetilde \lambda_{(e)} \in[-1,1]$ and obtained the corresponding \ac{QNM} frequency for both polarizations, $\omega^\pm$. As seen in section~\ref{subsubsec:EM_eikonal}, in the Eikonal limit, for Schwarzschild \acp{BH}, the imaginary part of $\omega_1$ vanishes. Thus, for $\ell\neq \infty$, we expect $\mathcal Im(\omega_1^\pm)$ to be small. This implies that, for Schwarzschild \acp{BH}, $\mathcal Im(\omega^\pm)$ will be particularly close to $\mathcal Im(\omega_0)$. Thus, in this regime, to appropriately estimate $\omega_1^\pm$, we must have very precise values of $\omega^\pm$ and $\omega_0$. In~\cite{Chen:2013ysa}, the authors solved equation~\eqref{eq:decoupled_maxwell_Sch_even} for many values of $\widetilde \lambda_{(e)}$, using a third order WKB approximation. This approach is only valid in the large $\ell$ limit. Although several results in the literature argue that WKB methods produce reasonable results even for $\ell = \mathcal O(1)$, the imprecision will necessarily grow the closer we get to $\ell=1$. In figure~\ref{fig:sch_limit_non_pertubative_imaginary_part}, we compare our results with theirs. Although we find rough agreement for $\ell=2,3$, for $\ell=1$ we see a large difference. In fact, the difference between the two results at $\widetilde \lambda_{(e)}=0$ is of the same order of magnitude as the maximum correction for $\widetilde \lambda_{(e)} \in [-1,1]$. This is not surprising given the discussion above.\\

\begin{table}[H]
\centering
\begin{tabular}{c|c|c|r|r|r}
\multicolumn{3}{c|}{Parameters} & \multicolumn{3}{c}{$100M\omega_1^\pm$}\\\hline
$\ell$&Pol.&Part.&\multicolumn{1}{l}{Direct}&\multicolumn{1}{|l}{Perturbative} &\multicolumn{1}{|l}{\cite{Chen:2013ysa}}\\\hline\hline
\multirow{4}{*}{1}&\multirow{2}{*}{-}&$\mathcal Re$&$1.0567713780$&$1.0567713780$&\multicolumn{1}{c}{-}\\
&&$\mathcal Im$&$0.0612235305$&$0.0612235305$&$-0.05$\\\cline{2-6}
&\multirow{2}{*}{+}&$\mathcal Re$&$-1.0567713780$&$-1.0567713780$&\multicolumn{1}{c}{-}\\
&&$\mathcal Im$&$-0.0612235305$&$-0.0612235305$&$-0.04$\\\hline

\multirow{4}{*}{2}&\multirow{2}{*}{-}&$\mathcal Re$&$1.7486904518$&$1.7486904518$&\multicolumn{1}{c}{-}\\
&&$\mathcal Im$&$0.0311341010$&$0.0311341010$&$0.03$\\\cline{2-6}
&\multirow{2}{*}{+}&$\mathcal Re$&$-1.7486904518$&$-1.7486904518$&\multicolumn{1}{c}{-}\\
&&$\mathcal Im$&$-0.0311341010$&$-0.0311341010$&$-0.03$\\\hline

\multirow{4}{*}{3}&\multirow{2}{*}{-}&$\mathcal Re$&$2.4648927703$&$-2.4648927703$&\multicolumn{1}{c}{-}\\
&&$\mathcal Im$&$0.0174061992$&$0.0174061992$&$0.02$\\\cline{2-6}
&\multirow{2}{*}{+}&$\mathcal Re$&$-2.4648927703$&$-2.4648927703$&\multicolumn{1}{c}{-}\\
&&$\mathcal Im$&$-0.0174061992$&$-0.0174061992$&$-0.02$\\\hline
\end{tabular}
\caption{In this table, we compare three values for the Schwarzschild limit of $\omega_1$. The \textit{Direct} column shows the correction estimated using equations~\eqref{eq:decoupled_maxwell_Sch_even} and ~\eqref{eq:central_derivative_omega}, taking $\widetilde \lambda_{(e)} = 10^{-9}$. The \textit{Perturbative} column contains $\omega_1$, computed using the perturbative approach developed in section~\ref{subsubsec:EM_even_Kerr}, evaluated at $a=0$. The last column has an estimation of this frequency deduced from the data in figure 4 of~\cite{Chen:2013ysa}. The results in the first two columns were truncated to $10$ significant digits, but we actually observed an absolute difference of at most $\sim 10^{-19}$. Furthermore, as expected, the corrections for the two polarizations are equal and opposite. As expected, due to the poor accuracy of the WKB approximation used in~\cite{Chen:2013ysa} at low $\ell$, there is no agreement in the last column for $\ell=1$. The situation improves with increasing $\ell$. \label{tab:compare_vector_results_Chen}}
\end{table}

We have proved, that at leading order in $\widetilde \lambda_{(e)}$, the \ac{EFT} corrections to the \ac{QNM} frequencies corresponding to each polarization should be equal and opposite. The results in figure~\ref{fig:sch_limit_non_pertubative_imaginary_part} seem to be consistent with this notion. To accurately evaluate this, we must specialize to the small $\widetilde \lambda_{(e)}$ regime. Assuming $\omega^\pm\l(\widetilde \lambda_{(e)}\r)$ is regular near $\lambda_{(e)}=0$, we have:
\begin{equation}
\begin{aligned}
    \omega_0 &= \frac{1}{2}\l(\omega^\pm\l(\widetilde \lambda_{(e)}\r) + \omega^\pm\l(-\widetilde \lambda_{(e)}\r)\r)+ \mathcal O(\widetilde \lambda_{(e)}^2)\,,\\
    \omega_1^\pm &= \frac{1}{2\widetilde \lambda_{(e)}}\l(\omega^\pm\l(\widetilde \lambda_{(e)}\r) - \omega^\pm\l(-\widetilde \lambda_{(e)}\r)\r) + \mathcal O(\widetilde \lambda_{(e)}^2)\,.
\end{aligned}\label{eq:central_derivative_omega}
\end{equation}
Plugging in $\widetilde \lambda_{(e)} = 10^{-9}$ we obtained very precise values for $\omega_0$ and $\omega_1^\pm$. The background component describes an electromagnetic \ac{QNM} of a Schwarzschild \ac{BH}. This has recently been calculated with very high precision, using pseudospectral methods in~\cite{Mamani:2022akq}. We find exact agreement between our results and the frequencies in table 1 of that paper. In table~\ref{tab:compare_vector_results_Chen}, we show the results for $\omega_1^\pm$. We find that $\abs{\omega_1^++\omega_1^-} \lesssim 10^{-19}$. In the next section, working perturbatively in $L/M$, we will generalize the results to Kerr \acp{BH}. Setting $a=0$, we get an independent result for $\omega_1^\pm$. Once again, we find agreement with at least 19 digits of precision. We can also estimate the value of $\omega_1^\pm$ using the data in~\cite{Chen:2013ysa}. We fit a second-degree polynomial to the data with small $\widetilde \lambda_{(e)}$, and take the derivative at $\widetilde \lambda_{(e)} =0$. Because there isn't a lot of data in this region, we get a very rough estimate. This is included in the last column of~\ref{tab:compare_vector_results_Chen}. As in the non-perturbative regime (see figure~\ref{fig:sch_limit_non_pertubative_imaginary_part}), the results are roughly consistent with ours for $\ell=2,3$, but there is no agreement at $\ell=1$.

\subsubsection{Kerr results \label{subsubsec:EM_Kerr_results}}

\begin{table*}[ht]
\begin{tabular}{c|c|r||r|r|r|r|r|r|r|r|r}
\multicolumn{3}{c||}{Parameters}&\multicolumn{9}{c}{$a/M$}\\\hline\hline
$\ell$&$m$&Component&$0$&$0.25$&$0.5$&$0.75$&$0.95$&$0.99$&$0.999$&$0.9999$&$0.999975$\\\hline\hline
\multirow{4}{*}{$1$}&\multirow{4}{*}{$-1$}&$M\mathcal Re(\omega_0)$&$0.248263$&$0.233724$&$0.222121$&$0.212538$&$0.205897$&$0.204656$&$0.204380$&$0.204352$&$0.204350$\\
&&$M\mathcal Im(\omega_0)$&$-0.092488$&$-0.092998$&$-0.092881$&$-0.092290$&$-0.091557$&$-0.091390$&$-0.091352$&$-0.091348$&$-0.091348$\\
&&$100\,M\mathcal Re(\omega_1)$&$1.056771$&$0.811775$&$0.600993$&$0.451404$&$0.380628$&$0.370092$&$0.367836$&$0.367612$&$0.367594$\\
&&$100\,M\mathcal Im(\omega_1)$&$0.061224$&$0.131237$&$0.129569$&$0.079621$&$0.035260$&$0.027951$&$0.026386$&$0.026231$&$0.026218$\\\hline
\multirow{4}{*}{$1$}&\multirow{4}{*}{$0$}&$M\mathcal Re(\omega_0)$&$0.248263$&$0.249619$&$0.253933$&$0.262079$&$0.272153$&$0.274311$&$0.274777$&$0.274823$&$0.274827$\\
&&$M\mathcal Im(\omega_0)$&$-0.092488$&$-0.092009$&$-0.090329$&$-0.086330$&$-0.078557$&$-0.075949$&$-0.075305$&$-0.075240$&$-0.075234$\\
&&$100\,M\mathcal Re(\omega_1)$&$1.056771$&$1.051645$&$1.033499$&$0.989413$&$0.915530$&$0.910598$&$0.912477$&$0.912739$&$0.912762$\\
&&$100\,M\mathcal Im(\omega_1)$&$0.061224$&$0.044916$&$-0.009798$&$-0.131062$&$-0.376146$&$-0.473005$&$-0.497676$&$-0.500190$&$-0.500400$\\\hline
\multirow{4}{*}{$1$}&\multirow{4}{*}{$1$}&$M\mathcal Re(\omega_0)$&$0.248263$&$0.267305$&$0.294091$&$0.337746$&$0.417719$&$0.463399$&$0.489671$&$0.497136$&$0.498655$\\
&&$M\mathcal Im(\omega_0)$&$-0.092488$&$-0.091007$&$-0.087677$&$-0.079456$&$-0.054429$&$-0.031292$&$-0.011609$&$-0.003805$&$-0.001886$\\
&&$100\,M\mathcal Re(\omega_1)$&$1.056771$&$1.304920$&$1.510306$&$1.547601$&$0.867720$&$0.143470$&$-0.167164$&$-0.116308$&$-0.074255$\\
&&$100\,M\mathcal Im(\omega_1)$&$0.061224$&$-0.087996$&$-0.307750$&$-0.521719$&$-0.236056$&$0.115480$&$0.138910$&$0.061506$&$0.033062$\\\hline
\multirow{4}{*}{$2$}&\multirow{4}{*}{$-2$}&$M\mathcal Re(\omega_0)$&$0.457596$&$0.425879$&$0.400193$&$0.378744$&$0.363875$&$0.361104$&$0.360490$&$0.360428$&$0.360423$\\
&&$M\mathcal Im(\omega_0)$&$-0.095004$&$-0.094954$&$-0.094378$&$-0.093493$&$-0.092647$&$-0.092467$&$-0.092426$&$-0.092422$&$-0.092422$\\
&&$100\,M\mathcal Re(\omega_1)$&$1.748690$&$1.428612$&$1.171122$&$0.970383$&$0.844319$&$0.822191$&$0.817339$&$0.816857$&$0.816817$\\
&&$100\,M\mathcal Im(\omega_1)$&$0.031134$&$0.082368$&$0.096079$&$0.087449$&$0.073683$&$0.070710$&$0.070038$&$0.069971$&$0.069966$\\\hline
\multirow{4}{*}{$2$}&\multirow{4}{*}{$2$}&$M\mathcal Re(\omega_0)$&$0.457596$&$0.498542$&$0.555399$&$0.646915$&$0.813050$&$0.909991$&$0.970549$&$0.990593$&$0.995284$\\
&&$M\mathcal Im(\omega_0)$&$-0.095004$&$-0.094089$&$-0.091131$&$-0.082512$&$-0.054668$&$-0.030053$&$-0.010655$&$-0.003482$&$-0.005264$\\
&&$100\,M\mathcal Re(\omega_1)$&$1.748690$&$2.132729$&$2.565175$&$2.919763$&$2.176113$&$0.930436$&$0.171548$&$0.036815$&$0.0170$\\
&&$100\,M\mathcal Im(\omega_1)$&$0.031134$&$-0.071956$&$-0.234223$&$-0.409970$&$-0.187044$&$0.057781$&$0.030259$&$0.002701$&$0.0023$\\\hline
\multirow{4}{*}{$3$}&\multirow{4}{*}{$-3$}&$M\mathcal Re(\omega_0)$&$0.656899$&$0.608409$&$0.569188$&$0.536482$&$0.513845$&$0.509631$&$0.508696$&$0.508603$&$0.508595$\\
&&$M\mathcal Im(\omega_0)$&$-0.095616$&$-0.095418$&$-0.094724$&$-0.093755$&$-0.092865$&$-0.092679$&$-0.092636$&$-0.092632$&$-0.092632$\\
&&$100\,M\mathcal Re(\omega_1)$&$2.464893$&$2.045951$&$1.713706$&$1.450022$&$1.278323$&$1.247517$&$1.240734$&$1.240059$&$1.240002$\\
&&$100\,M\mathcal Im(\omega_1)$&$0.017406$&$0.067761$&$0.090180$&$0.095180$&$0.092051$&$0.090966$&$0.090706$&$0.090680$&$0.090678$\\\hline
\multirow{4}{*}{$3$}&\multirow{4}{*}{$3$}&$M\mathcal Re(\omega_0)$&$0.656899$&$0.719584$&$0.806776$&$0.947457$&$1.204076$&$1.355490$&$1.451973$&$1.484571$&$1.492257$\\
&&$M\mathcal Im(\omega_0)$&$-0.095616$&$-0.094858$&$-0.092021$&$-0.083325$&$-0.054758$&$-0.029860$&$-0.010596$&$-0.003476$&$-0.001753$\\
&&$100\,M\mathcal Re(\omega_1)$&$2.464893$&$2.991711$&$3.640023$&$4.310268$&$3.631857$&$1.918234$&$0.593560$&$0.181262$&$0.089684$\\
&&$100\,M\mathcal Im(\omega_1)$&$0.017406$&$-0.073354$&$-0.214206$&$-0.373040$&$-0.198956$&$0.002294$&$0.009856$&$0.001438$&$0.000387$\\\hline

\end{tabular}

\caption{Numerical values for the \ac{QNM} frequencies and corresponding \ac{EFT} correction of electromagnetic modes. All the included digits are significant.\label{tab:QNM_frequencies_vector}}
\end{table*}

\begin{figure*}[ht]
    \subfloat[\label{subfig:vector_QNM_line_plot_Re_l=m_background}]{
    \includegraphics[width=0.457\textwidth]{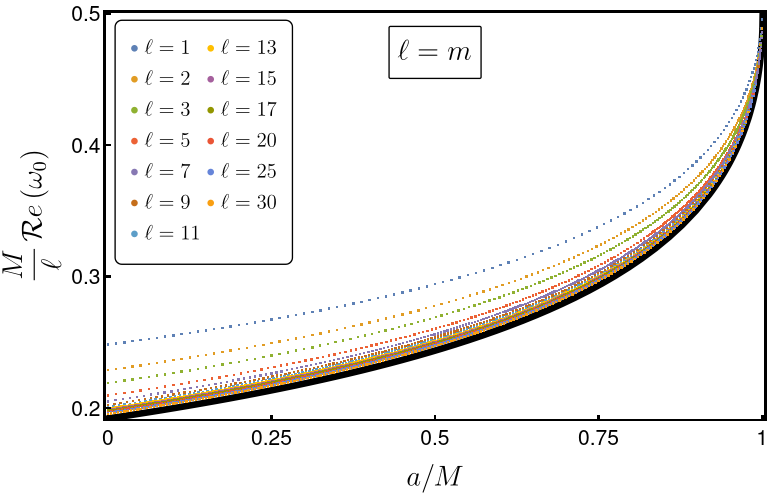}}\hskip 2em
    \subfloat[\label{subfig:vector_QNM_line_plot_Im_l=m_background}]{
    \includegraphics[width=0.457\textwidth]{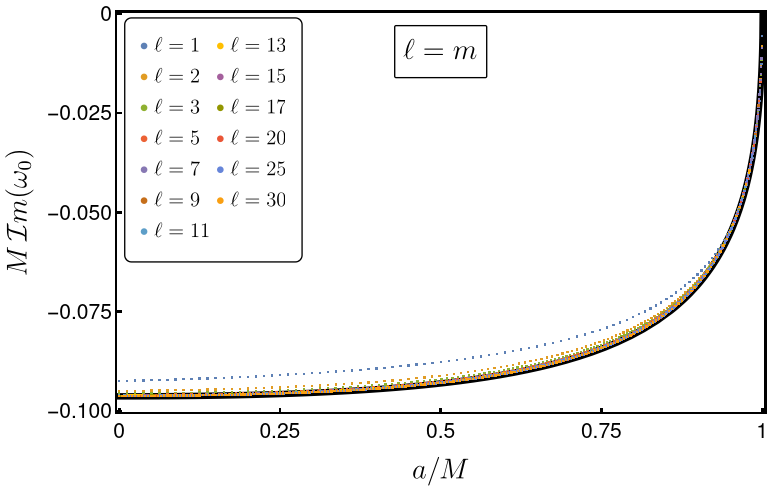}}

    \subfloat[\label{subfig:vector_QNM_line_plot_Re_l=m_correction}]{
    \includegraphics[width=0.457\textwidth]{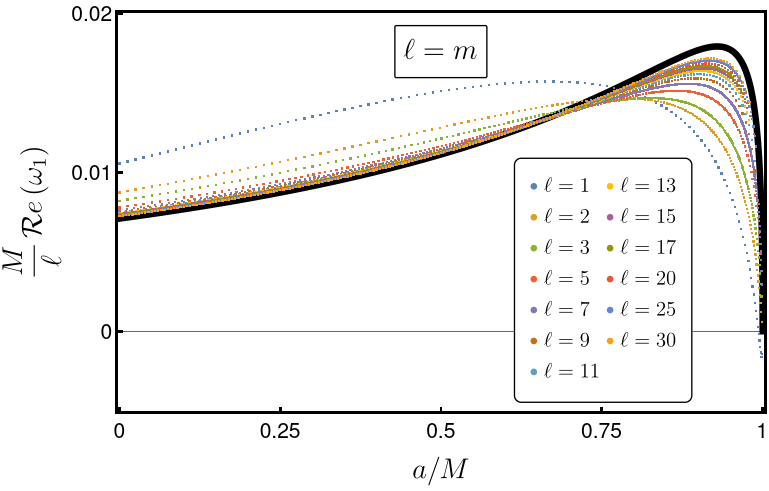}}\hskip 2em
    \subfloat[\label{subfig:vector_QNM_line_plot_Im_l=m_correction}]{
    \includegraphics[width=0.457\textwidth]{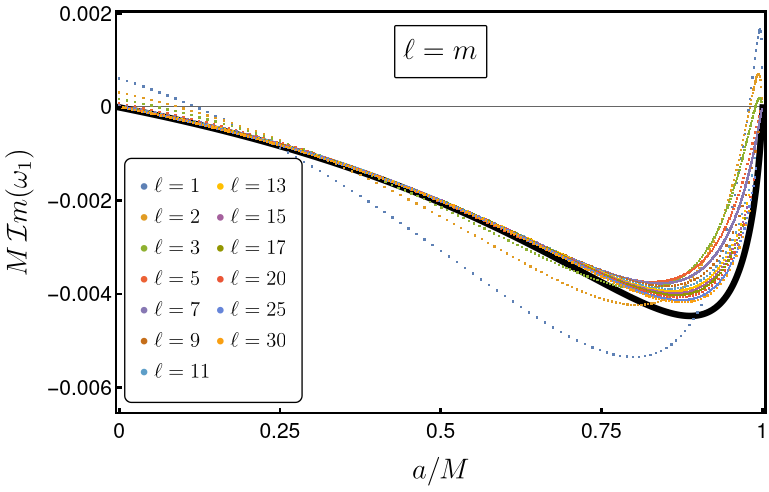}}

\caption{Here we show the \ac{QNM} frequency $\ell=m$ of electromagnetic modes, and corresponding \ac{EFT} correction for multiple values of $\ell$. Panels a) and b) contain the real and imaginary parts of the background frequency, whereas in panels c) and d) we have the corresponding \ac{EFT} corrections. The solid dark lines represent the Eikonal limit prediction derived in section~\ref{subsubsec:EM_eikonal}. As expected, with increasing $\ell$, the \ac{QNM} frequencies approach the large $\ell$ limit line. Furthermore, in agreement with figures~\ref{fig:vector_QNM_line_plot} and~\ref{fig:vector_QNM_complex_plot}, in the near extremal limit, $M \omega_0\rightarrow \ell/2$ and $M \omega_1\rightarrow 0$.\label{fig:vector_QNM_line_plot_l=m}}
\end{figure*}

\begin{figure*}[ht]
    \subfloat[\label{subfig:vector_der_log_Re_freq_log_l_l=m_background}]{
    \includegraphics[width=0.457\textwidth]{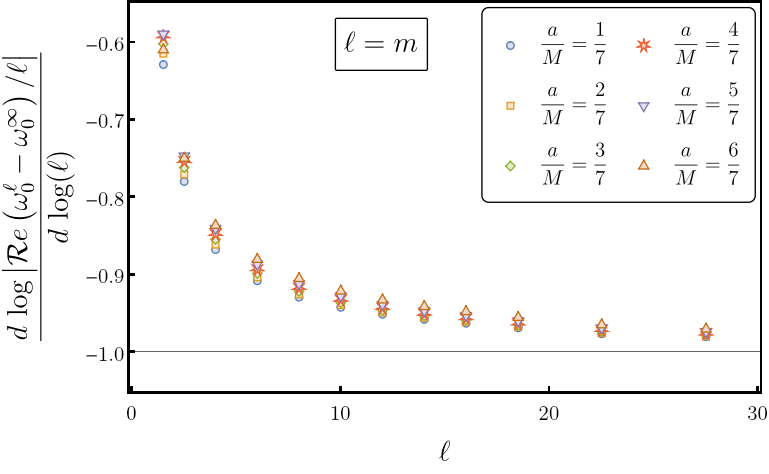}}\hskip 2em
    \subfloat[\label{subfig:vector_der_log_Im_freq_log_l_l=m_background}]{
    \includegraphics[width=0.457\textwidth]{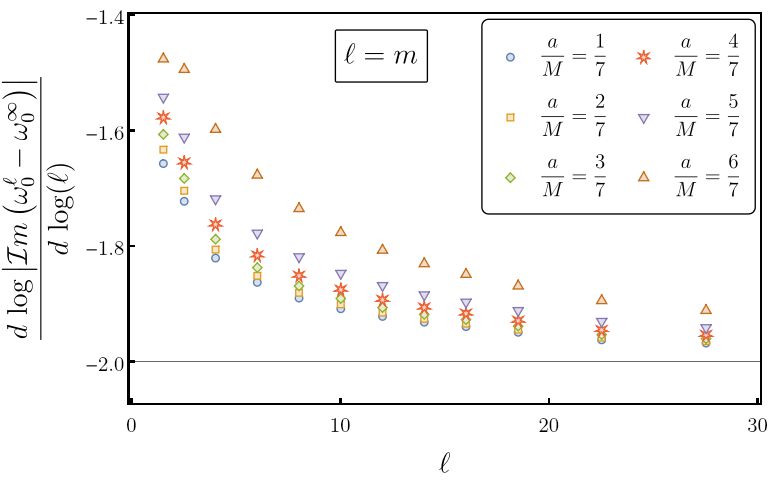}}

    \subfloat[\label{subfig:vector_der_log_Re_freq_log_l_l=m_correction}]{
    \includegraphics[width=0.457\textwidth]{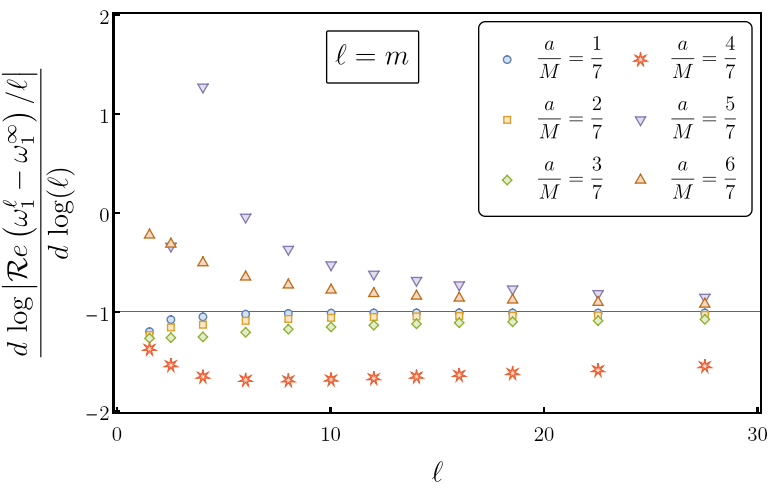}}\hskip 2em
    \subfloat[\label{subfig:vector_der_log_Im_freq_log_l_l=m_correction}]{
    \includegraphics[width=0.457\textwidth]{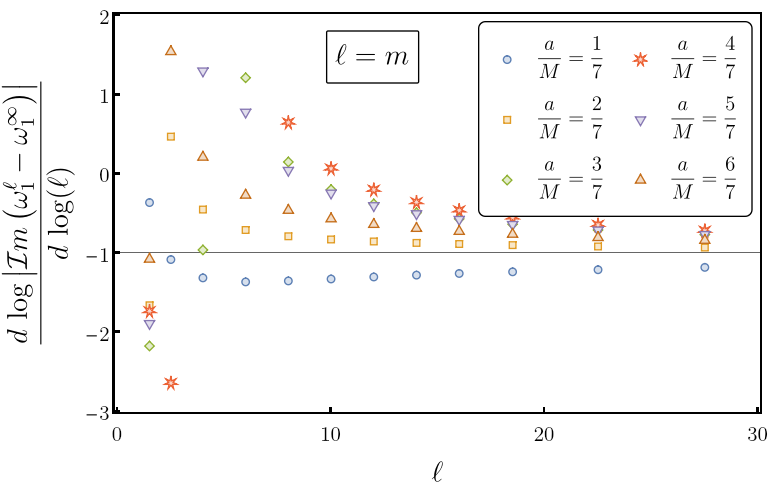}}

\caption{In this image we study how quickly the \ac{QNM} frequencies approach the Eikonal limit. On the y-axis, we see the logarithmic derivative, $\gamma= \dd\log\abs{\omega^\ell-\omega^\infty}/\dd \log(\ell)$.  Panels $a)$ and $b)$, show the convergence of the real and imaginary part of the background frequency, whereas $c)$ and $d)$ have the real and imaginary part of the \ac{EFT} correction. Because we expect $\mathcal Re \omega$ to grow linearly with $\ell$, we normalized the difference diving by $\ell$. We have $M\l(\omega^\ell-\omega^\infty\r)\sim\ell^{-\gamma}$. From equation~\eqref{eq:QNM_Eikonal_correspondence} we expect that the real part of the background frequency has $\gamma\rightarrow-1$, while the imaginary part should have $\gamma\rightarrow-2$. This is consistent with the results in the figure~\ref{fig:vector_QNM_line_plot_l=m}. We did not derive any theoretical results regarding the expected convergence speed of the \ac{EFT} correction. Notwithstanding, we find that for both the real and imaginary parts of $\omega_1$, the convergence rate is consistent with $\ell^{-1}$. Note that there is erratic behaviour of the \ac{EFT} correction for low $\ell$ values, particularly for $a/M=3/7$ and $a/M= 4/7$. This happens, because, as seen in figure~\ref{fig:vector_QNM_line_plot_l=m}, $\omega_1^\ell$ crosses $\omega_1^\infty$, in this region, taking the logarithm of the difference to infinity. \label{fig:vector_der_log_freq_log_l_l=m}}
\end{figure*}

\begin{figure*}[ht]
    \subfloat[\label{subfig:vector_QNM_complex_plot_background}]{
    \includegraphics[width=0.457\textwidth]{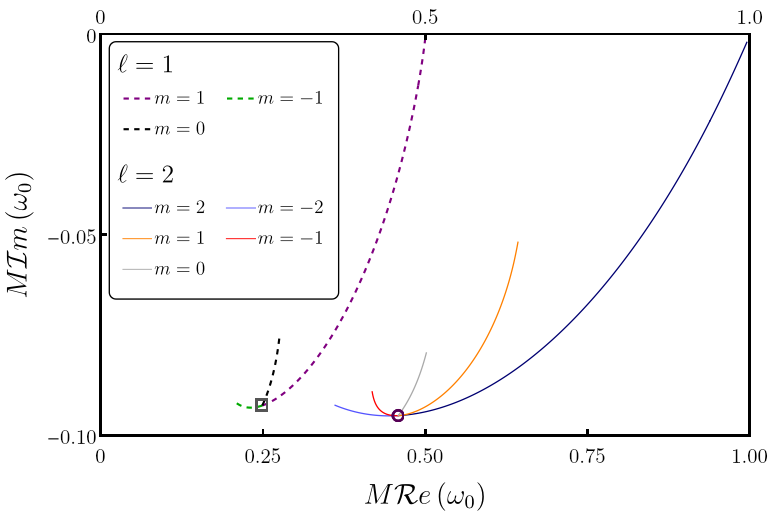}}\hskip 2em
    \subfloat[\label{subfig:vector_QNM_complex_plot_correction}]{
    \includegraphics[width=0.457\textwidth]{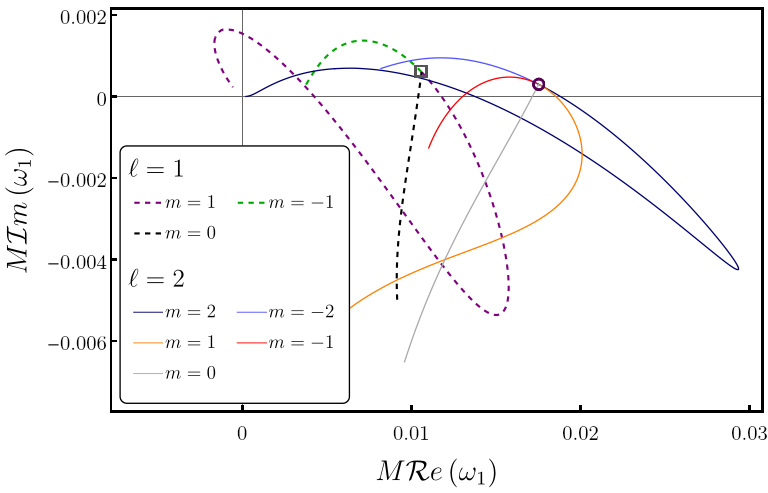}}

\caption{Here, we show a parametric plot of the lowest lying electromagnetic \ac{QNM} frequencies. The left panel shows the background component, while on the right we have the corresponding \ac{EFT} correction. The plot markers show the Schwarzschild limit, obtained by solving the \ac{EOM} in section~\ref{subsubsec:EM_even_Schwarzschild}. The parametric lines show the solution to the \ac{EOM} in section~\ref{subsubsec:EM_even_Kerr}, for \acp{BH} with $a/M \in(0, 0.999975)$. We find great agreement between the two results. Qualitatively, the frequencies share most features of scalar \acp{QNM}. In the Schwarzschild limit, modes with the same $\ell$ share the same frequency and correction. In the near-extremal limit, for $\ell=m$ modes, the background frequency approaches $\ell/2$, whereas the correction approaches $0$.\label{fig:vector_QNM_complex_plot}}
\end{figure*}

\begin{figure*}[ht]
    \subfloat[\label{subfig:slowly_damped_near_extremal_vector_QNM_background}]{
    \includegraphics[width=0.457\textwidth]{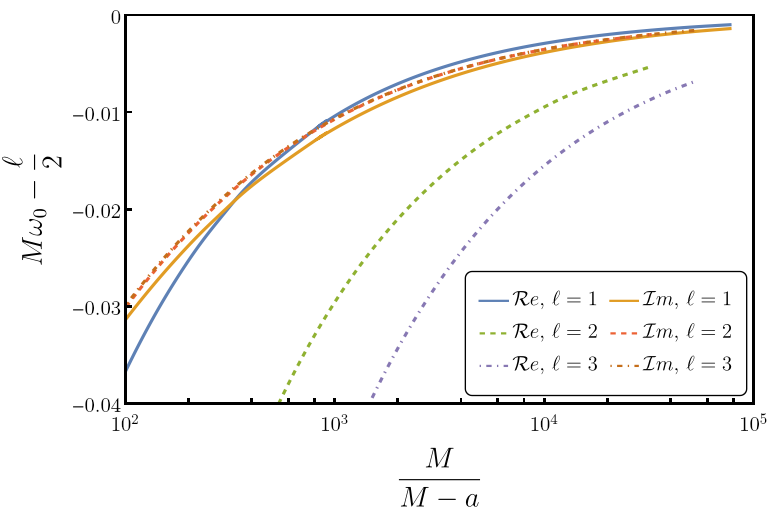}}\hskip 2em
    \subfloat[\label{subfig:slowly_damped_near_extremal_vector_QNM_correction}]{
    \includegraphics[width=0.457\textwidth]{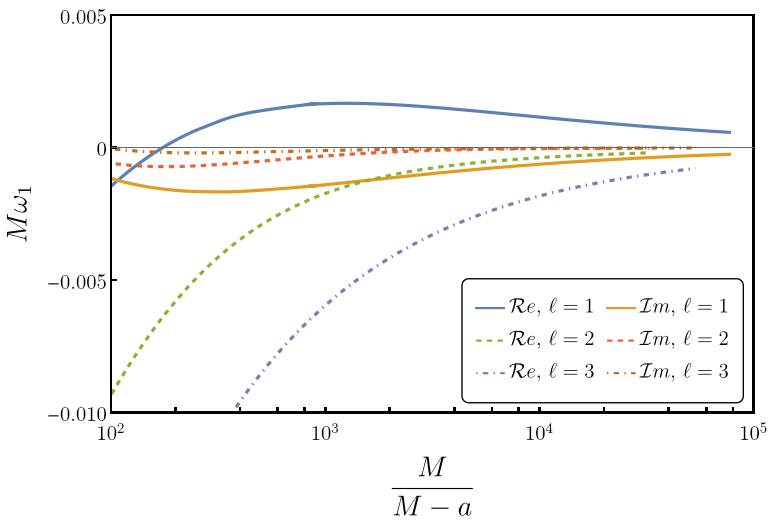}}

\caption{Here, we show the near extremal behaviour of $\ell=m$ electromagnetic \ac{QNM} frequencies. On the left panel, we see the background components, while the right panel shows the corresponding \ac{EFT} corrections. As seen for scalar \acp{QNM}, in the limit $a\rightarrow M$, $M \omega_0$ approaches $\ell/2$, while $M \omega_1\rightarrow0$. Contrary to the behaviour at low spin, in the near-extremal limit, background modes with larger $\ell$ have a larger imaginary part. In this figure, the lowest lying \ac{QNM} is the $\ell=m=3$ mode. \label{fig:slowly_damped_near_extremal_vector_QNM}}
\end{figure*}

To solve equation~\eqref{eq:corrected_Teukolsky_final}, and obtain the \ac{QNM} frequencies corresponding corrections, we employed pseudospectral methods, as described in the appendix~\ref{appendix:numerics}. The main results are summarized in figure~\ref{fig:vector_QNM_line_plot} and table~\ref{tab:QNM_frequencies_vector}. We will discuss qualitatively these results below, but we will first establish their consistency and accuracy. In figure~\ref{fig:convergence_plot}, we computed $\omega_1$ for a \ac{BH} with $a= 0.999975M$ at varying grid sizes. We proved that, with increasing grid size, the numerical accuracy of $\omega_1$ increased exponentially. This is in line with the expectation for pseudo-spectral methods.\\

One of our main consistency checks was done in the Schwarzschild limit. In section~\ref{subsubsec:EM_even_Schwarzschild}, we derived the \ac{EOM} for the \acp{QNM} in this regime, following an approach that is not perturbative in $\widetilde \lambda_{(e)}$. In table~\ref{tab:compare_vector_results_Chen}, we compared $\omega_1$ derived through that approach, with the solutions of equation~\eqref{eq:corrected_Teukolsky_final}, with $a=0$. We found perfect agreement within the numerical accuracy of both approaches. Due to spherical symmetry, in this regime, \acp{QNM} and corresponding corrections are independent of $m$, with the degeneracy being broken for $a>0$. This can be seen in figures~\ref{fig:vector_QNM_line_plot} and~\ref{fig:vector_QNM_complex_plot}.\\

On the other hand, our results are consistent with the Eikonal limit prediction of section~\ref{subsubsec:EM_eikonal}. In figures~\ref{fig:vector_QNM_line_plot_l=m} and~\ref{fig:vector_QNM_line_plot_l=-m} we show, accordingly, the $\ell=m$ and $\ell=-m$ \ac{QNM} frequencies of Kerr \acp{BH} for a large range of $\ell$ values, together with the analytic large $\ell$ limit prediction. We find that, with increasing $\ell$, the numerical curves converge to the analytic prediction. To quantify the convergence rate, in figures~\ref{fig:vector_der_log_freq_log_l_l=m} and~\ref{fig:vector_der_log_freq_log_l_l=-m}, we used a logarithmic derivative approach. We find that the real part of the background frequency converges at a rate of order $\ell^{-1}$, whereas the imaginary part converges as $\ell^{-2}$. This is consistent with the prediction in~\cite{Dolan:2010wr}. Although we had no results regarding the expected convergence rate for $\omega_1$, we found that the real and imaginary parts approach the Eikonal limit as $\ell^{-1}$.\\

Qualitatively, our results are very similar to the ones obtained for the scalar case (see section~\ref{subsec:KG_results}). The crucial difference is the splitting of the \ac{QNM} frequencies into two families, corresponding to the two possible polarizations of light. We have:
\begin{equation}
    \omega^\pm = \omega_0 \pm (-1)^{m+\ell +1} \l(\frac{L}{M}\r)^2\sqrt{\lambda_{(e)}^2 + \lambda_{(o)}^2}\,\omega_1 \,.\label{eq:QNM_family_split_results}
\end{equation}
We see that parity even and parity odd \acp{EFT} corrections change the \ac{QNM} frequencies in the exact same way. Furthermore, we find that the corrections corresponding to the two polarizations are equal and opposite. This only happens because, in the \ac{RHS} of equation~\eqref{eq:corrected_Teulkosky_compact}, we were able to cancel all terms proportional to $\Phi_i$, leaving only terms involving $\bar{\Phi}_i$. Otherwise, we should expect the corrected frequencies to not be centered around $\omega_0$. This seems to be the case in~\cite{Cano:2021myl,Cano:2023tmv,Cano:2023jbk}. Presumably because the \ac{EFT} corrects the background spacetime, the authors find that $\omega_1^++\omega_1^-\neq0$. Notwithstanding, the authors find that the mixing between parity even and parity odd \acp{QNM} is given by the square root of the sum, in agreement with~\eqref{eq:QNM_family_split_results}.\\

Modes with $\ell=m$, also dubbed \textit{slowly damped \acp{QNM}}, have background frequencies, with the least negative imaginary part. This means they are the slowest decaying modes, thus dominating the spectrum of the \ac{BH} ringdown. From these, for most of the parameter space, the $\ell=m=1$ is the lowest lying, However, at $a\approx 0.96M$,  $\mathcal Im(\omega_0)$ for $\ell=m=2$ and $\ell=m=3$ cross the value for $\ell=m=1$, and these become the slowest decaying \acp{QNM}.\footnote{From table~\ref{tab:QNM_frequencies_vector}, we see that the $\ell=m=3$ has the largest imaginary part of the two, thus being the lowest lying \ac{QNM}}. We did not see this tendency for modes with $\ell>3$, but it could be present sufficiently close to extremality. Coincidentally, except very close to extremality, for each value $\ell$, $\omega_1$ has the largest absolute real and imaginary part when $\ell=m$. These hit a maximum for $a \sim 0.75 M$. Thus. these modes seem to be the most relevant if we wish to experimentally find an upper bound for $\lambda$.

In figure~\ref{fig:vector_QNM_complex_plot}, we show the data of figure~\ref{fig:vector_QNM_line_plot} in a parametric form, as a function of $a/M$. The plot markers represent \ac{QNM} frequencies in the Schwarzschild limit, as obtained in section~\ref{subsubsec:EM_Schwarzschild_results}, whereas the plot lines correspond to the \ac{QNM} frequency when $a\neq0$. There is a perfect agreement between the two. Particularly relevant, is the shape of the slowly damped \acp{QNM}. Their parametric lines cover a large region of the complex plane, varying particularly quickly in the near extremal lime. For these modes, should we exclude $a\gtrsim 0.99M$, we would lose approximately the last tenth of the parametric lines. Thus, in figure~\ref{fig:slowly_damped_near_extremal_vector_QNM}, we zoom in this region of parameter space, computing the \acp{QNM} for \acp{BH} with $a/M \in(0,99, 0.999975)$. As predicted in~\cite{Hod:2008zz}, as $a\rightarrow M$, the background \ac{QNM} frequencies seem to converge polynomially to $\ell/2$. Analogously, in the same limit, we find that $M\omega_1$ approaches $0$ at a similar convergence rate.\footnote{In the $\ell=m=1$ and $\ell=m=2$ cases, modes seem to converge slightly quicker than expected, but, for $\ell=m=3$, the convergence rate is consistent with the expectation: $\omega_0^{\ell=m=3}- \ell/2 \sim \sqrt{M-a}$. We think that, by getting even closer to extremality, we should get the correct convergence rate. It was harder to quantify the convergence rate of $\omega_1$ as a logarithm derivative approach did not seem to converge for $a\lesssim 0.999975 M$.} Furthermore, we find that the absolute value of $M \mathcal Re \omega_1$ grows with $\ell$. This is consistent with the prediction of~\cite{Hod:2008zz} for the background modes. Finally, note that at $a\sim 0.992$, $M \mathcal Re\, \omega_1$ crosses $0$.

\section{Discussion \label{sec:discussion}}

In this paper, we computed the correction to \ac{BH} \acp{QNM}, under scalar and electromagnetic, leading order \ac{EFT}  corrections. In the electromagnetic case, we proved that the leading order parity even and parity odd \ac{EFT} corrections have the same effect on \ac{QNM} frequencies. Taking $\lambda{(e)/(o)}$ to be the coupling parameter that control parity even / odd corrections, the combined effect of the two is controlled by $\sqrt{\lambda_{(e)}^2+\lambda_{(o)}^2}$. Furthermore, we found that the degeneracy between the two polarizations of light is broken, splitting the \acp{QNM} into two families. At leading order, the corrections to the corresponding frequencies were shown to be equal and opposite.\\

Electromagnetic \acp{QNM} from a \ac{BH} merger have not yet been detected. Thus, from an observational perspective, our results are not very relevant. However, as seen in~\cite{Cano:2021myl,Cano:2023tmv,Cano:2023jbk}, the main features generalize to pure gravity \acp{EFT} corrections. There, the degeneracy breaking between the two possible polarizations is particularly relevant. In theory, if we know the precise direction of a \ac{GW} source, combining the signal from several \ac{GW} detectors, we may disentangle the two \ac{GW} tensor polarizations, see~\cite{Will:2014kxa}. Thus, if both families of \acp{QNM} are excited in a \ac{BH} merger, we may be able to disentangle the \ac{QNM} spectrum into contributions corresponding to each polarization, and use the difference to infer an upper bound for the size of the \ac{EFT} correction.\\ 

We used, scalar, and electromagnetic \acp{QNM} as toy models to learn how to, in general, compute the \ac{EFT} corrections to gravitational \acp{QNM} of Kerr \acp{BH}. There, the challenge is that not only the \ac{EOM} are corrected, but the \textit{background} stationary metric is also corrected. \cite{Cano:2021myl,Cano:2023tmv,Cano:2023jbk} have overcome this challenge, by working perturbatively in the \ac{BH} spin. In future work, we expect that it is possible to extend our method to compute the gravitational \acp{QNM} of \acp{BH} with arbitrary spin. The correction to the background metric can be determined numerically and then, the resulting \ac{EOM} can be decoupled and solved with an extension of both our methods.\\

\section{Acknowledgements}
The author thanks the contribution of his supervisor Harvey S. Reall, in motivating, discussing, and reviewing the contents of this paper, Andy Zhao, for initial calculations regarding the correction to electromagnetic quasinormal modes in the Schwarzschild limit, and Jorge E. Santos for helpful discussions regarding the numerical approach. The author also thanks David Baker, Rifath Khan, Frank Qu, Gonçalo Regado, and Bilyana Tomova for fruitful physics discussions. The author is jointly funded by the IOA (University of Cambridge), the University of Cambridge Trust, and King's College (University of Cambridge).\\

\appendix

\section{Numerical methods \label{appendix:numerics}}
\begin{figure*}[ht]
    \subfloat[\label{subfig:vector_QNM_line_plot_Re_l=-m_background}]{
    \includegraphics[width=0.457\textwidth]{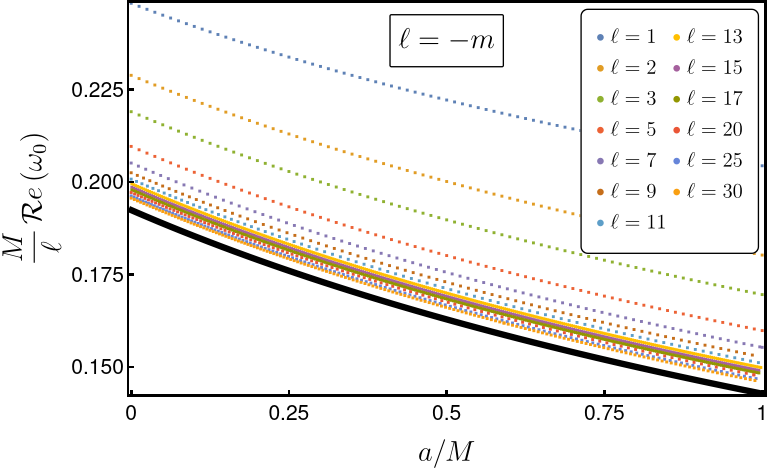}}\hskip 2em
    \subfloat[\label{subfig:vector_QNM_line_plot_Im_l=-m_background}]{
    \includegraphics[width=0.457\textwidth]{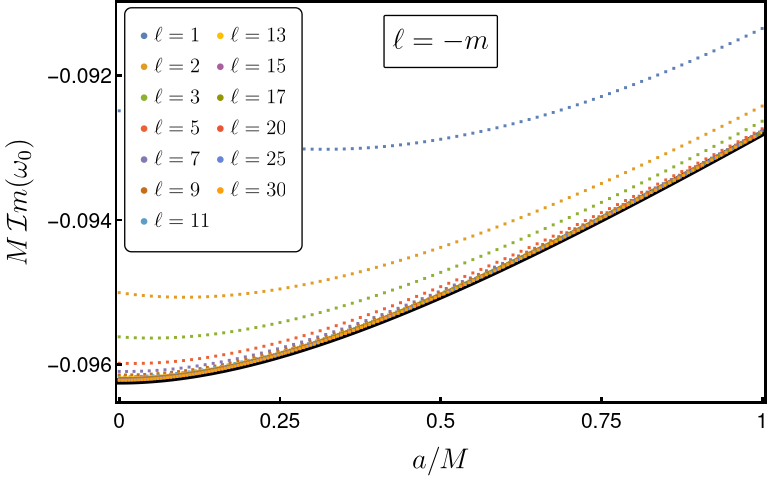}}

    \subfloat[\label{subfig:vector_QNM_line_plot_Re_l=-m_correction}]{
    \includegraphics[width=0.457\textwidth]{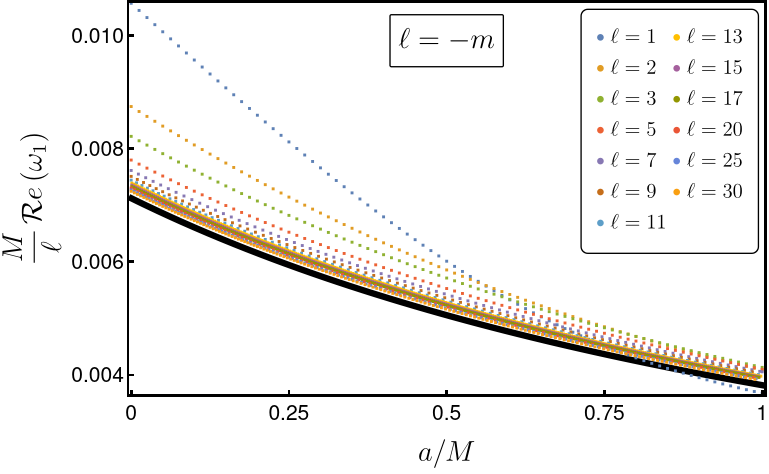}}\hskip 2em
    \subfloat[\label{subfig:vector_QNM_line_plot_Im_l=-m_correction}]{
    \includegraphics[width=0.457\textwidth]{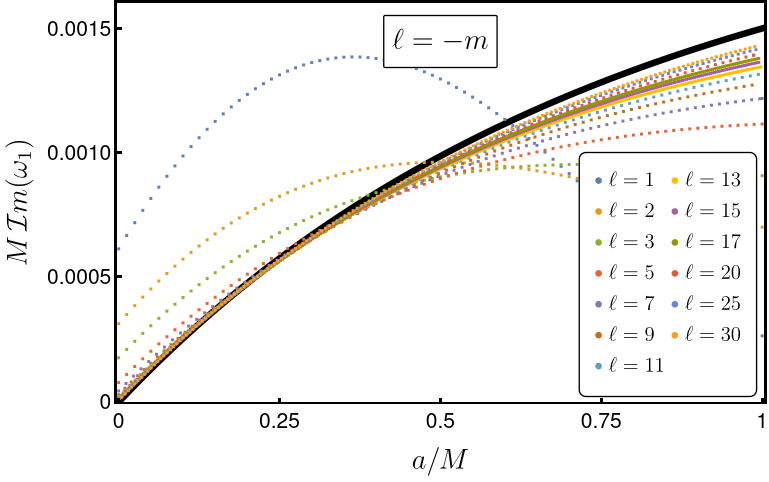}}

\caption{Here we show the \ac{QNM} frequencies of $\ell=-m$ electromagnetic modes for several values of $\ell$. In panels $a)$ and $b)$, we have the real and imaginary part of the background frequency, with panels $c)$ and $d)$ showing the corresponding correction. The thick dark line shows the Eikonal limit prediction. As expected, the \ac{QNM} frequencies converge to the asymptotic prediction. \label{fig:vector_QNM_line_plot_l=-m}}
\end{figure*}

In this appendix, we will describe the numerical methods used throughout the paper. We will be using collocation methods, as described in~\cite{Dias:2010eu,Miguel:2020uln,Dias:2015nua,Boyd:2000}. Effectively, these map continuous \acp{ODE} and \acp{PDE} problems into finite-dimensional linear algebra problems.\\

The final \ac{EOM} for scalar \acp{QNM} in section~\ref{sec:KG_modes} and electromagnetic \acp{QNM} in section~\ref{sec:EM_modes} can be written as:
\begin{subnumcases}{\hskip -.5cm\label{eq:EOM_general}}
    L^{(0)}(\omega_0)\,\Phi^{(0)}=0\,,\label{eq:EOM_background_general}\\
    L^{(0)}(\omega_0)\,\Phi^{(1)}+\omega_1 \delta L \Phi^{(0)} = L^{(1)}\Phi^{(0)}\,.\label{eq:EOM_correction_general}
\end{subnumcases}
As seen in equation~\eqref{eq:teukolsky_rx}, $L^{(0)}$ separates as the sum of two \ac{ODE} operators and, as such, $\Phi^{(0)}(r,x) = R(r)\Theta(x)$: 
\begin{equation}
    L^{(0)}\Phi^{(0)}(r,x) = \Theta(x)L^{(0)}_rR(r)+R(r)L^{(0)}_x\Theta(x)\,.
\end{equation}
\ac{QNM} frequencies are values of $\omega_{0/1}$ such that there are solutions $\Phi^{(0/1)}$ that are smooth in the \ac{PDE}'s domain. In section~\ref{subsec:pseudospectral_methods}, we will explain the main features of the numerical methods used. Then, in section~\ref{subsec:numerics_background}, we explain how to use this to solve equation~\eqref{eq:EOM_background_general}, and then, in section~\ref{subsec:numerics_perturbation}, we will solve equation~\eqref{eq:EOM_correction_general}.

\begin{figure*}[ht]
    \subfloat[\label{subfig:vector_der_log_Re_freq_log_l_l=-m_background}]{
    \includegraphics[width=0.457\textwidth]{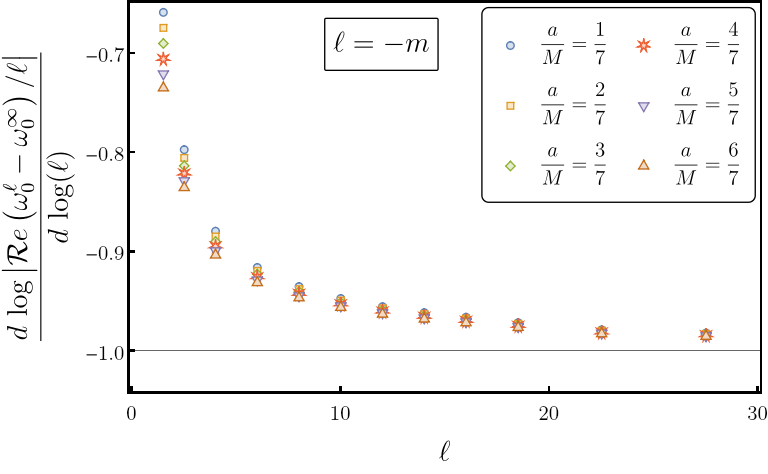}}\hskip 2em
    \subfloat[\label{subfig:vector_der_log_Im_freq_log_l_l=-m_background}]{
    \includegraphics[width=0.457\textwidth]{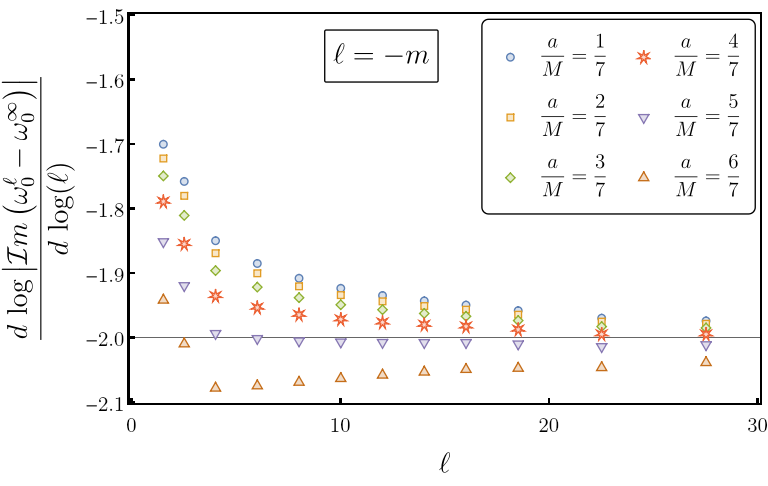}}

    \subfloat[\label{subfig:vector_der_log_Re_freq_log_l_l=-m_correction}]{
    \includegraphics[width=0.457\textwidth]{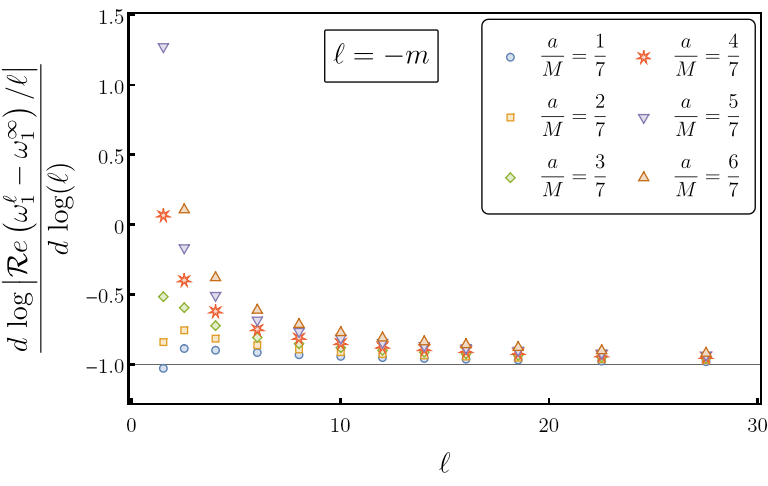}}\hskip 2em
    \subfloat[\label{subfig:vector_der_log_Im_freq_log_l_l=-m_correction}]{
    \includegraphics[width=0.457\textwidth]{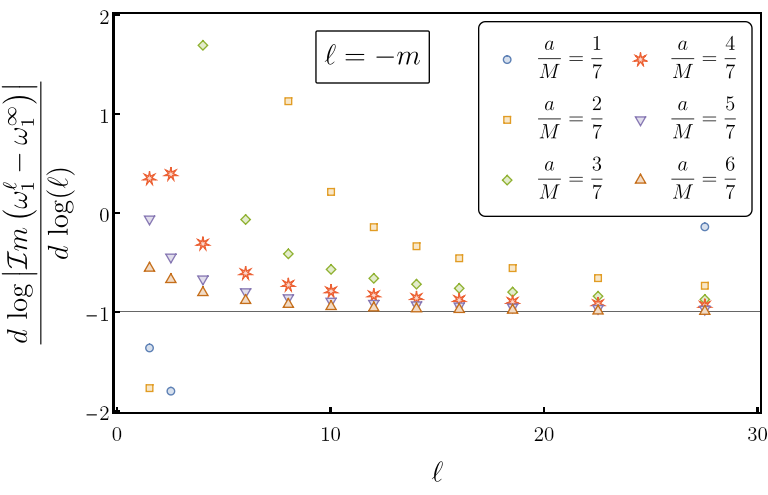}}

\caption{In this image, we study how quickly do $\ell=-m$ \ac{QNM} frequencies converge to the Eikonal limit prediction. As in figure~\ref{fig:vector_der_log_freq_log_l_l=m}, we find that the real and imaginary parts of the background converge as $\ell^{-1}$, and $\ell^{-2}$. This is consistent with the prediction in equation~\eqref{eq:QNM_Eikonal_correspondence}. Furthermore, we find that the real and imaginary of the correction converge to the theoretical prediction as $\ell^{-1}$.\label{fig:vector_der_log_freq_log_l_l=-m}}
\end{figure*}

\subsection{Pseudo-spectral methods \label{subsec:pseudospectral_methods}}

Take a set of functions $\{T_n\}_{n=1}^N$ that are orthonormal with respect to some inner product defined in a given domain. A common approach to solving \ac{ODE} or \ac{PDE} equations, is to expand its solution in this functional basis. Then, by understanding the behaviour of differential operators on the basis of functions, we can map the continuous differential equation onto a discrete problem. In general, we expect that by increasing $N$, higher frequency features will be captured and the discrete solution converges to the true continuous solution. In the literature, this approach is usually known as the Galerkin, or spectral method. Another possible approach is to sample the differential equation's solution in a discrete collocation grid $\{x_n\}_{n=1}^{N-1}$, that spans its domain. Then, we can map the differential operators into matrix operators, reducing the continuous problem into a simpler finite-dimensional version. If $\{T_n\}_{n=1}^N$ is a set of polynomials in ascending order of degree, there is a correspondence between the two methods. Choosing $\{x_n\}_{n=1}^{N-1}$ to be the $N-1$ roots of $\{T_n\}_{n=1}^N$, the solutions derived from the Galerkin and collocation methods are identical. Hence, collocation methods are many times known in the literature as \textit{pseudo-spectral} methods.\\

It has been shown that, in general, that when $\phi_n$ are the Chebyshev polynomials, pseudo-spectral methods have exponential convergence properties. Furthermore, because the corresponding grid heavily samples values close to the boundaries of the domain, it is especially suited to \acp{BVP}, see~\cite{Boyd:2000}. Throughout this paper, we will work with this basis. $T_n$ will denote the $n^{th}$ Chebyshev polynomial, whereas $\{x_i\}$ encodes the corresponding Chebyshev grid:
\begin{equation}
    x_i = \frac{1}{2}\l(1 - \cos\l(\frac{i \pi}{N}\r)\r),\qquad i=0,\cdots,N\,.\label{eq:define_cheb_grid}
\end{equation}
Here, the $N-1$ values in the middle of the grid are the roots of $T_{N-2}$, whereas $x_0$ and $x_N$ are needed to enforce the \acp{BC} of our problem.\\

Lets first understand how to use this approach to discretize \acp{ODE}. Take:
\begin{equation}
    L(x) f(x) = 0\,,
\end{equation}
where
\begin{equation}
    L(x):=\sum_n L^{(n)}(x) \dv[n]{x}\,,\label{eq:define_ODE_operator}
\end{equation}
is some differential operator, containing functions of $x$ and total derivatives with respect to $x$. To obtain the corresponding discrete version, we start by choosing some Chebyshev grid $x_i$. Then, sample $f(x)$ in the Chebyshev grid, defining $f_i:= f(x_i)$. Differentiation is a linear operation, thus, it must have a matrix representation in the discrete space. There must be some operator $D$ such that:
\begin{equation}
    f'(x_i) = \sum_j D_{ij} f_j\,.
\end{equation}
In fact, for a Chebyshev grid, the expression of $D$ can be found in section 6 of~\cite{Trefethen:2000}. The discretization of $L(x)$ can then be obtained by replacing $\dv[n]{x}$ with $D^n$ and $L_n(x)$ with diagonal matrices $\hat L^{(n)}_{ij}:=L^{(n)}(x_i) \mathbb I_{ij}$, where $\mathbb I$ is the identity matrix. Then, the discrete representation is:
\begin{equation}
    \sum_{j} \hat L_{ij} f_j := \sum_{n,k,j} \hat L^{(n)}_{ik}(D^n)_{kj} f_j\,.
\end{equation}
Treating $f_i$ as an unknown, we can solve the equation directly using standard methods.\\

The generalization to \acp{PDE} only requires small changes. Take
\begin{equation}
    L(x^1,\cdots,x^n)\,\Phi(x^1,\cdots,x^n)=0\,,\label{eq:general_PDE}
\end{equation}
where $L$ is a \ac{PDE} operator with $n$ variables. Start by defining $n$ Chebyshev grids $\{x^k_i\}$, that need not have the same size. The discretization of $\Phi$ will be a $n$-tensor 
\begin{equation}
    \Phi_{i_1,\cdots,i_n} :=\Phi(x_{i_1}^1,\cdots,x_{i_n}^n)\,.
\end{equation}
Then, the discrete version of partial derivative operators will be:
\begin{equation}
    \pdv{x^k}\rightarrow \underbrace{\mathbb I\otimes \cdots \otimes\mathbb I}_{k-1}\otimes D\otimes\underbrace{\mathbb I \otimes \cdots\otimes \mathbb I}_{n-k}\,.
\end{equation}
Mapping the remaining factors to "diagonal tensors", the discretization of~\eqref{eq:general_PDE} will take the form:
\begin{equation}
    \hat L_{i_1,\cdots,i_n}^{j_1,\cdots,j_n}\Phi_{j_1,\cdots,j_n}=0\label{eq:general_PDE_discrete}
\end{equation}
where the Einstein summation convention was assumed.\\

When using pseudo-spectral methods, it is important to quantify the size of the numerical error. Because these methods converge exponentially with the grid size, we can simply solve the \ac{EOM} using two different grid sizes. The absolute difference between the values obtained for each is a good estimate of the error size. In this paper, in most cases, we set the second grid to be roughly $10\%$ bigger than the first.\\

\subsection{Solving the background equation \label{subsec:numerics_background}}

Following the discretization procedure, we can now solve the \ac{EOM} and find the \acp{QNM}. We will first describe how to solve the background equation, equation~\eqref{eq:EOM_background_general}. As argued before, this equation separates into a system of two \acp{ODE}. Choosing a Chebyshev grid with size $N_x$ in the angular direction and a grid with size $N_r$ in the radial direction, the discrete representation of equation~\eqref{eq:EOM_background_general} is:
\begin{subnumcases}{\hskip -.5cm\label{eq:discrete_EOM_background}}
    A_x(J,\,m,\,\mathcal A_{\ell m},\, \omega_0) \cdot \underline \Theta=0\,,\label{eq:discrete_EOM_background_angular}\\
    A_r(J,\,m,\,\mathcal A_{\ell m},\, \omega_0) \cdot \underline R=0\,.\label{eq:discrete_EOM_background_radial}
\end{subnumcases}
Here $A_x$ and $A_r$ are matrix operators, $J$ and $m$ are parameters of the problem (see equations~\eqref{eq:define_J} and ~\eqref{eq:define_S}), $\mathcal A_{\ell m}$ and $\omega_0$ are to be treated as eigenvalues, and $\underline R$ and $\underline \Theta$ are the corresponding eigenvectors. Throughout the rest of this section, all underlined quantities are to be understood as vectors. The two equations depend on both eigenvalues, hence the system is not completely decoupled. Added difficulty comes from the fact that the second equation is quadratic in $\omega$. \\

To alleviate this, we can start by solving the \ac{EOM} in the Schwarzschild limit. As argued in section~\ref{subsec:master_equation}, in this limit, the angular equation can be solved analytically, yielding the spin-weighted spherical harmonics. $\mathcal A_{\ell m}$ reduces to $l(l+1) - s(s+1)$. The symmetry of the problem also removes the $m$ dependence from the second equation. Following section III.C of~\cite{Dias:2015nua}, we used a linearization approach to reduce~\eqref{eq:discrete_EOM_background_radial} to a standard eigenvalue problem. The resulting discrete eigenvalue problem can be inserted in a numeric eigenvalue solver. We used \textit{Mathematica}'s \textit{Eigensystem} method, selecting the \textit{Arnoldi} method. \footnote{Most eigenvalues obtained in this manner are not actual eigenvalues of the continuous system, to find the true frequencies the computation must be repeated with multiple grid sizes, keeping only the ones that are consistent throughout.} To achieve accurate results, the use of \textit{Mathematica}'s arbitrary precision capabilities was crucial. In general, we used at least 200 digits of precision.

To tackle the case with arbitrary spin, we used the Newton-Raphson approach, (see section VI.A of~\cite{Dias:2015nua}). This is an iterative method to find the roots of any map $f : \mathbb R^d \rightarrow \mathbb R^d$. Define $\underline x_0$ such that:
\begin{equation}
    f(\underline x_0)=0 \label{eq:define_NRap_eq}
\end{equation}

Taylor expanding $f(\underline x)$ around $\underline x_0$, we get:
\begin{equation}
    f(\underline x) =  f(\underline x_0) + \delta f(\underline x) \cdot \delta \underline x +  \mathcal O\l(\delta \underline x ^2\r),
\end{equation}
where $\delta \underline x := (\underline x - \underline x_0)$ and $\delta f(\underline x) := \pdv{f(\underline x)}{\underline x}$. By definition $f(\underline x_0) = 0$, thus, we can approximate $\delta \underline x$ with:
\begin{equation}
    \delta \underline x = \l(\delta f(\underline x)\r)^{-1}\,\cdot\,f(\underline x)\,.
\end{equation}
We update $\underline x \rightarrow \underline x - \delta \underline x$, and find a new estimate of $\delta \underline x$, iterating the process  until convergence. As long as the initial seed is close enough to $\underline x_0$, this method has quadratic convergence. However, if the seed is too distant from $\underline x_0$, the method will either not converge, or converge to a different root of $f$.\\

To apply this method to~\eqref{eq:discrete_EOM_background}, we must first fix some leftover \textit{gauge freedom}. The vectors $\underline \Theta$ and $\underline R$ can be rescaled by constant factors, thus, we must supplement the \ac{EOM} with two normalization conditions. We choose two vectors $\underline n^{(x)}$ and $\underline n^{(r)}$, and require:
\begin{equation}
\begin{cases}
    \underline n^{(x)}\,\cdot\, \underline \Theta = 1\,,\\
    \underline n^{(r)}\,\cdot\, \underline R = 1\,.
\end{cases}\label{eq:define_normalization}
\end{equation}

Taking $\underline x =\l(\underline \Theta, \underline R, \omega_0, \mathcal A_{\ell m}\r)$, we use equations~\eqref{eq:discrete_EOM_background} and~\eqref{eq:define_normalization} to define $f$. The Newton-Raphson update equation will be:
\begin{widetext}
\begin{equation}
\begingroup
\setlength \arraycolsep{5pt}
\begin{pmatrix}
    A_x & \pdv{\omega_0}A_x \,\cdot\,\underline \Theta & \pdv{\mathcal A_{\ell m}}A_x \,\cdot\,\underline \Theta\\
    A_r & \pdv{\omega_0}A_r \,\cdot\,\underline R & \pdv{\mathcal A_{\ell m}}A_r \,\cdot\,\underline R\\
    \underline n^{(x)} & 0 & 0 \\
    \underline n^{(r)} & 0 & 0 
\end{pmatrix}\cdot
\endgroup
\begin{pmatrix}
    \underline{\delta \Theta}^T\\
    \underline{\delta R}^T\\
    \delta \omega_0\\
    \delta \mathcal A_{\ell m}
\end{pmatrix}\,=\,
\begin{pmatrix}
    \l(A_x \cdot \underline \Theta\r)^T\\
    \l(A_r \cdot \underline R\r)^T\\
    \underline n^{(x)} \cdot \underline \Theta -1\\
    \underline n^{(r)} \cdot \underline R -1   
\end{pmatrix}
\end{equation}
\end{widetext}

Now, to find the \textit{background} \ac{QNM} frequency for a background of any spin, we just need an accurate enough initial guess. We take the Schwarzschild limit frequency and use it as a seed for a very slowly rotating \ac{BH}. After convergence, we can use the resulting frequency to obtain the \ac{QNM} of a \ac{BH} with a slightly larger spin. We iterate this, increasing $J$ gradually, until we cover the whole range of Kerr \acp{BH}, $J\in[0,1)$.\\

Deciding how much to increase $J$ at each iterative step, is subtle. To aid in this decision, and speed-up convergence, we can estimate $\dd \underline x_0 / \dd J$. Including $J$ in equation~\eqref{eq:define_NRap_eq}, we have $f(\underline x_0, J) = 0$. Thus, we can treat $\underline x_0$ as a function of $J$, parametrizing the level curves of $f$. Along these curves, we have:
\begin{equation}
    0=\dv{J}f\l(\underline x_0(J), J\r) = \pdv{f}{\underline x_0}\, \cdot\,\dv{x_0}{J} + \pdv{f}{J}\,.
\end{equation}
Solving with respect to $\dd \underline x_0 / \dd J$ we get:
\begin{equation}
    \dv{\underline x_0}{J}= -\l(\delta f\r)^{-1}\,\cdot \, \pdv{f}{J}\,.
\end{equation}
Plugging this in the definition of the derivative, we can estimate $\underline x_0(J+ \delta J)$:
\begin{equation}
    \underline x_0(J+ \delta J) \simeq  -\underline x_0(J)-\l(\delta f\r)^{-1}\,\cdot \, \pdv{f}{J}\,\delta J\,.
\end{equation}
In our calculations, we fixed the size of $\delta\mathcal Im(\omega_0)$, deducing the corresponding size of $\delta J$.

\subsection{Solving the perturbation equation  \label{subsec:numerics_perturbation}}

Using the background \acp{QNM}, we can now solve the perturbation equation, equation~\eqref{eq:EOM_correction_general}. Following the prescription in section~\ref{subsec:pseudospectral_methods} we obtain the corresponding discrete version. Because the \ac{EOM} is a 2 dimensional \ac{PDE}, $\Phi^{(0/1)}$ are represented by $N_x\cross N_r$ matrices, and $L^{(0)}$, $L^{(1)}$ and $\delta L$ will be represented by 4-tensors with size $N_x\cross N_r\cross N_x\cross N_r$. Fixing normalization freedom as was done~\eqref{eq:define_normalization}, we get:
\begin{subnumcases}{\hskip -.5cm \label{eq:define_discretization_perturbation}}
    L^{(0)}_{i j k l} \Phi_{kl}^{(1)} +\omega_1 \delta L^{(1)}_{i j k l} \Phi_{kl}^{(0)} = \mathcal L^{(1)}_{i j k l}\Phi_{kl}^{(0)},\label{eq:define_discretization_perturbation_EOM}\\[.5em]
    \underline n^{(x)}_i \cdot\l(\Phi^{(0)}_{ij} + \lambda \Phi^{(1)}_{ij}\r) \cdot \underline n^{(r)}_j = 1\label{eq:define_discretization_perturbation_normalization}
\end{subnumcases}
where we assumed the Einstein summation convention. These tensors depend on the same parameters as the background and on $(\omega_0, \mathcal A_{\ell m})$. Furthermore, the separability condition at $0$ order translates to $\Phi^{(0)} = \underline \Theta \otimes \underline R$. Using equation~\eqref{eq:define_normalization}, the normalization condition~\eqref{eq:define_discretization_perturbation_normalization} simplifies to $\underline n^{(x)} \cdot\Phi^{(1)} \cdot \underline n^{(r)} = 0$. Using \textit{Mathematica}'s \textit{Flattening} operator, we get a vector representation of $\Phi^{(0/1)}$. This in turn implies a matrix representation of the 4-tensors above. In this basis, the discrete \ac{EOM} take the form of a linear system:
\begin{equation}
    \mathcal M \cdot \underline x =  \underline b\,, \label{eq:linear_solve}
\end{equation}
where:
\begin{equation}
\begin{aligned}
    &\underline x = \l(\underline {\Phi^{(1)}},\,\omega_1\r)\,,\\[.7em]
    &\mathcal M = \begin{pmatrix}
    L^{(0)}& \underline{\l(\delta L ^{(1)}\cdot \Phi^{(0)}\r)}^T\\[.5em]
    \underline{n^{(r)}\otimes n^{(x)}} & 0
    \end{pmatrix}\,,\\[.7em]
    &\underline b^T = \begin{pmatrix}
    \underline {\l(L^{(1)}\cdot \Phi^{(0)}\r)}^T\\[.5em]
    0
    \end{pmatrix}\,.
\end{aligned}
\end{equation}
Underlined quantities are $N_x\cdot N_r$ vectors whereas not underlined quantities are scalars or $(N_x\cdot N_r)\cross(N_x\cdot N_r)$ matrices, depending on the context. The system~\eqref{eq:linear_solve} can then directly be solved with \textit{Mathematica}'s \textit{LinearSolve} method. In general $\mathcal M$ is non-singular, thus, we expect that there will be a single $\omega_1$ corresponding to each background frequency $\omega_0$.\\

Chebyshev differentiation matrices are very dense, thus, in general, the discrete systems we obtain have very high condition number. Thus, to solve the \ac{EOM} we must work with very high-precision arithmetic. This is particularly relevant in the near extremal limit, where the condition number increases very fast. $\mathcal M$ and $b$ are functions of $\omega_0$ and $\mathcal A_{\ell m}$, thus the system's precision is determined by the precision of these eigenvalues. To accurately solve the \ac{EOM} in the near extremal regime, we needed these eigenvalues to have at least 70 digits of precision. To achieve such feat, we had to solve the background \ac{EOM} using very large Chebyshev grids. In the most extreme scenario, we used $N_x = 250$ and $N_r=650$. When solving the background \ac{EOM}, this implies that each step of the Newton Raphson method involves solving a $\sim 900\cross 900$ system of equations, which is feasible with current computing capabilities.  However, when discretizing the first order \ac{PDE}, rank of $\mathcal M$ will be $N_r\cdot N_x\sim 160 000$. Even though this matrix is not very dense, solving this system would be very challenging. To work around this issue, we can use a clever form of interpolation to reduce the rank of $\mathcal M$. As mentioned in section~\ref{subsec:pseudospectral_methods}, there is a correspondence a Chebyshev collocation grid and the spectral decomposition in a Chebyshev polynomial basis. In fact, these coefficients are related by a linear map that can be found in~\cite{Boyd:2000}. To project the perturbation equations into a smaller Chebyshev grid, we find their spectral representation, truncate the spectral sum at a lower order, and then, map it back to a Chebyshev collocation grid.

\bibliography{references}

\end{document}